\documentclass[a4paper,fleqn,usenatbib]{mnras}

\usepackage[T1]{fontenc}
\usepackage{ae,aecompl}

\usepackage{rotating}

\usepackage{graphicx}	
\usepackage{amsmath}	
\usepackage{amssymb}	

\newcommand{\Msun}{$M_{\odot}$}
\newcommand{\Rsun}{$R_{\odot}$}
\newcommand{\kms}{km\,s$^{-1}$}
\newcommand{\vs}{$v \sin i$}
\newcommand{\teff}{$T_{\rm eff}$}
\newcommand{\lgg}{$\log\,{g}$}

\newcommand{\Bmean}{$\langle B \rangle$}
\newcommand{\dOmega}{${\rm d} \Omega$}

\title[Evolution of magnetic fields in solar-type stars]
      {The evolution of surface magnetic fields in young solar-type stars II: The early main sequence (250-650 Myr)\thanks{
Based on observations obtained at the Canada-France-Hawaii Telescope (CFHT) which is operated by the National Research Council of Canada, the Institut National des Sciences de l''Univers of the Centre National de la Recherche Scientique of France, and the University of Hawaii. The observations at the Canada-France-Hawaii Telescope were performed with care and respect from the summit of Maunakea which is a significant cultural and historic site. 
Also based on observations obtained at the Bernard Lyot Telescope (TBL, Pic du Midi, France) of the Midi-Pyr\'en\'ees Observatory, which is operated by the Institut National des Sciences de l'Univers of the Centre National de la Recherche Scientifique of France. }
}

\author[Folsom et al.]{C.P. Folsom$^{1,2,3}$\thanks{\tt colin.folsom@irap.omp.eu}, J. Bouvier$^{1}$, P. Petit$^{2,3}$, A. L\`ebre$^{4}$, L. Amard$^{4}$, A. Palacios$^{4}$, 
\newauthor
J. Morin$^{4}$, J.-F. Donati$^{2,3}$, A.A. Vidotto$^{5}$\\
$^1$Univ. Grenoble Alpes, CNRS, IPAG, F-38000 Grenoble, France \\
$^2$Universit\'e de Toulouse, UPS-OMP, IRAP, Toulouse, France\\
$^3$CNRS, Institut de Recherche en Astrophysique et Planetologie, 14, avenue Edouard Belin, F-31400 Toulouse, France\\
$^4$LUPM, Universit\'e de Montpellier, CNRS, Place Eug\`ene Bataillon, 34095, France\\
$^5$School of Physics, Trinity College Dublin, Dublin, Ireland\\
}

\date{Accepted XXX. Received YYY; in original form ZZZ}

\pubyear{2017}

\begin{document}
\label{firstpage}
\pagerange{\pageref{firstpage}--\pageref{lastpage}}
\maketitle

\begin{abstract}
There is a large change in surface rotation rates of sun-like stars on the pre-main sequence and early main sequence. Since these stars have dynamo driven magnetic fields, this implies a strong evolution of their magnetic properties over this time period. The spin-down of these stars is controlled by interactions between stellar winds and magnetic fields, thus magnetic evolution in turn plays an important role in rotational evolution. We present here the second part of a study investigating the evolution of large-scale surface magnetic fields in this critical time period. We observed stars in open clusters and stellar associations with known ages between 120 and 650 Myr, and used spectropolarimetry and Zeeman Doppler Imaging to characterize their large-scale magnetic field strength and geometry. We report 15 stars with magnetic detections here. These stars have masses from 0.8 to 0.95 $M_{\odot}$, rotation periods from 0.326 to 10.6 days, and we find large-scale magnetic field strengths from 8.5 to 195 G with a wide range of geometries. We find a clear trend towards decreasing magnetic field strength with age, and a power-law decrease in magnetic field strength with Rossby number. There is some tentative evidence for saturation of the large-scale magnetic field strength at Rossby numbers below 0.1, although the saturation point is not yet well defined. Comparing to younger classical T Tauri stars, we support the hypothesis that differences in internal structure produce large differences in observed magnetic fields, however for weak lined T Tauri stars this is less clear.
\end{abstract}


\section{Introduction}

The rotational evolution of solar-type stars, from the pre-main sequence to nearly the age of the sun, is increasingly well characterized \citep[for a recent review seen][]{Bouvier2013-rotation-evol-review}.  Stars early on the pre-main sequence (PMS) strongly interact with their disks, and this regulates their rotation rates, despite accretion and contraction.  Stars eventually stop strongly interacting with their disks while still on the PMS, and by conservation of angular momentum spin up.  Stars also lose angular momentum, by the interaction of stellar winds and magnetic fields.  This is a slower process than spin up due to contraction, thus once stars reach the main sequence they begin spinning down. 

The magnetic evolution of young solar-type stars is less well characterized.  In particular direct observations of the large-scale magnetic field are needed.  The strong rotational evolution likely affects the dynamo generated magnetic fields in these stars, thus there should be large changes in the stellar magnetic properties.  Additionally, the spin-down of these stars is controlled by the magnetic field, thus to fully understand the angular momentum loss, we must have well characterized large-scale magnetic properties \citep[e.g.][]{Vidotto2011-MHD-wind-V374Peg, Matt2012-magnetic-breaking-formulation, Reville2015-MHD-wind-torque, Reville2016-3D-MHD-wind-Jdot}.  Indeed, for studies of stellar winds and angular momentum loss, it is the large-scale component of the magnetic field that is important \citep{Jardine2017-wind-lowres-mag, See2017-spindown-ZDImaps}.

Studies of individual young solar-like stars, using spectropolarimetry to measure large-scale magnetic fields, have been performed for a number of stars.  The earliest studies focused on very rapid rotators with strong magnetic fields \citep[e.g][]{Donati1999-ABDor-mag-geom, Donati2003-sempol-monitoring-cool-active}.  More recently a number of slower rotators with weaker magnetic fields have been investigated \citep[e.g][]{Petit2008-sunlike-mag-geom, Marsden2011-hd141943-sempol-hd101412, Jeffers2014-epsEri-mag-var, Waite2015-2-young-solar-B, BoroSaikia2015-HNPeg, doNascimento2016-kappaCeti-mag-wind, Hackman2016-young-solarlike-small, Waite2017-young-solar-EKDra}.  However many of these stars are field objects and have poorly determined ages, and they span a wide range of masses and spectral types, making an inhomogeneous sample.

Long term variability in the large-scale magnetic fields of G and K stars has been studied using spectropolarimetry for a number of stars \citep[e.g.][]{Donati2003-sempol-monitoring-cool-active, Jeffers2011-nonSolar-HD171488-3epochs, Jeffers2014-epsEri-mag-var, Mengel2016-tauBoo-mag-update, BoroSaikia2016-61CygA-solar-like, Scalia2017-B-linear-regression-cool-stars}.  In some cases cyclical variability in the large-scale magnetic field has been found \citep[e.g.][]{Mengel2016-tauBoo-mag-update, BoroSaikia2016-61CygA-solar-like}, but in other cases no clear periodicity is yet known.  This variability amounts to a factor of a couple in magnetic field strength, and often large changes in magnetic geometry.   Despite this variability, trends in large-scale magnetic field strength and geometry with mass and rotation period have been found using large samples of stars \citep[e.g.][]{Donati2009-ARAA-magnetic-fields, Vidotto2014-magnetism-age-rot, Folsom2016-Toupies1}, and multi-epoch studies of stars have suggestions of further trends \citep[e.g.][]{See2016-mag-geom-cycles-PolTor-Rossby}. 

A study compiling literature magnetic field results was performed by \citet{Vidotto2014-magnetism-age-rot}.  They found a clear trend of decreasing magnetic field with age, although the somewhat heterogeneous sample produced significant scatter.  They also found trends in decreasing magnetic field with rotation period and Rossby number.  This is expected in the context of stellar spin-down, where the older stars rotate more slowly, and thus have weaker dynamos.  \citet[][the first paper in this series]{Folsom2016-Toupies1} began a study focusing on a more well defined sample, with ages established from clusters or co-moving groups, and found broadly similar results to \citet{Vidotto2014-magnetism-age-rot}.  \citet{Rosen2016-mag-young-solar-twins} performed a study on a small sample of stars, although with multiple epochs of observation for most stars in their sample.  They also found similar results, although with much of the scatter in their sample apparently driven by intrinsic long-term variability of the large-scale magnetic fields.  

We present here the second set of results from our ongoing study of magnetic fields in young solar-type stars.  The first set of results from this study was published in \citet{Folsom2016-Toupies1} (henceforth Paper I).  The whole study focuses on stars between 20 and 650 Myr old, and 0.7 to 1.2 $M_{\odot}$, while this paper focuses on a subset of older stars with a narrower range of mass.  We use spectropolarimetry to directly detect Zeeman splitting in polarized spectra.  With time a series of rotationally modulated spectra, we can use Zeeman Doppler Imaging (ZDI) to invert the polarized spectra and reconstruct the large-scale stellar magnetic field strength and geometry.  We observe a relatively large number of stars in order to investigate trends in magnetic field with age, rotation rate, and Rossby number, and to overcome scatter due to long term variability in the magnetic fields.  In this paper we specifically focus on completing the older part of our sample, from 250 to 650 Myr old, with three additional stars in AB Dor (120 Myr old). 

This work is part of the `TOwards Understanding the sPIn Evolution of Stars' (TOUPIES) project\footnote{http://ipag.osug.fr/Anr\_Toupies/}.  In particular, observations from the Canada France Hawaii Telescope are from the large program `History of the Magnetic Sun'.

\section{Observations}
\label{observations}

Observations for this study were obtained at the Canada France Hawaii Telescope (CFHT) using the ESPaDOnS instrument (\citealt{Donati2003-ESPaDOnS-descript}; see also \citealt{Silvester2012-data-paper}),
and at the T\'elescope Bernard Lyot (TBL) at the Observatoire du Pic du Midi, France, using the Narval instrument \citep{Auriere2003-Narval-early}.  Both instruments are high resolution \'echelle spectropolarimeters, and Narval is a direct copy of ESPaDOnS.  Both instruments have a Cassegrain mounted polarimeter module, connected by optical fiber to a bench mounted, cross dispersed \'echelle spectrograph.  They have a resolution of R$\sim$65000 and cover the wavelength range from 3700 to 10500 \AA.
Observations were obtained using spectropolarimetric mode, which provides simultaneous Stokes $V$ (circularly polarized) and $I$ (total intensity) spectra.   Observations were reduced using the Libre-ESpRIT package \citep{Donati1997-major}, as in Paper I.  

A series of observations were obtained for each star, with a goal of 15 observations distributed evenly over a couple rotation cycles of the star.  
Observations of a single star were typically obtained within a two week period, to limit the possibility of intrinsic variations in the magnetic field during the observations.  This is the same general observing strategy as in Paper I.  However, due to varying observing conditions, for some targets fewer observations were obtained, or the time span of the observations was longer.  
A minimum target peak S/N in the reduced $V$ spectra of 100 (per spectral pixel) was adopted, although for stars with weaker magnetic fields a higher target S/N was used.  This was achieved for almost all observations, except for a few cases of observations obtained in poor weather conditions.  For a couple targets (BD-072388 and HD 6569) exposure times were increased during the observing run, to ensure we obtained consistent detections.   A summary of the observations is presented in Table \ref{observations-table}.

\begin{table*}
\centering
\caption{Summary of observations obtained. Exposure times are for a full sequence of 4 sub-exposures, 
and the S/N values are the peak in the $V$ spectrum (per 1.8 \kms\ spectral pixel, typically near 730 nm). 
For three stars the exposure times were modified during the set of observations to ensure an adequate S/N (for BD-072388 and HD 6569), or to make efficient use of time when the target S/N was exceeded (for Mel 25-151). }
\begin{tabular}{lcc@{~ }cccccc}
\hline
Object          & Assoc.   & RA          & Dec.        & Dates of             &Telescope &Integration& Num & S/N   \\
                &          &             &             & Observations         &Semester   & Time (s) & Obs & Range \\
\hline
BD-072388       & AB Dor   & 08 13 50.99 & -07 38 24.6 & 16 Jan - 25 Jan 2016 & CFHT 15B &  496,  992 & 20 & 126-224 \\
HIP 10272       & AB Dor   & 02 12 15.41 & +23 57 29.5 & 16 Oct - 01 Nov 2014 & TBL 14B  &  400 & 13 &  80-165 \\
HD 6569         & AB Dor   & 01 06 26.15 & -14 17 47.1 & 18 Sep - 29 Sep 2015 & CFHT 15B &  576, 1152 & 12 &  77-170 \\
HH Leo          & Her-Lyr  & 11 04 41.47 & -04 13 15.9 & 06 Mar - 26 May 2015 & TBL 15A  & 1520 & 14 & 225-352 \\
EP Eri          & Her-Lyr  & 02 52 32.13 & -12 46 11.0 & 24 Oct - 01 Nov 2014 & TBL 14B  &  160 &  9 & 158-265 \\
EX Cet          & Her-Lyr  & 01 37 35.47 & -06 45 37.5 & 01 Sep - 27 Sep 2014 & TBL 14B  &  800 & 13 & 158-264 \\
AV 2177         & Coma Ber & 12 33 42.13 & +25 56 34.1 & 09 Apr - 19 Jun 2014 & CFHT 14A & 3600 & 21 & 124-212 \\
AV 1693         & Coma Ber & 12 27 20.69 & +23 19 47.5 & 24 Mar - 09 Apr 2015 & CFHT 15A & 5608 & 15 & 170-327 \\
AV 1826         & Coma Ber & 12 28 56.43 & +26 32 57.4 & 09 Apr - 19 Jun 2014 & CFHT 14A & 3600 & 22 & 115-199 \\
TYC 1987-509-1  & Coma Ber & 11 48 37.71 & +28 16 30.6 & 24 Mar - 01 Apr 2015 & CFHT 15A & 6740 &  9 & 312-330 \\
AV 523          & Coma Ber & 12 12 53.24 & +26 15 01.5 & 16 Feb - 02 Mar 2016 & CFHT 16A & 3920 & 30 & 108-179 \\
Mel 25-151      & Hyades   & 05 05 40.38 & +06 27 54.6 & 13 Jan - 28 Jan 2016 & CFHT 15B & 3080, 2584 & 15 & 280-207 \\
Mel 25-43       & Hyades   & 04 23 22.85 & +19 39 31.2 & 17 Nov - 02 Dec 2015 & CFHT 15B & 1880 & 13 & 219-293 \\
Mel 25-21       & Hyades   & 04 16 33.48 & +21 54 26.9 & 18 Sep - 01 Oct 2015 & CFHT 15B & 1420 & 14 & 136-223 \\
Mel 25-179      & Hyades   & 04 27 47.04 & +14 25 03.9 & 17 Nov - 02 Dec 2015 & CFHT 15B & 2200 & 12 & 266-303 \\
Mel 25-5        & Hyades   & 03 37 34.98 & +21 20 35.4 & 18 Sep - 01 Oct 2015 & CFHT 15B & 2120 & 14 & 179-263 \\
\hline
\end{tabular} 
\label{observations-table} 
\end{table*}

\subsection{Sample Selection}

The sample of stars in this paper followed the same selection criteria as in Paper I, however in here we focus on stars in the age range from 250 to 650 Myr, with three additional targets of interest in AB Dor (120 Myr).  Targets were selected from lists of members in nearby stellar associations or clusters, and only stars with published rotation periods were used.  In this paper we focused on a mass range from 0.8-0.95 \Msun, and attempted to cover the full range of periods available in the associations. 
Relatively bright targets were selected to ensure we could meet our S/N targets.  

The targets in this study are from the AB Dor association \citep[120 Myr][]{Luhman2005-ABDor-age, Barenfeld2013-ABDor-memb-age}, the Her-Lyr association \citep[257 Myr][]{Lopez-Santiago2006-HerLyr-ABDor-assoc, Eisenbeiss2013-HerLyr-age}, the Coma Ber cluster \citep[584 Myr][]{CollierCameron2009-ComaBer-periods, Delorme2011-periods-Hyades-Praesepe}, and the Hyades \citep[625 Myr][]{Perryman1998-Hyades-age-dist}.  They span a range of effective temperatures from 4700 K to 5400 K and masses from 0.8 to 0.95 \Msun.  This relatively narrow range in mass provides a sample with relatively consistent internal structure.  Rotation periods range from 0.326 days to 10.5 days, although almost all stars in this study rotate slower than 6 days, while in Paper I most stars rotated faster than 6 days.  Combined, these two studies provide a wide range of rotation rates and Rossby numbers.  

Individual stars are discussed in Appendix \ref{Individual Targets}, and the physical parameters of the stars are summarized in \ref{fundimental-param-table}.

\section{Fundamental physical parameters}
\label{fund-param}

\subsection{Spectroscopic analysis}
\label{spectrum-fitting}

\subsubsection{Primary analysis}
The physical atmospheric parameters \teff, \lgg, \vs, and microturbulence ($\xi$) were derived for all stars in the sample.  This was done by directly fitting synthetic spectra to our observed Stokes $I$ spectra.  The initial analysis was done using the {\sc Zeeman} spectrum synthesis program \citep{Landstreet1988-Zeeman1,Wade2001-zeeman2_etc}, using the same methodology as Paper I.  

The observed spectra were first normalized to their continuum.  A low order polynomial was fit through carefully chosen continuum points in the observations, then the observations were divided by that continuum polynomial, as in Paper I.  

Stellar parameters were derived by fitting synthetic spectra to observed spectra, though $\chi^2$ minimization, using metallic lines.  The synthetic spectra were produced with {\sc Zeeman} \citep{Landstreet1988-Zeeman1,Wade2001-zeeman2_etc}, and fit to observations using the Levenberg-Marquardt procedure of \citet{Folsom2012-HAeBe-abundances} \citep[see also][]{Folsom2013-PhD-thesis}.  Atomic data were extracted from the Vienna Atomic Line Database \citep[VALD][]{Ryabchikova1997-VALD-early, Kupka1999-VALD, Ryabchikova2015-VALD3}, through `extract stellar' requests.  Model atmospheres from the MARCS grid \citep{Gustafsson2008-MARCS-grid} were used.  A comparison between fitting results with MARCS and {\sc atlas9} \citep{Kurucz1993-ATLAS9etc} models, in this parameter range, showed that these models produced results consistent to much less than the uncertainties.  

Solar chemical abundances were assumed for this analysis, except for the Hyades targets.  Since all stars in this study are relatively young and near the sun, they likely have very nearly solar abundances.  
The Hyades is well established to have a mildly enhanced metallicity of [Fe/H] = 0.13, with a star to star dispersion of $\sim$0.05 \citep{Perryman1998-Hyades-age-dist, Paulson2003-Hyades-abun, Heiter2014-clusters-metallicity}.  Thus in our primary spectroscopic analysis, we assume a metallicity of [Fe/H] = 0.13 for the spectrum synthesis.  However the exact value of [Fe/H] has a relatively small impact on the derived stellar parameters, at the level of the quoted uncertainties or smaller.  

Spectra were fit independently in five spectral windows (6000-6100, 6100-6276, 6314-6402, 6402-6500, and 6600-6700 \AA\ excluding telluric features and Balmer lines).  The results of these five independent fits were averaged to produce the final best values, and the standard deviation of these results was taken as the final uncertainty.  This allows for a robust inclusion of systematic errors, such as errors in atomic data or continuum placement.  If we consider the differences in results for the different windows to be driven primarily by random errors, then a better uncertainty estimate would be the standard error on the mean (standard deviation divided by the square root of the number of windows).  This would scale our formal uncertainties down by a factor of $\sim$0.45.  However to be cautious and ensure we account for the range of possible systematic errors, we report standard deviation here.

\subsubsection{Secondary analysis and comparison}

A second independent analysis of the stellar parameters was performed, as in Paper I, which was crosschecked against the analysis with {\sc Zeeman}.  This used spectrum synthesis from MARCS models of stellar atmospheres, and fit for lithium abundances ($A_{\rm Li}$) and metallicity in addition to \teff, \lgg, \vs, and microturbulence values.  This analysis also proceeded by directly fitting synthetic spectra to observations though $\chi^2$.  

\begin{figure*}
  \centering
  \includegraphics[width=6.5in]{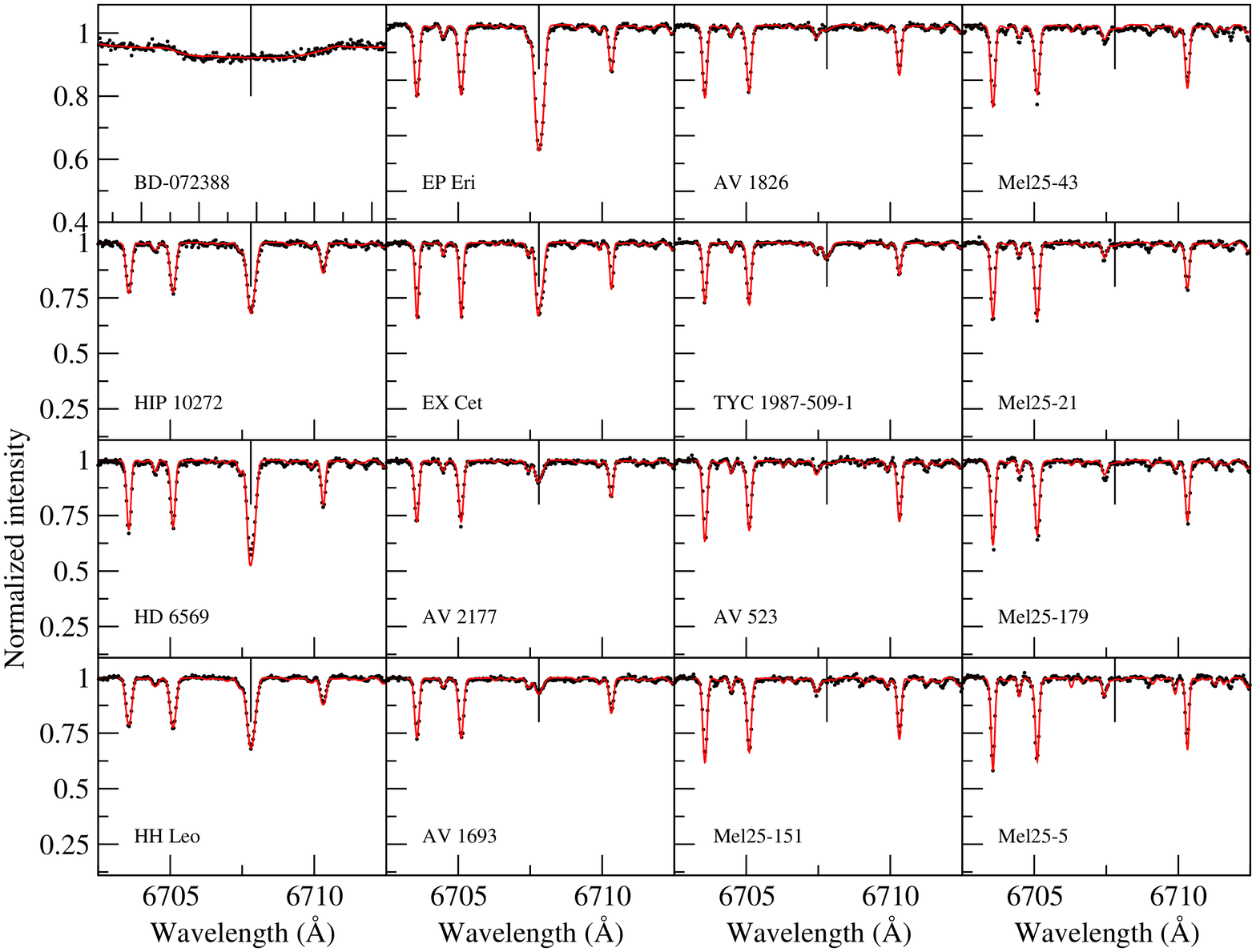}
  \caption{The region around the 6707.8 \AA\ lithium line (indicated with a vertical tick) for the stars in the sample. Observations are points and the best fit synthetic spectra are over-plotted with smooth lines. }
  \label{fig-Li-line}
\end{figure*}

This analysis used the region around the 6707.8 \AA\ lithium line, with checks from regions around the Ca IR triplet and H$\beta$.  It used synthetic spectra from the TurboSpectrum code \citep{Alvarez1998-TurboSpectrum-etc}, atomic data from VALD with some modifications, and the fitting procedure discussed in \citet{CantoMartins2011-Li-abun-M67}.  Sample fits to the 6707.8 \AA\ lithium line are provided in Fig.~\ref{fig-Li-line}.  While the underlying model atmospheres in both analyses are similar, the spectral regions and spectrum synthesis tools are entirely independent, thus the more important systematic uncertainties are independent.  

In this analysis metallicity was included as a free parameter.  For most stars we find metallicities consistent with zero, supporting the assumption used in the previous analysis, with an average (excluding Hyades) members of [Fe/H] = 0.036, relative to an uncertainty of 0.05.  However, for several Hyades members we find enhanced metallicities by [Fe/H] of +0.1 to +0.15.  

The agreement between the two analyses is generally good, with our values usually consistent within $1\sigma$.  For \lgg, and microturbulence our values always differ by less than $2\sigma$, and the majority agree within $1\sigma$.  For \vs, three of the stars (HIP 10272, HD 6569, and Mel25-151) disagree by a little over $2\sigma$ (a little over 1 \kms), however the rest show better agreement with the majority (12 stars) better than $1\sigma$.  Thus we don't consider this marginal disagreement serious.
\teff\ is the most sensitive parameter to systematic errors in metallicity or continuum normalization, relative to the formal uncertainties.  However, all stars agree in \teff\ within  $2\sigma$, and the majority agree within $1\sigma$.

\begin{table*}
  \caption{Derived fundamental parameters for the stars in our sample.  $P_{\rm rot}$ are the adopted rotation periods, containing a mix of literature and our spectropolarimetric periods, as discussed in Appendix~\ref{Individual Targets}.  Radial velocities ($v_r$) are the averages and standard deviations of our observations.  Lithium abundances are in the form $\log(N_{\rm Li}/N_{\rm H})+12$.
  Distance references: $^1$ \citet{van_Leeuwen2007-Hipparcos_book}, $^2$ \citet{Torres2008-youngNearbyAssoc}, $^3$ \citet{Gaia2016-data-release1} }  
\begin{sideways}
\begin{tabular}{lccccccccc}
\hline
Star           & Assoc.    & Age            & $P_{\rm rot}$         &  \teff          &  \lgg            &  \vs              &  $\xi$           & $v_r$               & $i$              \\
               &           & (Myr)          &  (days)             &  (K)            &                  &  (km/s)           &  (km/s)          &  (km/s)             & ($^{\circ}$)     \\
\hline
BD-072388      & AB Dor    & $ 120 \pm 10 $ & $0.32595\pm 0.0005$ & $ 5121 \pm 137$ & $ 4.44 \pm 0.18$ & $126.1 \pm 1.7  $ & $ 1.26 \pm 0.40$ & $26.14   \pm 0.56 $ & $ 38^{+13}_{-13}$ \\
HIP10272       & AB Dor    & $ 120 \pm 10 $ & $ 6.13  \pm 0.03  $ & $ 5281 \pm 79 $ & $ 4.61 \pm 0.15$ & $ 7.17 \pm 0.21 $ & $ 1.05 \pm 0.33$ & $ 0.490  \pm 0.028$ & $ 55^{+20}_{-20}$ \\
HD 6569        & AB Dor    & $ 120 \pm 10 $ & $ 7.13  \pm 0.05  $ & $ 5118 \pm 95 $ & $ 4.56 \pm 0.08$ & $ 5.25 \pm 0.29 $ & $ 1.14 \pm 0.09$ & $ 7.940  \pm 0.029$ & $ 77^{+13}_{-15}$ \\
HH Leo         & Her-Lyr   & $ 257 \pm 46 $ & $ 5.915 \pm 0.017 $ & $ 5402 \pm 73 $ & $ 4.62 \pm 0.11$ & $ 6.70 \pm 0.27 $ & $ 1.31 \pm 0.31$ & $ 18.931 \pm 0.035$ & $ 67^{+19}_{-9 }$ \\
EP Eri         & Her-Lyr   & $ 257 \pm 46 $ & $ 6.76  \pm 0.20  $ & $ 5125 \pm 87 $ & $ 4.53 \pm 0.13$ & $ 5.37 \pm 0.38 $ & $ 1.25 \pm 0.21$ & $ 18.273 \pm 0.036$ & $ 85^{+5 }_{-30}$ \\
EX Cet         & Her-Lyr   & $ 257 \pm 46 $ & $ 7.15  \pm 0.10  $ & $ 5326 \pm 63 $ & $ 4.65 \pm 0.13$ & $ 2.83 \pm 0.48 $ & $ 1.12 \pm 0.28$ & $ 11.827 \pm 0.034$ & $ 28^{+8 }_{-6 }$ \\
AV 2177        & Coma Ber  & $ 584 \pm 10 $ & $ 8.98  \pm 0.12  $ & $ 5316 \pm 61 $ & $ 4.69 \pm 0.12$ & $ 4.62 \pm 0.31 $ & $ 1.00 \pm 0.25$ & $-1.30 \pm 1.26$ (SB1) & $ 77^{+14}_{-22}$ \\
AV 1693        & Coma Ber  & $ 584 \pm 10 $ & $ 9.05  \pm 0.10  $ & $ 5372 \pm 63 $ & $ 4.69 \pm 0.10$ & $ 4.77 \pm 0.31 $ & $ 1.06 \pm 0.26$ & $ 0.628  \pm 0.024$ & $ 75^{+15}_{-20}$ \\
AV 1826        & Coma Ber  & $ 584 \pm 10 $ & $ 9.34  \pm 0.08  $ & $ 5098 \pm 76 $ & $ 4.67 \pm 0.35$ & $ 4.66 \pm 0.35 $ & $ 1.12 \pm 0.2 $ & $ 0.744  \pm 0.090$ (SB1) & $ 65^{+25}_{-18}$ \\
TYC 1987-509-1 & Coma Ber  & $ 584 \pm 10 $ & $ 9.43  \pm 0.10  $ & $ 5379 \pm 68 $ & $ 4.68 \pm 0.11$ & $ 4.88 \pm 0.30 $ & $ 1.09 \pm 0.26$ & $ 0.663  \pm 0.020$ & $ 67^{+15}_{-15}$ \\
AV 523         & Coma Ber  & $ 584 \pm 10 $ & $ 11.1  \pm 0.20  $ & $ 4769 \pm 74 $ & $ 4.61 \pm 0.14$ & $ 3.89 \pm 0.26 $ & $ 1.03 \pm 0.20$ & $ 0.509  \pm 0.032$ & $ 50^{+10}_{-10}$ \\
Mel25-151      & Hyades    & $ 625 \pm 50 $ & $ 10.41 \pm 0.10  $ & $ 4920 \pm 73 $ & $ 4.43 \pm 0.10$ & $ 4.83 \pm 0.33 $ & $ 1.27 \pm 0.26$ & $37.98   \pm 0.24 $ (SB1) & $ 52^{+12}_{-12}$ \\
Mel25-43       & Hyades    & $ 625 \pm 50 $ & $ 9.90  \pm 0.10  $ & $ 5121 \pm 71 $ & $ 4.51 \pm 0.17$ & $ 4.01 \pm 0.28 $ & $ 0.88 \pm 0.18$ & $37.78   \pm 0.16 $ (SB1) & $ 46^{+13}_{-9 }$ \\
Mel25-21       & Hyades    & $ 625 \pm 50 $ & $ 9.73  \pm 0.20  $ & $ 5236 \pm 88 $ & $ 4.39 \pm 0.18$ & $ 3.65 \pm 0.38 $ & $ 1.03 \pm 0.20$ & $38.285  \pm 0.023$ & $ 53^{+18}_{-11}$ \\
Mel25-179      & Hyades    & $ 625 \pm 50 $ & $ 9.70  \pm 0.10  $ & $ 5023 \pm 55 $ & $ 4.46 \pm 0.12$ & $ 4.04 \pm 0.28 $ & $ 1.15 \pm 0.19$ & $39.684  \pm 0.053$ (SB1) & $ 65^{+17}_{-17}$ \\
Mel25-5        & Hyades    & $ 625 \pm 50 $ & $ 10.57 \pm 0.10  $ & $ 4916 \pm 97 $ & $ 4.35 \pm 0.22$ & $ 3.28 \pm 0.34 $ & $ 1.01 \pm 0.19$ & $31.618  \pm 0.024$ & $ 61^{+29}_{-14}$ \\
\hline
Star           & Distance              & $L$             & $R$             & $M$                 & $\tau_{\rm conv}$   &  Rossby             & $A_{\rm Li}$     & ${\rm d}\Omega$      & ${\rm d}\Omega / \Omega_{\rm eq}$\\
               & (pc)                  & ($L_\odot$)      & ($R_\odot$)      & ($M_\odot$)          &    (days)         &  number             & (dex)          & (rad/day)              &    \\
\hline
BD-072388      & $93.0  \pm 18.6$ $^2$ & $0.38 \pm 0.08$ & $0.78 \pm 0.09$ & $0.85^{+0.05}_{-0.04}$ & $22.5^{+5.5}_{-2.2}$ &$0.014^{+0.002}_{-0.003}$ & $2.7 \pm 0.1$ & $0.16^{+0.23}_{-0.23}$ n & $0.008^{+0.012}_{-0.012}$ \\
HIP10272       & $36.6  \pm 1.6 $ $^1$ & $0.45 \pm 0.10$ & $0.80 \pm 0.08$ & $0.90^{+0.04}_{-0.04}$ & $20.2^{+3.0}_{-1.0}$ & $0.30^{+0.02 }_{-0.04 }$ & $2.2 \pm 0.1$ & $0.20^{+0.10}_{-0.10}$ m & $0.20^{+0.10}_{-0.10}$ \\
HD 6569        & $45.8  \pm 0.5 $ $^3$ & $0.36 \pm 0.01$ & $0.76 \pm 0.03$ & $0.85^{+0.04}_{-0.04}$ & $23.3^{+1.2}_{-1.2}$ & $0.32^{+0.02 }_{-0.02 }$ & $2.0 \pm 0.1$ & $0.30^{+0.25}_{-0.50}$ n & $0.34^{+0.28}_{-0.57}$ \\
HH Leo         & $25.9  \pm 0.3 $ $^3$ & $0.54 \pm 0.02$ & $0.84 \pm 0.03$ & $0.95^{+0.05}_{-0.05}$ & $18.1^{+2.6}_{-0.9}$ & $0.33^{+0.02 }_{-0.04 }$ & $2.5 \pm 0.1$ & $0.10^{+0.02}_{-0.02}$ D & $0.10^{+0.02}_{-0.02}$ \\
EP Eri         & $10.35 \pm 0.04$ $^1$ & $0.30 \pm 0.06$ & $0.72 \pm 0.08$ & $0.85^{+0.04}_{-0.05}$ & $22.5^{+3.1}_{-1.1}$ & $0.30^{+0.03 }_{-0.04 }$ & $2.9 \pm 0.1$ & $< 0.2$	          n &  $< 0.2$ \\
EX Cet         & $24.0  \pm 0.2 $ $^3$ & $0.46 \pm 0.02$ & $0.86 \pm 0.05$ & $0.90^{+0.04}_{-0.04}$ & $20.1^{+1.0}_{-1.0}$ & $0.36^{+0.02 }_{-0.02 }$ & $2.3 \pm 0.1$ & -	            & - \\
AV 2177        & $82.3  \pm 2.6 $ $^3$ & $0.43 \pm 0.03$ & $0.78 \pm 0.03$ & $0.90^{+0.04}_{-0.04}$ & $20.1^{+1.0}_{-1.0}$ & $0.45^{+0.03 }_{-0.03 }$ & $1.6 \pm 0.1$ & $0.05^{+0.05}_{-0.02}$ m & $0.07^{+0.07}_{-0.03}$ \\
AV 1693        & $84.7  \pm 2.0 $ $^3$ & $0.52 \pm 0.03$ & $0.83 \pm 0.03$ & $0.90^{+0.05}_{-0.05}$ & $20.1^{+1.0}_{-2.3}$ & $0.45^{+0.07 }_{-0.03 }$ & $1.4 \pm 0.1$ & $0.22^{+0.10}_{-0.08}$ D & $0.32^{+0.14}_{-0.12}$ \\
AV 1826        & $87.4  \pm 3.4 $ $^3$ & $0.39 \pm 0.03$ & $0.80 \pm 0.04$ & $0.85^{+0.04}_{-0.04}$ & $22.6^{+3.6}_{-1.1}$ & $0.41^{+0.03 }_{-0.06 }$ & $0.7 \pm 0.2$ & $0.09^{+0.04}_{-0.03}$ D & $0.13^{+0.06}_{-0.04}$ \\
TYC 1987-509-1 & $88.3  \pm 2.4 $ $^3$ & $0.52 \pm 0.03$ & $0.83 \pm 0.03$ & $0.90^{+0.05}_{-0.05}$ & $20.1^{+1.0}_{-2.4}$ & $0.47^{+0.07 }_{-0.03 }$ & $1.0 \pm 0.2$ & $0.07^{+0.20}_{-0.15}$ n & $0.11^{+0.30}_{-0.23}$ \\
AV 523         & $85.7  \pm 2.6 $ $^3$ & $0.24 \pm 0.02$ & $0.72 \pm 0.03$ & $0.80^{+0.04}_{-0.05}$ & $26.2^{+2.6}_{-1.3}$ & $0.42^{+0.03 }_{-0.04 }$ & $0.2 \pm 0.2$ & $0.00^{+0.20}_{-0.20}$ n & $0.00^{+0.35}_{-0.35}$ \\
Mel25-151      & $52.0  \pm 8.2 $ $^1$ & $0.35 \pm 0.11$ & $0.82 \pm 0.13$ & $0.85^{+0.05}_{-0.05}$ & $26.9^{+6.0}_{-1.8}$ & $0.39^{+0.03 }_{-0.07 }$ & $0.2 \pm 0.3$ & $0.03^{+0.06}_{-0.07}$ n & $0.05^{+0.10}_{-0.12}$ \\
Mel25-43       & $60.6  \pm 4.8 $ $^1$ & $0.38 \pm 0.08$ & $0.79 \pm 0.08$ & $0.85^{+0.05}_{-0.04}$ & $22.6^{+3.6}_{-1.1}$ & $0.44^{+0.03 }_{-0.06 }$ & $0.2 \pm 0.2$ & $0.18^{+0.16}_{-0.16}$ n & $0.28^{+0.25}_{-0.25}$ \\
Mel25-21       & $51.1  \pm 0.7 $ $^3$ & $0.56 \pm 0.02$ & $0.91 \pm 0.04$ & $0.90^{+0.05}_{-0.05}$ & $24.2^{+1.5}_{-3.7}$ & $0.40^{+0.08 }_{-0.03 }$ & $0.8 \pm 0.3$ & $0.20^{+0.15}_{-0.13}$ m & $0.31^{+0.23}_{-0.20}$ \\
Mel25-179      & $49.1  \pm 0.7 $ $^3$ & $0.40 \pm 0.02$ & $0.84 \pm 0.03$ & $0.85^{+0.04}_{-0.04}$ & $27.6^{+1.9}_{-5.3}$ & $0.35^{+0.09 }_{-0.03 }$ & $0.2 \pm 0.3$ & $0.10^{+0.08}_{-0.08}$ n & $0.15^{+0.12}_{-0.12}$ \\
Mel25-5        & $46.2  \pm 0.7 $ $^3$ & $0.43 \pm 0.02$ & $0.91 \pm 0.04$ & $0.85^{+0.05}_{-0.04}$ & $29.5^{+5.8}_{-3.0}$ & $0.36^{+0.04 }_{-0.06 }$ & $0.0 \pm 0.4$ & $-0.17^{+0.18}_{-0.18}$ n& $-0.29^{+0.30}_{-0.30}$ \\
\hline
\end{tabular} 
\end{sideways}
\label{fundimental-param-table} 
\end{table*}

\subsection{H-R diagram and evolutionary tracks}
\label{H-Rdiagram}

The stars in our sample were placed on a Hertzsprung-Russell (H-R) diagram, in order to derive masses. Luminosities were derived as in Paper I, based on J-band photometry from 2MASS \citep{Cutri2003-2MASS}.  The bolometric correction of \citet{Pecaut2013-PMS-BC-withJ} was used, together with our \teff.  Reddening was assumed to be negligible, since our targets are all near the sun ($<100$ pc).  

To derive distances to the stars in the sample we used parallax measurements from the Gaia Data Release 1 \citep{Gaia2016-data-release1} when possible.  For a few stars, Gaia parallaxes were not yet available, so we used Hipparcos parallax measurements \citep{van_Leeuwen2007-Hipparcos_book}.  When both values were available for a target the parallaxes were consistent but the Gaia values were more precise. 
The Hyades cluster has a known distance \citep[e.g. 46.3 pc, tidal radius $\sim$10 pc][]{Perryman1998-Hyades-age-dist}, however there are significant star to star differences in the measured parallax, thus we prefer the individual parallax measurements, which are relatively precise. 
The Coma Ber cluster also has a known distance of 86.7 pc \citep{vanLeeuwen2009-cluster-distances}, and a radius of $\sim$9.1 pc (based on the $\sim 6^\circ$ radius of the cluster).  However, these targets all have precise Gaia parallaxes, thus we use the more precise Gaia values, although they are all consistent with the cluster distance.  
BD-072388 (in AB Dor) does not have a Hipparcos parallax and does not yet have a Gaia parallax, so we used the dynamical distance from \citet{Torres2008-youngNearbyAssoc} and arbitrarily assumed a 20\% uncertainty on the value.  With these distances we derive absolute luminosities in Table \ref{fundimental-param-table}. 
For three stars, BD-072388, HIP 10272, and Mel25-43, there are significant uncertainties in their luminosity due to binarity.  These three stars all fall significantly above their association isochrones, as noted in Appendix \ref{Individual Targets}, using luminosities from photometry.  Thus for these targets we estimate their luminosity by fixing them to their association isochrones, as this is likely more accurate.  
Stellar radii were derived from the Stefan-Boltzmann law, using our \teff\ and luminosities.

\begin{figure}
  \centering
  \includegraphics[width=3.3in]{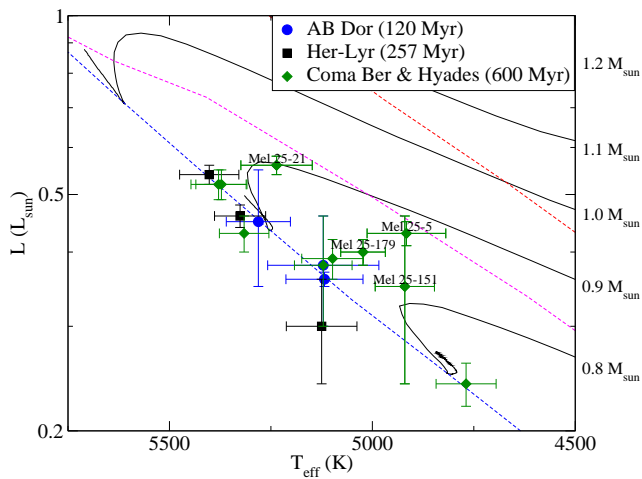}
  \caption{H-R diagram of the stars in this study.  Evolutionary tracks are from \citet{Amard2016-eol-traks-rotating}, plotted in 0.1\Msun\ increments for the masses labeled on the right (in \Msun).  Isochrones are shown for 24 Myr, 42 Myr, and the ZAMS, as in Paper I.  The stars are grouped by age and association, as indicated.   }
  \label{fig-hr-diagram}
\end{figure}

Masses were derived by comparing with a grid of evolutionary tracks (c.f.\ Fig.~\ref{fig-hr-diagram}).  The evolutionary tracks were computed with the STAREVOL V3.30 stellar evolution code, as discussed in \citet{Amard2016-eol-traks-rotating}, and are the same tracks described in Paper I. 
These evolutionary tracks assumed initial solar abundances, since the associations have nearly solar abundances and the stars are too young to have undergone significant chemical evolution \citep{VianaAlmeida2009-SACY-abun-young-assoc,Biazzo2012-abun-3nearby-assoc}.  A constant mixing length was used since neither the range of metallicities nor masses is large enough to significantly affect this approximation. 
The stars Mel25-5, Mel25-21 and Mel25-179 fall above the their cluster isochrone (and the ZAMS).  This may be due to binarity, indeed Mel25-179 is an SB1 in our observations, alternately a small overestimate in the distance could cause this, and using the Hyades distance of \citet{Perryman1998-Hyades-age-dist} is sufficient to bring Mel25-21 and Mel25-179 onto the ZAMS.  These evolutionary tracks were also used to derive convective turnover times for the stars.  The convective turnover time at one pressure scale height above the base of the convective envelope was used, as discussed in Paper I.  These, combined with the rotation periods of the stars, were used to compute Rossby numbers ($R_{o} = P_{\rm rot}/\tau_{\rm conv}$).  

\section{Spectropolarimetric analysis}

\subsection{Least squares deconvolution}
\label{Least squares deconvolution}

\begin{figure*}
  \centering
  \includegraphics[width=6.8in]{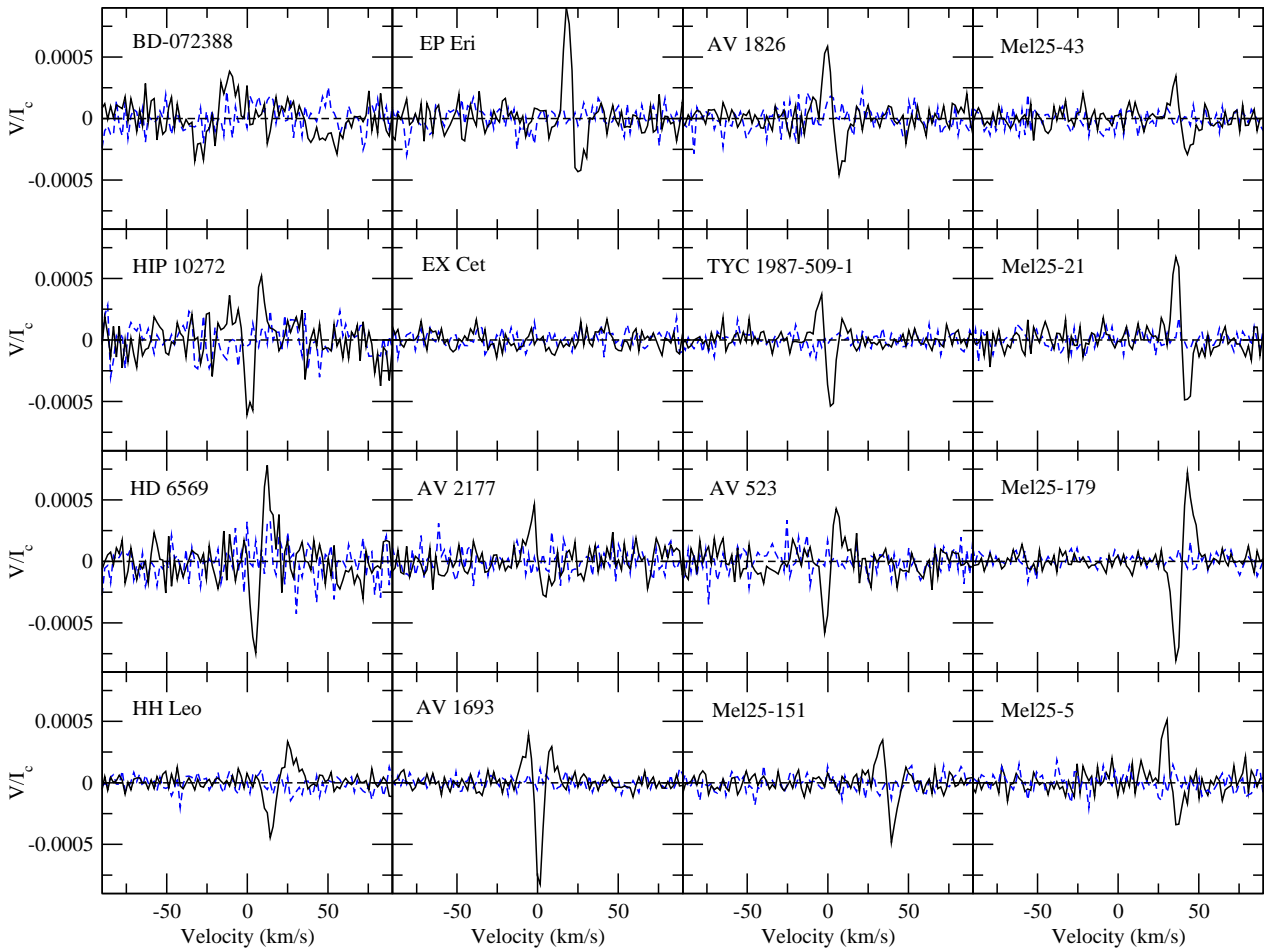}
  \caption{Sample LSD Stokes $V$ profiles for the stars in this study.  The associated diagnostic null profile for each observation is plotted in the background as a dashed line, with a second horizontal dashed line indicating zero.  }
  \label{fig-lsd-grid}
\end{figure*}

The signature of the Zeeman effect in Stokes $V$ is typically quite weak for solar-like stars, and undetectable in individual lines for any practical S/N.  Therefore, we used the multi-line technique Least Squares Deconvolution  \citep[LSD;][]{Donati1997-major,Kochukhov2010-LSD} to produce a pseudo-average line profile with much higher S/N.  The LSD procedure used here was identical to that from Paper I.  The same line masks were used, based on data from the VALD using `extract stellar' requests, and rounded to the nearest 500 K in \teff.  The same normalization parameters for LSD were also used, specifically a line depth of 0.39, Land\'e factor of 1.195, and a wavelength of 650~nm.  Sample LSD profiles for each star in this study are presented in Fig.~\ref{fig-lsd-grid}.

\subsection{Longitudinal magnetic field measurements}
\label{longitudinal-magnetic}

Longitudinal magnetic field ($B_l$) measurements were made, as in Paper I, for all observations.  This quantity represents the disk averaged line of sight component of the magnetic field.  These values were primarily used to investigate the rotation period of the star, since $B_l$ should vary smoothly as the star rotates.  However, they can also provide an estimate of the strength and degree of axisymmetry of the global stellar magnetic field.  This was measured with Eq.~2 from Paper I \citep[e.g.][]{Rees1979-magnetic-cog}.  This requires a wavelength and Land\'e factor, and the normalizing values from our LSD analysis were used (Sect.~\ref{Least squares deconvolution}).  The resulting $B_l$ measurements, phased with rotation period, are presented in Figs.~\ref{fig-bz} and \ref{fig-bz2}, and the maximum  absolute value of $B_l$ and full amplitude of variability for each star is reported in Table \ref{table-mag-param}.

We find peak $B_l$ between 20 and 7 G for most stars in the sample.  For BD-072388 we find a peak $B_l$ of 320 G, which is much stronger than the rest of the sample, but consistent with the stars much shorter rotation period.  For EX Cet we find no magnetic field, and $B_l$ is consistent with zero with uncertainties between 1.5 and 2 G (usually 1.7 G).  Thus $B_l$ must remain below 5 G, with a $3\sigma$ confidence.  EX Cet clearly has the weakest $B_l$ of the sample, despite having a \teff\ and a literature rotation period in the middle of the sample's range.  Unless the literature rotation period is incorrect (the star has the lowest \vs\ in the sample, hinting at a possible error) we have no explanation for the weakness of the magnetic field.  

For every star in the sample we performed a period analysis using $B_l$.  This was done as in Paper I, by fitting sinusoids with a grid of periods to the data by minimizing $\chi^2$, and constructing a periodogram in $\chi^2$.  When an adequate fit to the data could not be achieved with a simple sine curve, due to more complex magnetic field topology, a higher order sinusoid was used (e.g. $\sin + \sin^2 ...$).  This accounts for a quadrupole component (for $\sin^2$) and an octupole component (if extended to $\sin^3$), and is equivalent to a Lomb-Scargle periodogram for a simple sine.   The results of this for individual stars are discussed in Appendix \ref{Individual Targets}.  The rotation periods for all the stars are consistent with the best literature periods, and we do not find any stars lacking accurate literature periods as we did in Paper I.  However, our periods are typically more uncertain than the literature values, due to the relatively short timespan of our observations.  For AV 523, we are able to resolve a possible ambiguity in the literature rotation period.  We need sinusoids beyond first order for AV 1693, AV 1862, TYC 1987-509-1, Mel 25-151, and Mel 25-179 to achieve an adequate fit to the data.  For EX Cet, we do not detect a magnetic field, and thus cannot derive a rotation period from this method.

\subsection{Radial velocity}
\label{Radial velocity}

Radial velocities were measured for all observations by fitting a Gaussian to the Stokes $I$ line profiles by $\chi^2$ minimization, and taking the center of the Gaussian of the to be radial velocity ($v_r$).  Uncertainties on individual $v_r$ measurements were taken from the covariance matrix of the $\chi^2$ fit.  While this method is potentially influenced by spots on the stellar surface, the influence of spots is useful for our study as it provides another way to check the rotation period of the star, and this study does not require extremely high precision velocimetry.  The $v_r$ value averaged over all observations of a target is reported in Table \ref{fundimental-param-table}, with the standard deviation of the $v_r$ values reported as an uncertainty.

For each star we checked for systematic variations in $v_r$ with Julian Date.  In particular, we looked for trends on longer timespans than the rotation period of the star.  This was done by plotting $v_r$ versus Julian Date and looking for cases were there were significant differences between the earlier and later $v_r$ measurements, based on the uncertainties for individual $v_r$ values.  For the stars AV 2177, AV 1826, Mel25-151, and Mel25-43, there are clear systematic trends in the $v_r$ measurements, strongly suggesting that they are the primaries of SB1 systems.  For Mel25-179 we find a similar but weaker trend, tentatively suggesting it may be an SB1.  This is reflected in the larger standard deviations in $v_r$ for these stars.  We do not have sufficient data to find an orbital solution, thus we do not subtract off their orbital motion in the reported $v_r$.  However, since the $v_r$ for these stars are consistent with their cluster $v_r$, the amplitude of variability is likely small and this likely does not introduce a large error in our reported values.  

We performed a period analysis on $v_r$, similar to the analysis used for $B_l$, as in Paper I.  The apparent variability in $v_r$ used here is assumed to be due to surface features on the star distorting line profiles, not due to actual motion of the star.  However, since most of these stars were less spotted than the stars in Paper I, this analysis was less useful.  Significant unambiguous periods were only found for EP Eri and BD-072388, both of which were consistent with literature.  For HIP 10277 and AV 1693 pairs of ambiguous minima were found that were also consistent with their literature periods.  For EX Cet, for which we have no constraint on the period from magnetic data, the $v_r$ data are not able to strongly constrain the rotation period either.

\section{Magnetic mapping}
\label{ZDI}

Magnetic mapping was done in two stages, first a preliminary map was made, to check the quality of the Stokes $V$ data and to ensure the stellar parameters were correct.  Then a more detailed search for an optimal rotation period and differential rotation value was made, around the rotation period from the literature and $B_l$.  This further refined the rotation period, and where possible derived a differential rotation estimate.  Then the final best magnetic map was made using these optimal parameters.  The search for differential rotation was not performed in Paper I, however some of the stars in this paper with datasets spanning a longer time period (particularly AV 1826, AV 2177, and HH Leo) required non-zero differential rotation to achieve an acceptable fit to the observations.  

For the ZDI analysis in this paper we developed a new code, which implements the same physical model and analysis principles as the code used in Paper I.  This code has the practical advantages of being easier to use and easier to modify in the future, however the scientific output of the two codes is identical.  The code used in Paper I was described in that paper and was based on the code of \citet{Donati2006-tauSco}, while the new code used here is described in Appendix \ref{ZDIpy}.  Both codes use Gaussian model Stokes $I$ line profiles and the weak field approximation for Stokes $V$ profiles.  They both use the spherical harmonics description of the magnetic field from \citet{Donati2006-tauSco}, and use the maximum entropy fitting routine from \citet{Skilling1984-max-entropy-regularisation} to find the regularized best fit solution.  

The two ZDI codes were extensively tested to ensure they produced identical results for identical input parameters.  Indeed, for every star in this sample we produced a ZDI map with both codes and compared them to ensure the results were identical.  Thus, despite changing the underlying ZDI code, the results from this paper and Paper I are homogeneous, since the performance of the two codes is identical. 

While ZDI has been used successfully for many years, concerns continue to be raised \citep[e.g.][]{Stift2012-ZDI-systematics-atmo}.  Indeed, ZDI maps do not represent a complete picture of a stellar magnetic field, but only the components of the field that are constrained observationally.  A wide range of studies have shown the general reliability of ZDI \citep[e.g.][]{Donati1997-ZDI-tests, Hussain2000-ZDI-code-comparison, Hussain2001-DOTS-descript, Kochukhov2002-MDI-intro2, Yadav2015-MHD-models-ZDI-recon}.  However, there are some potential systematic trends that need to be considered.  In particular, the resolution of the map is dependent on the \vs\ of the star (e.g. Morin 2010), and we provide a discussion of this in Appendix \ref{Trends in magnetic geometry and resolution}.  There is the possibility of cross-talk between radial and azimuthal magnetic field, at least for some inclinations, when only Stokes $V$ is used.  The map is also somewhat sensitive to the degree of regularization used, mostly for the amount of energy in higher degree harmonics, although this also has a small impact on the total magnetic field strength.  While ZDI may contain some biases, we are using a consistent methodology across our sample, and the same basic methodology as the BCool \citep[][Petit et al. in prep.]{Marsden2014-Bcool-survey1}, MaPP \citep{Donati2008-BPTau-ZDI}, and MaTYSSE \citep{Donati2014-LkCa4-wTTs-mag-planet} samples.  This crucially provides results that can be directly compared for a large number of stars at different evolutionary stages.  

The input parameters for the ZDI model were the same as in Paper I.  Specifically, the model line used the normalizing Land\'e factor and wavelength from LSD, a Gaussian line full width at half maximum of 7.8 \kms\ ($1\sigma$ width of 3.2 \kms) was used (see Paper I), and the line strength was set by fitting the central line depth of the $I$ LSD profile for each star.  The stellar model again used a linear limb darkening law with a coefficient of 0.75.  For computing disk integrated model lines, the stellar surface was modeled using 2000 surface elements.
The spherical harmonic expansion was carried out to 15th degree in $l$, although for most stars in the sample the higher degrees are unnecessary, since they are unresolved in the observations due to the low \vs.
We find very little information (with values close to zero) in the higher degree harmonics, confirming that we are not reconstructing spurious smaller scale magnetic field.  A uniform maximum $l$ degree was used to provide a more uniform analysis of the sample.  
As in Paper I, a uniform surface brightness was assumed. Since the Stokes $I$ line profile variability is very weak or undetectable in these stars (except for BD-072388, c.f.\ the standard deviation of radial velocities in Table \ref{fundimental-param-table}) the stars are not strongly spotted and this approximation should not affect the results.  

Inclinations of the stellar rotation axis relative to the line of sight ($i$) were, when possible, derived from our measured \vs\ (Sect.\ \ref{spectrum-fitting}), radius (Sect.\ \ref{H-Rdiagram}), and rotation period.  However, in cases where the radius has a large uncertainty, or \vs\ is very small (significantly below the instrumental resolution) this becomes unreliable.  In these cases we used ZDI to derive an inclination angle.  For this we generated ZDI maps for a grid of inclinations, and selected the map with the best maximum entropy.  Then using this entropy as a target, we performed ZDI with a fixed target entropy and variable minimum $\chi^2$ \citep[as in][]{Petit2002-diff-rot-DI}, for the same grid of inclinations, and selected the model with a minimum $\chi^2$.  Generally these two inclinations agreed, however the curve of $\chi^2$ as a function of inclination allows us to derive formal uncertainties on the inclination.  Uncertainties were taken to be the variation in $i$ around the minimum needed to produce a $1\sigma$ difference according to $\chi^2$ statistics.  This approach allowed for a sensible target entropy, and allowed us to check that the maximum in entropy for a target $\chi^2$, and minimum in $\chi^2$ for a target entropy, are consistent.  Details of the derivation of $i$ are given in Appendix \ref{Individual Targets} for stars where the ZDI method was used, and our adopted values are given in Table \ref{fundimental-param-table}.

\subsection{Rotation period and differential rotation}
\label{Rotation period and differential rotation}
In order to verify and possibly refine the rotation periods of the stars, we performed a rotation period search using ZDI, initially assuming no differential rotation.  This proceeded by assuming a grid of rotation periods, and performing ZDI for each assumed rotation period, similar to in Paper I.  From this a periodogram in entropy and rotation period can be constructed, and the period that produces the maximum entropy can be selected.  While the assumption of no differential rotation at this stage may be inaccurate, this allows us to efficiently explore a wide range of periods, since we only have one dimension of parameter space to search.  Thus we can ensure we find a global maximum, not just a local maximum. 

Then we repeat this analysis, but rather than maximizing entropy for a fixed target $\chi^2$ as done by \citet{Skilling1984-max-entropy-regularisation}, we can minimize $\chi^2$ for a fixed target entropy as done by \citet{Petit2002-diff-rot-DI}.  The target entropy used is the previous global maximum, and this produces a curve of $\chi^2$ across the parameter space.  From the change in $\chi^2$ around the minimum we can define a confidence region at $1\sigma$ \citep[e.g.][]{numerical-recipes-Fortran}, and we use the extent of that region as our formal uncertainty.  First performing the search in entropy for fixed $\chi^2$ allows us to chose an appropriate target entropy, for the search in $\chi^2$ at fixed entropy.  Thus we get a periodogram in $\chi^2$, with formal uncertainties on the period.

In order to further verify the rotation periods, and when possible to derive differential rotation estimates, we used a second search based on ZDI, simultaneously probing rotation period and differential rotation following the method of \citet{Petit2002-diff-rot-DI}.  In this we assume a solar-like differential rotation law in the form
\begin{equation}
  \Omega(\theta) = \Omega_{\rm eq} + {\rm d} \Omega \sin^2 \theta ,
\end{equation}
where $\Omega(\theta)$ is the angular frequency at latitude $\theta$, $\Omega_{\rm eq}$ is the angular frequency at the equator, and ${\rm d} \Omega$ is the difference in angular frequency between the equator and pole.  

A ZDI fit is performed for each point in a grid of $\Omega_{\rm eq}$ and ${\rm d} \Omega$, using a range of periods around the global best period found in the previous analysis.  This produces a map of maximum achievable entropy in the $\Omega_{\rm eq}$ - ${\rm d} \Omega$ parameter space, and from this we can select the pair of parameters that produce the global maximum entropy.
Similar to the simple period analysis, we repeat this analysis but minimizing $\chi^2$ for a fixed target entropy as done by \citet{Petit2002-diff-rot-DI}.  Again, a $\chi^2$ contour around the minimum provides a confidence region at $1\sigma$, the extent of which defines our formal uncertainty. 

This analysis only produced reliable differential rotation values for some stars in our sample.  This approach requires observations at similar rotation phases but on different rotation cycles.  Larger differences in time between observations provides more sensitivity, as long as there has not been significant intrinsic evolution of the magnetic field.  This approach requires good S/N to detect changes in line profiles due to differential rotation, and it requires a reasonably large number of observations.
Thus, due to the limited time span of our observations, no reliable value of differential rotation could be found for EP Eri, TYC 1987-509-1, AV 523, Mel25-151, Mel25-43, Mel25-179, and Mel25-5, all of which have observations covering less than 1.5 rotation cycles.  Limitations from the S/N do not allow us to detect differential rotation in BD-07 2388 and HD 6569.  Marginally significant values of differential rotation were found for AV 2177 and HIP 10272, limited by S/N, and for Mel25-21, limited by phases with repeated observations.  More reliable differential rotation values were found for HH Leo, AV 1693, and AV 1826, aided by the relatively long time span over which the observations were obtained.  
A detailed discussion of the attempted differential rotation measurements is reported in Appendix \ref{Individual Targets}, and the values found are summarized in Table \ref{fundimental-param-table}. 

\begin{figure}
  \centering
  \includegraphics[width=3.3in]{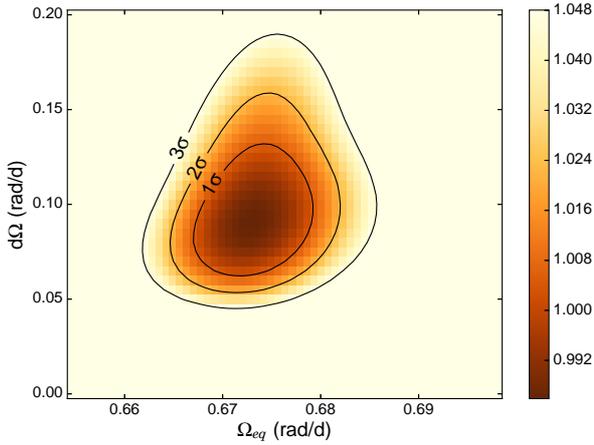}
    \caption{Sample reduced $\chi^2$ map, as a function of rotation frequency and differential rotation, for AV 1826.  Contours corresponding to $1\sigma$, $2\sigma$, and $3\sigma$ confidence levels, calculated from the changes in $\chi^2$ from the minimum, are show.  A well defined non-zero value of ${\rm d}\Omega$ is found.  }
  \label{fig-sample-diff-rot-search}
\end{figure}

\subsection{ZDI Results}

\begin{figure}
  \centering
  \includegraphics[width=3.in]{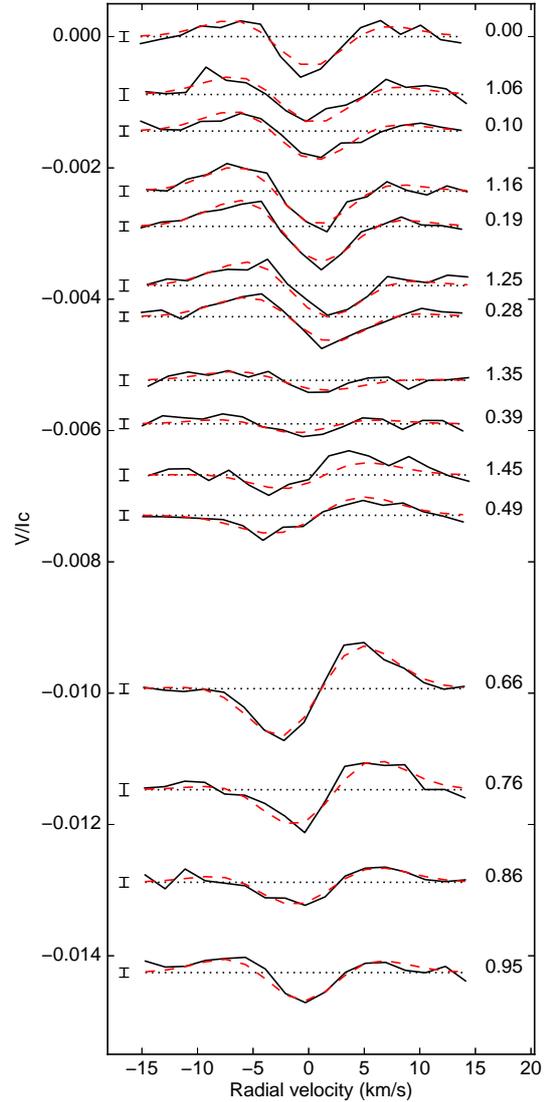}
  \caption{Sample ZDI fit for Mel25-151.  Solid lines are the observed Stokes $V$ LSD profiles, and dashed lines are the best fit synthetic ZDI line profiles.  The line profiles are shifted vertically by their rotation phase, and labeled by rotation cycle.  Error bars for the observations are given on the left.  }
  \label{fig-sample-zdi-fit}
\end{figure}

The final magnetic maps derived for the stars in this paper are presented in Figs.~\ref{fig-zdi-maps} and \ref{fig-zdi-maps2}, and a sample ZDI fit to $V$ LSD profiles is provided in Fig.~\ref{fig-sample-zdi-fit}.  We find a wide range of magnetic field strengths and geometries.
In order to effectively compare this large number of stars, we parameterize the magnetic field in a number of ways, with those parameters given in Table \ref{table-mag-param}.  For the global large-scale magnetic strength, we consider the unsigned (magnitude of the vector) field averaged over the surface of the star ($\langle B \rangle$).  To describe the geometry we consider the square of the magnetic field in different components, which is proportional to the magnetic energy, averaged over the surface of the star.  In the spherical harmonic description of \citet{Donati2006-tauSco} the $\alpha_{l,m}$ and $\beta_{l,m}$ terms are poloidal components, while the $\gamma_{l,m}$ terms are toroidal components, and we consider terms with $m = 0$ to be the axisymmetric components (about the rotation axis).  For geometry independent of field strength, we consider ratios of these components, and refer to them as fractions of energy, since magnetic energy is proportional to $B^2$.  We include the dipolar ($l=1$) quadrupolar ($l=2$) and octupolar ($l=3$) components of the poloidal field in Table \ref{table-mag-param}.  Some energy is present in higher degree spherical harmonics for some maps, however those are more sensitive to the spacial resolution of the maps, and for most stars this is enough to capture most of the poloidal energy.
We also include the axisymmetry of the total magnetic field, just the poloidal part of the field, and just the toroidal part of the field.  

BD-072388 has by far the strongest and most complex magnetic field in this paper, with an average surface field of 195 G.  This is likely due to it having by far the shortest rotation period, driving a much stronger dynamo.  However, BD-072388 has a similar strength and morphology to LO Peg in Paper I, which is a similarly fast rotator.  The rest of the sample has somewhat more similar field strengths, with surface average value from 34 to 8.5 G.  
The stars generally have significant toroidal components to their fields (e.g. HIP10272 at 68\% total energy and EP Eri at 77\% total energy, the weakest toroidal field being Mel25-21 at 20\% total energy), but it is never completely dominant.  The toroidal magnetic field components are generally axisymmetric, while the poloidal field components are generally less than 50\% axisymmetric (except for HD 6569).  In comparison to Paper I, the magnetic fields are on average weaker, due to these older stars rotating more slowly and hence having weaker dynamos.

\begin{table*}
\centering
\caption{Derived magnetic properties for the stars in our sample.  The maximum disk integrated longitudinal magnetic field is in column 2, and the amplitude of variability in the longitudinal field is in column 3.  The surface averaged large-scale magnetic field strength from the ZDI map is in column 4, and the maximum field value from the ZDI map is in column 5.  The remaining columns present the percent of the magnetic energy in different components of the field (poloidal, toroidal, dipolar, quadrupolar, octupolar and axisymmetric), as percentages of the total, poloidal, or toroidal field.  For EX Cet we do not detect a magnetic field, the limits on $B_{l}$ are $3\sigma$ upper limits.  }
\begin{sideways}
\begin{tabular}{lcccccccccccccc}
\hline
Star            & Assoc.    &$B_{l, {\rm max}}$&$B_{l, {\rm range}}$&$\langle B \rangle$& $|B_{\rm peak}|$ & pol. & tor. & dip. & quad. & oct.  & axisym.    & axisym.  &  axisym.   & axisym.\\
                &           & (G)       & (G)       & ZDI (G)   & ZDI (G)   & (\%tot)   & (\%tot)   & (\%pol)   & (\%pol)   & (\%pol)  &  (\%tot)   & (\%pol)  &  (\%tor)   & (\%dip) \\
\hline
BD-072388       & AB Dor    & 320       & 440       & 195.5     & 1015.7    & 39.7      & 60.3      & 35.0      & 7.7       & 6.7      & 62.5       & 34.7     & 80.8       & 81.5    \\
HIP10272        & AB Dor    & 18        & 28        & 21.2      & 40.2      & 32.0      & 68.0      & 83.0      & 10.9      & 2.3      & 74.6       & 29.8     & 95.7       & 34.0    \\
HD 6569         & AB Dor    & 20        & 19        & 25.0      & 48.6      & 60.0      & 40.0      & 88.5      & 7.3       & 3.3      & 85.2       & 76.2     & 98.7       & 79.4    \\
HH Leo          & Her-Lyr   & 18        & 33        & 28.9      & 66.2      & 45.9      & 54.2      & 57.0      & 18.3      & 8.7      & 49.2       & 1.9      & 89.3       & 2.6     \\
EP Eri          & Her-Lyr   & 10        & 11        & 34.3      & 82.3      & 22.4      & 77.6      & 65.5      & 30.7      & 0.8      & 77.3       & 21.4     & 93.4       & 2.9     \\
AV 2177         & Coma Ber  & 10        & 15        & 10.3      & 28.9      & 49.7      & 50.3      & 59.6      & 20.9      & 8.1      & 47.1       & 8.7      & 85.0       & 6.4     \\
AV 1693         & Coma Ber  & 13        & 20        & 33.7      & 71.0      & 50.5      & 49.5      & 31.3      & 51.1      & 10.8     & 51.9       & 17.7     & 86.8       & 42.8    \\
AV 1826         & Coma Ber  & 14        & 25        & 25.1      & 57.8      & 41.7      & 58.3      & 31.6      & 38.1      & 21.7     & 63.6       & 26.9     & 89.9       & 74.9    \\
TYC 1987-509-1  & Coma Ber  & 11        & 16        & 25.0      & 62.8      & 55.7      & 44.3      & 35.4      & 27.2      & 9.8      & 59.4       & 38.1     & 86.3       & 74.0    \\
AV 523          & Coma Ber  & 10        & 12        & 22.8      & 56.3      & 32.9      & 67.1      & 39.9      & 24.8      & 24.0     & 78.3       & 48.0     & 93.1       & 76.4    \\
Mel25-151       & Hyades    & 15        & 23        & 23.7      & 74.5      & 42.2      & 57.8      & 49.5      & 22.1      & 8.7      & 63.4       & 34.5     & 84.5       & 45.3    \\
Mel25-43        & Hyades    & 7         & 14        & 8.5       & 18.5      & 61.3      & 38.7      & 71.8      & 22.3      & 4.3      & 36.2       & 0.6      & 92.5       & 0.4     \\
Mel25-21        & Hyades    & 14        & 18        & 12.7      & 43.9      & 80.0      & 20.0      & 71.8      & 14.0      & 6.0      & 31.0       & 22.4     & 65.5       & 29.5    \\
Mel25-179       & Hyades    & 17        & 26        & 26.0      & 63.1      & 52.5      & 47.5      & 70.3      & 18.4      & 9.5      & 61.0       & 34.6     & 90.1       & 43.9    \\
Mel25-5         & Hyades    & 11        & 15        & 13.0      & 32.8      & 35.8      & 64.2      & 73.4      & 16.4      & 7.2      & 69.6       & 19.9     & 97.4       & 24.1    \\
EX Cet          & Her-Lyr   & <5        & <10       &           &           &           &           &           &           &          &            &          &            &         \\
\hline
\end{tabular} 
\end{sideways}
\label{table-mag-param} 
\end{table*}

\section{Discussion}

\begin{figure*}
  \centering
  \includegraphics[width=3.4in]{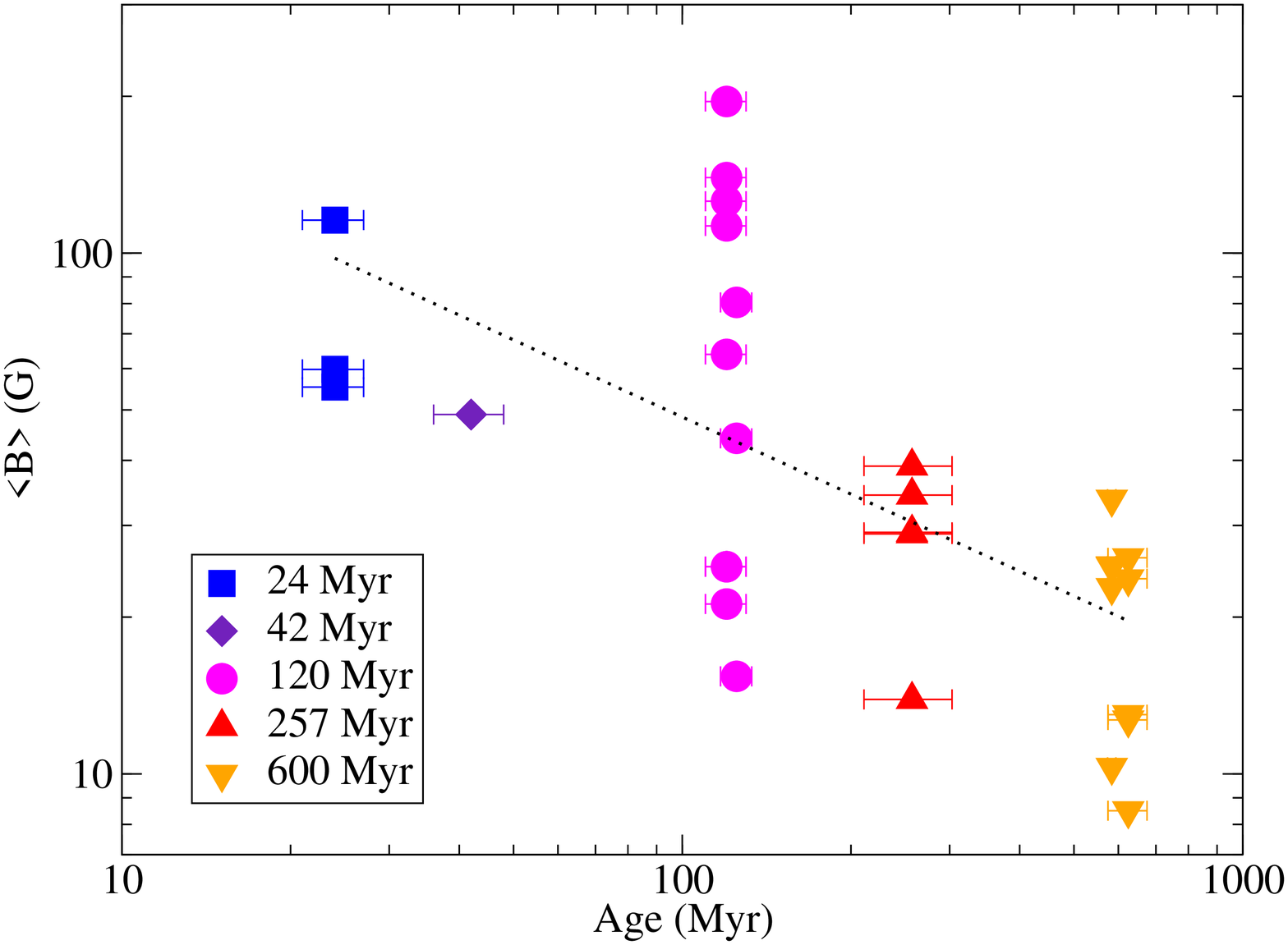}
  \includegraphics[width=3.4in]{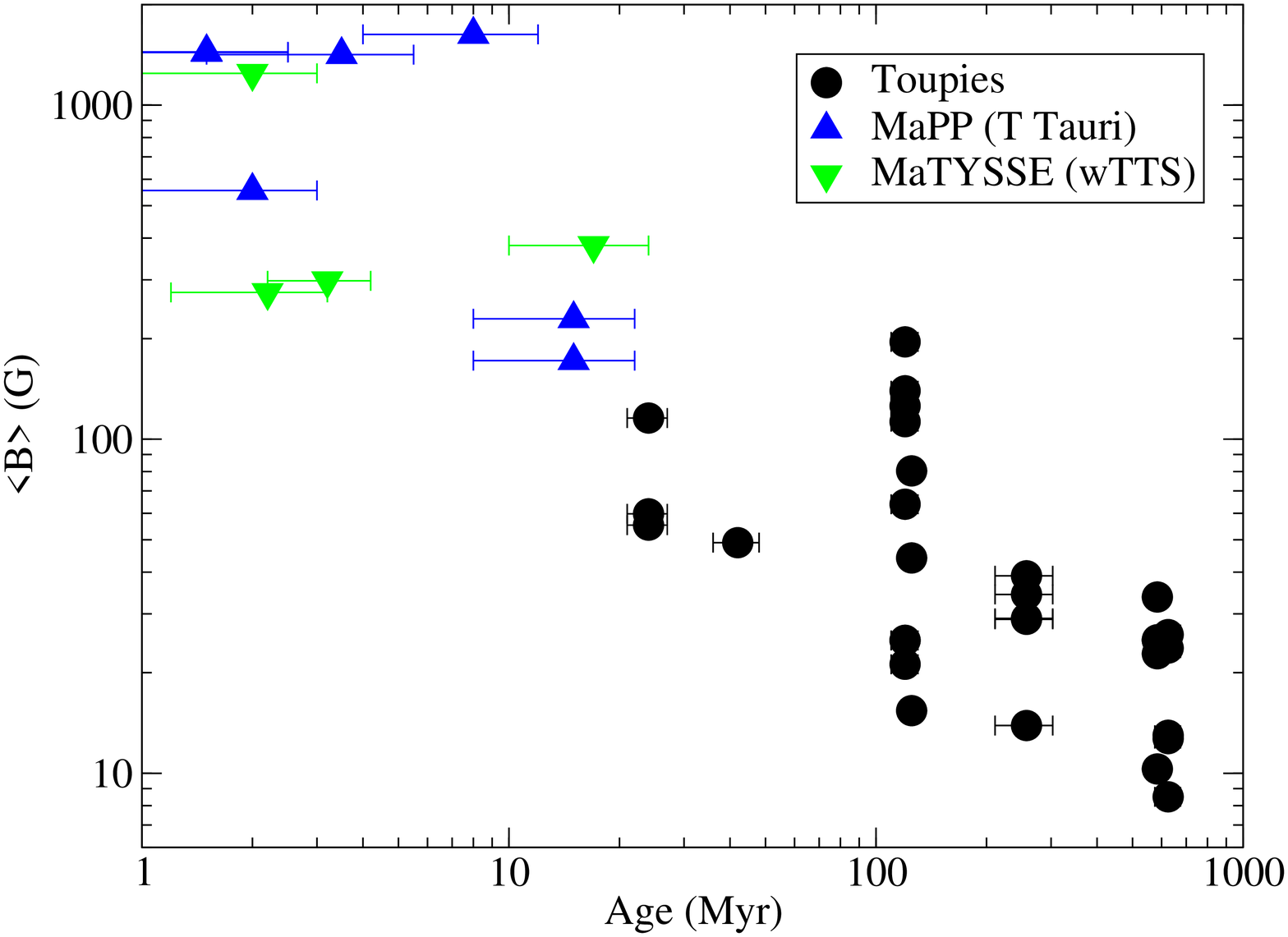}
  \caption{Mean large-scale magnetic field from ZDI as a function age for the stars in our study (left) and compared with some literature results (right).  The dotted line is a power law fit. }
  \label{fig-trend-age-B}
\end{figure*}

\begin{figure}
  \centering
  \includegraphics[width=3.3in]{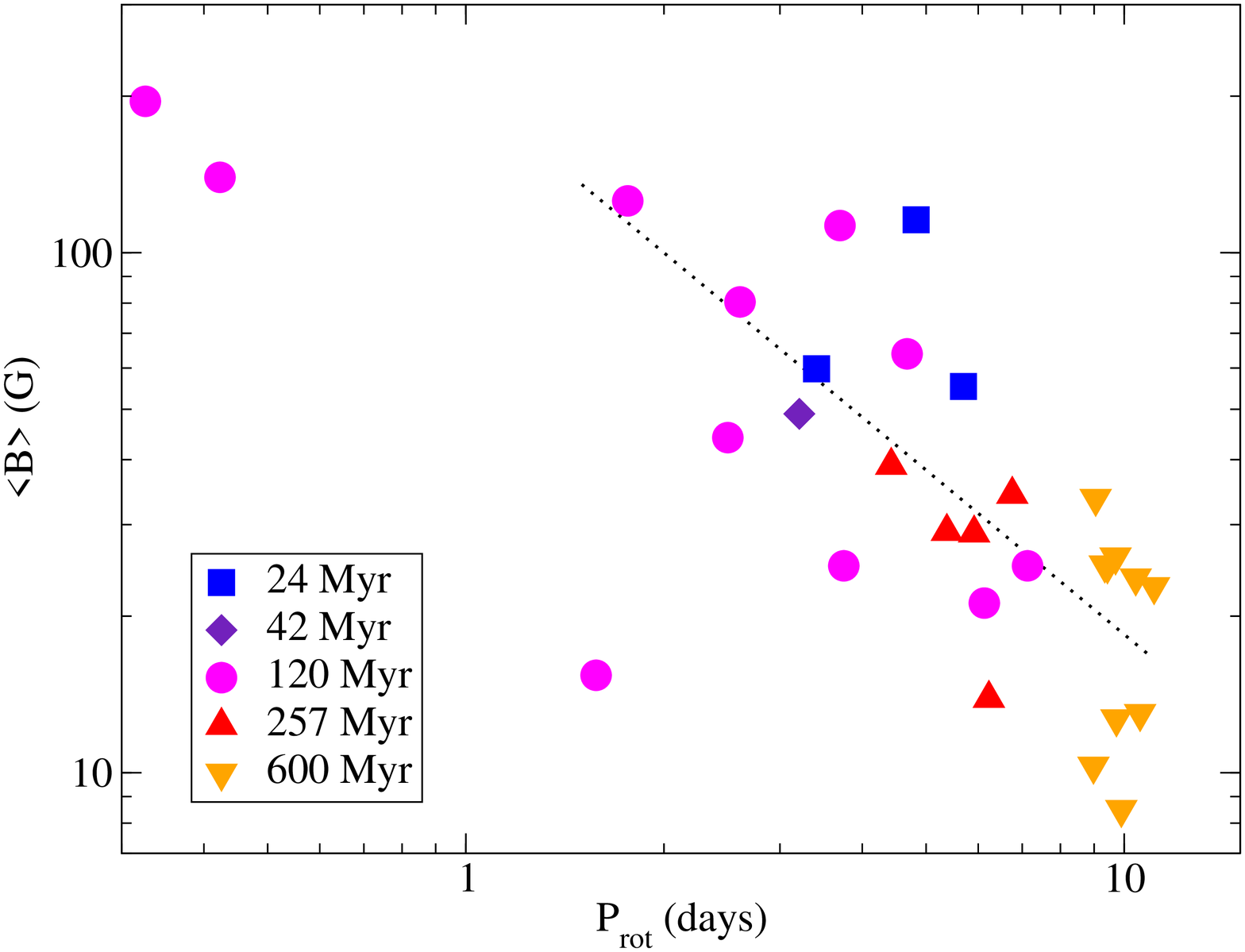}
  \caption{Mean large-scale magnetic field from ZDI as a function rotation period for the stars in our study.  The dotted line is a power law fit. }
  \label{fig-trend-period-B}
\end{figure}

\begin{figure*}
  \centering
  \includegraphics[width=3.4in]{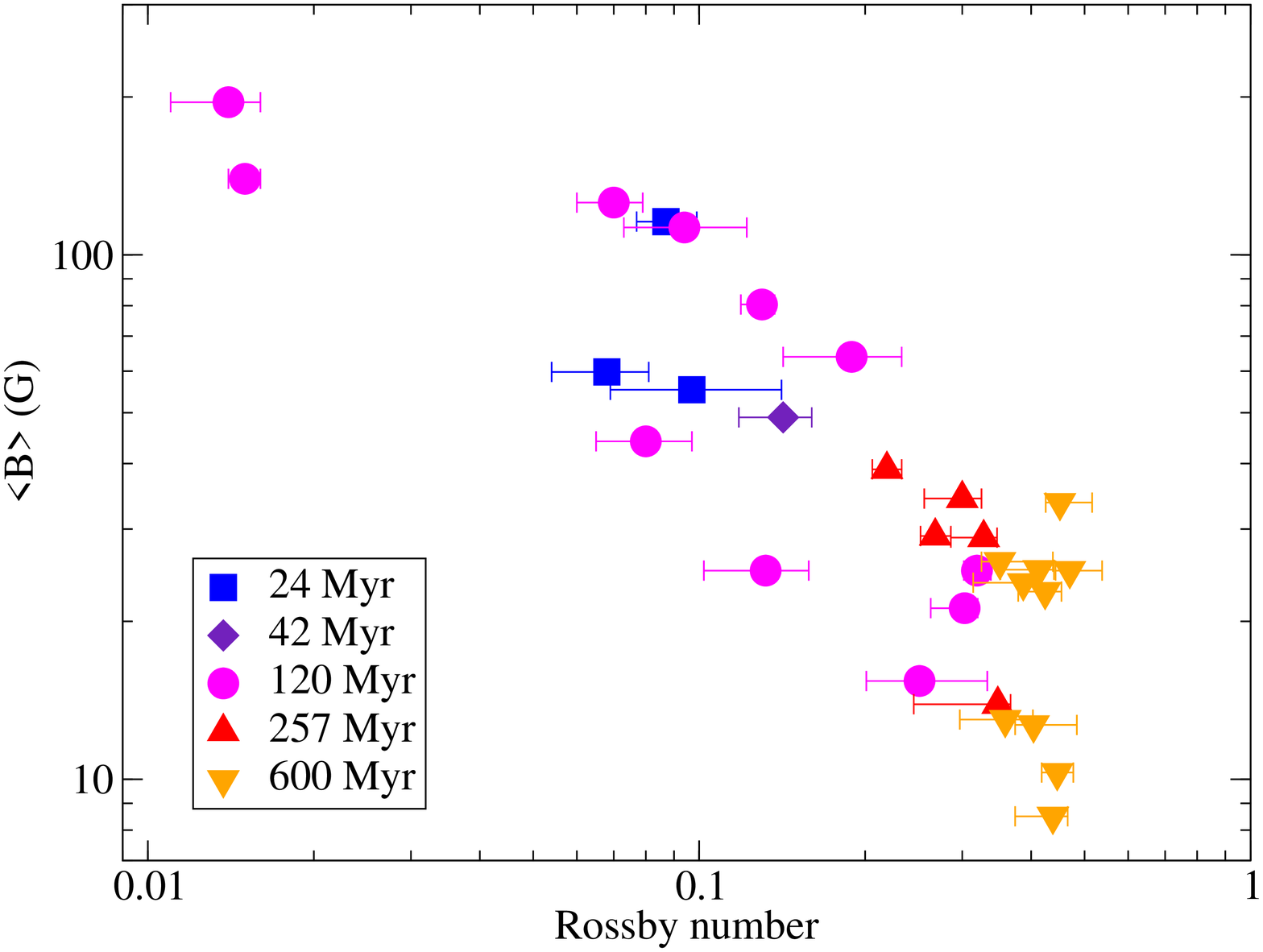}
  \includegraphics[width=3.4in]{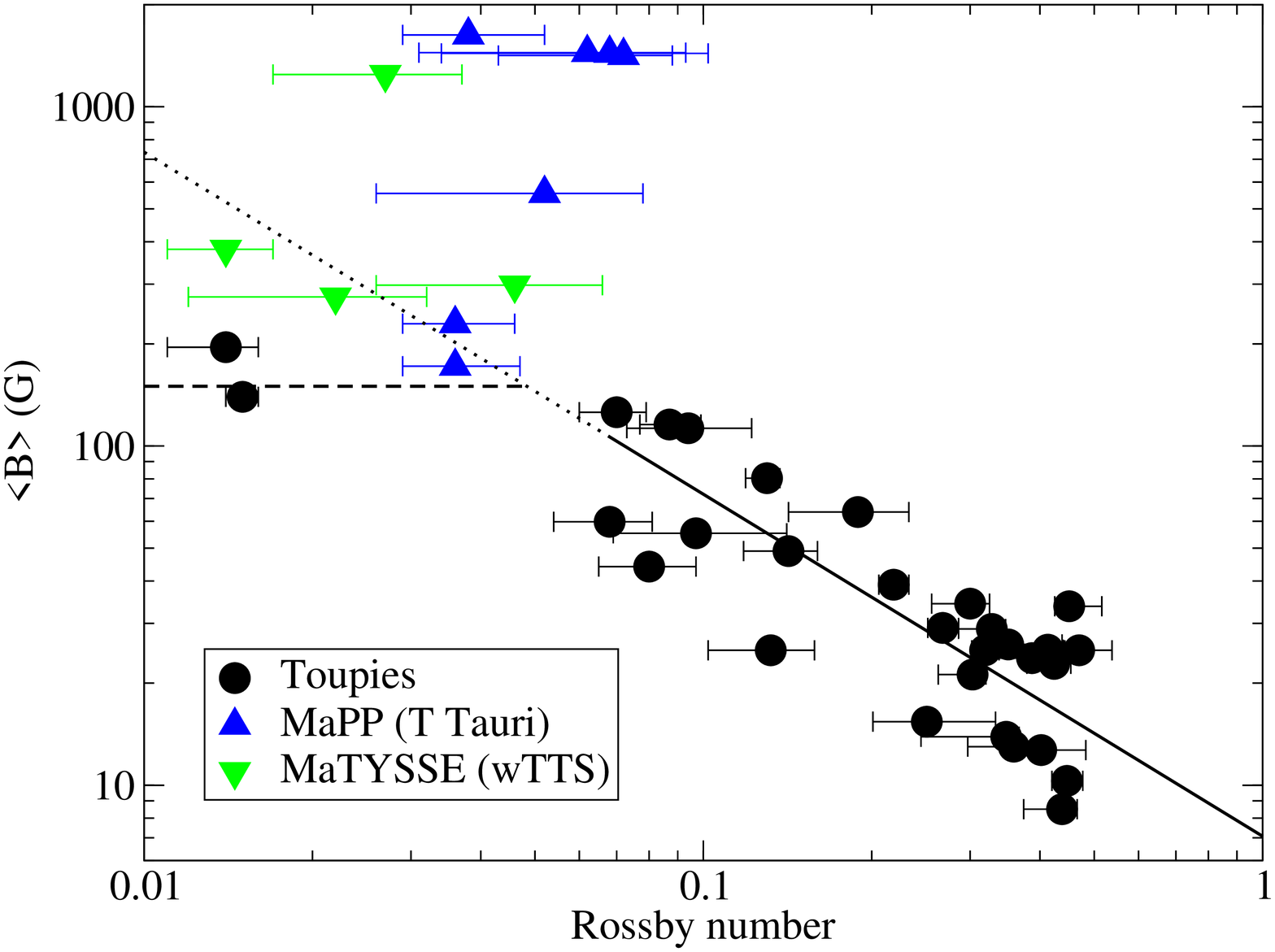}
  \caption{Mean large-scale magnetic field from ZDI as a function Rossby number for the stars in our study (left) and compared with some literature results (right).   In the left panel, a power law fit to the data is presented (solid line), an extrapolation of this to lower Rossby numbers (dashed line) and a hypothetical saturation level (dashed line) are also shown.  }
  \label{fig-trend-rossby-B}
\end{figure*}

The expanded sample of stars with magnetic properties derived here strengthens many of the trends we found in Paper 1.  We again find a clear decreasing trend in the average large-scale magnetic field strength (the unsigned magnetic field strength from our maps averaged over the surface of the star, \Bmean) with age, shown in Fig.~\ref{fig-trend-age-B}.  We also find a decreasing trend with rotation period, shown in Fig.~\ref{fig-trend-period-B}.  Having older, slower rotating stars in our sample improves these correlations.  There is also a decreasing trend in \Bmean\ with Rossby number, shown in Fig.~\ref{fig-trend-rossby-B}, which provides a tighter correlation than simply rotation period. 

The trend we find in the average large-scale magnetic field strength can be described by a power law:
$\langle B \rangle = (466 \pm 290) t ^{-0.49 \pm 0.12}$, for age $t$ in Myr, based on the Toupies sample (Fig.~\ref{fig-trend-age-B} left). Including T Tauri stars from the MaPP and MaTYSSE projects would produce an exponent of $-0.68 \pm 0.05$.    
This is consistent within $1.5\sigma$ with the trend found by \citet{Vidotto2014-magnetism-age-rot}, who found an exponent of $-0.655 \pm 0.045$.  The ages of the stars in our sample are much more accurate than in \citet{Vidotto2014-magnetism-age-rot}.  However, due to the large range of magnetic fields found around an age of $\sim$120 Myr, the scatter in our relationship is similar.
\citet{Rosen2016-mag-young-solar-twins} studied six young solar analogues using ZDI and also found a decreasing trend in \Bmean\ with age.  Their results are consistent with our trend, although the trend is much clearer here due to the larger sample size.

In rotation period we find a power law trend in \Bmean\ of:
$\langle B \rangle = (207 \pm 71) P_{\rm rot} ^{-1.05 \pm 0.19}$ with a saturation below periods of 1 or 2 days (Fig.\ \ref{fig-trend-period-B}). 
This trend has an exponent slightly smaller than \citet{Vidotto2014-magnetism-age-rot}, who found an exponent of $-1.32 \pm 0.14$, but it is consistent within $1.5\sigma$. 

In Rossby number we find a power law trend with \Bmean\ of:
$\langle B \rangle = (8.4 \pm 1.8) R_o ^{-0.89 \pm 0.13}$.   
This assumes saturation for values below 0.06 (Fig.\ \ref{fig-trend-rossby-B}), however the exact saturation value of Rossby number is not strongly constrained, and could be as high as 0.1.  We also note that the convective turnover time depends on how deep in the convective envelope this is calculated.  Using a different choice of depth will shift all Rossby numbers (as discussed in Paper I), and would lead to a somewhat different saturation value.
This trend is qualitatively consistent with the trend found by \citet{Vidotto2014-magnetism-age-rot}, however the exponent we find is smaller by roughly $2.5\sigma$ than their value of $-1.38 \pm 0.14$.  This could partly be due to us including stars near the saturated regime, with Rossby numbers between 0.06 and 1.0.  However repeating the power law fit restricting it to $R_o > 0.1$, we still find an exponent of -0.90, which is not enough for a good agreement.
Our two studies use different sources for convective turnover times, which could contribute to this discrepancy. 
The scatter in our trend of \Bmean\ with Rossby number is much smaller than the trend from \citet{Vidotto2014-magnetism-age-rot}, since our sample is much more homogeneous.  Thus our power law fit may in fact be closer to the correct value.
Interestingly the exponents for both the trends in $R_o$ and $P_{\rm rot}$ are close to -1.0.  

The very fast rotator BD-072388, together with LO Peg from Paper I, supports the hypothesis that we are seeing a saturation of the large-scale magnetic field strength due to increasing rotation period.  This star has a magnetic field of similar strength to LO Peg in Paper I, with a qualitatively similar complex geometry.  Both BD-072388 and LO Peg have magnetic field strengths similar to stars with rotation periods around 2 days and Rossby numbers around 0.1, despite having much shorter rotation periods and smaller Rossby numbers (0.3-0.4 days and $R_o$ 0.01-0.02).  The star AB Dor, while slightly more massive than  BD-072388 and LO Peg ($M \sim 1.0$ $M_{\odot}$, $P \sim 0.514$ d), has been studied using ZDI \citep{Donati1999-ABDor-mag-geom, Donati2003-sempol-monitoring-cool-active, Hussain2007-ABDor-mag-x-ray}, and was found to have a similar large-scale magnetic strength (\Bmean\ $\sim 125$ G), and a similar complex geometry.
The star LQ Hya ($M \sim 0.8$ $M_{\odot}$, $P \sim 1.60$ d) is less confidently in the saturated regime by Rossby number, but also has a strong (\Bmean\ $\sim 100$ G) complex magnetic field \citep{Donati2003-sempol-monitoring-cool-active}, which is comparable to BD-072388 and LO Peg. 
These four stars are consistent with the saturation of the large-scale magnetic field due to rapid rotation.

Saturation of magnetic proxies, such as X-ray emission, at low Rossby number are well established \citep[e.g.][]{Noyes1984-CaHK-Rossby, Pizzolato2003-Xray-saturation-rossby, Wright2011-x-ray-rossby-relations}.
However, those proxies are only indirectly related to the large-scale magnetic field, by several physical processes (they depend on small-scale magnetic field, a filling factor, and magnetic reconnection or chromospheric heating), thus it is not clear that the large-scale magnetic field should behave similarly.
The behavior of the large-scale component of the field is perhaps the most direct observational constraint for dynamo simulations.  Saturation of the large-scale magnetic field at low Rossby number, due to changing convective properties, has been observed in comparisons of mostly convective M-dwarfs to K-stars \citep[e.g.][]{Morin2008-Mdwarf-topo, Donati2009-ARAA-magnetic-fields, Vidotto2014-magnetism-age-rot}.  However, this is due to the growth of the convective zone to dominate the star.  Thus an independent constraint is the saturation of the large-scale magnetic field due to rapid rotation, for stars with approximately the same size of convective envelope.  

\begin{figure}
  \centering
  \includegraphics[width=3.3in]{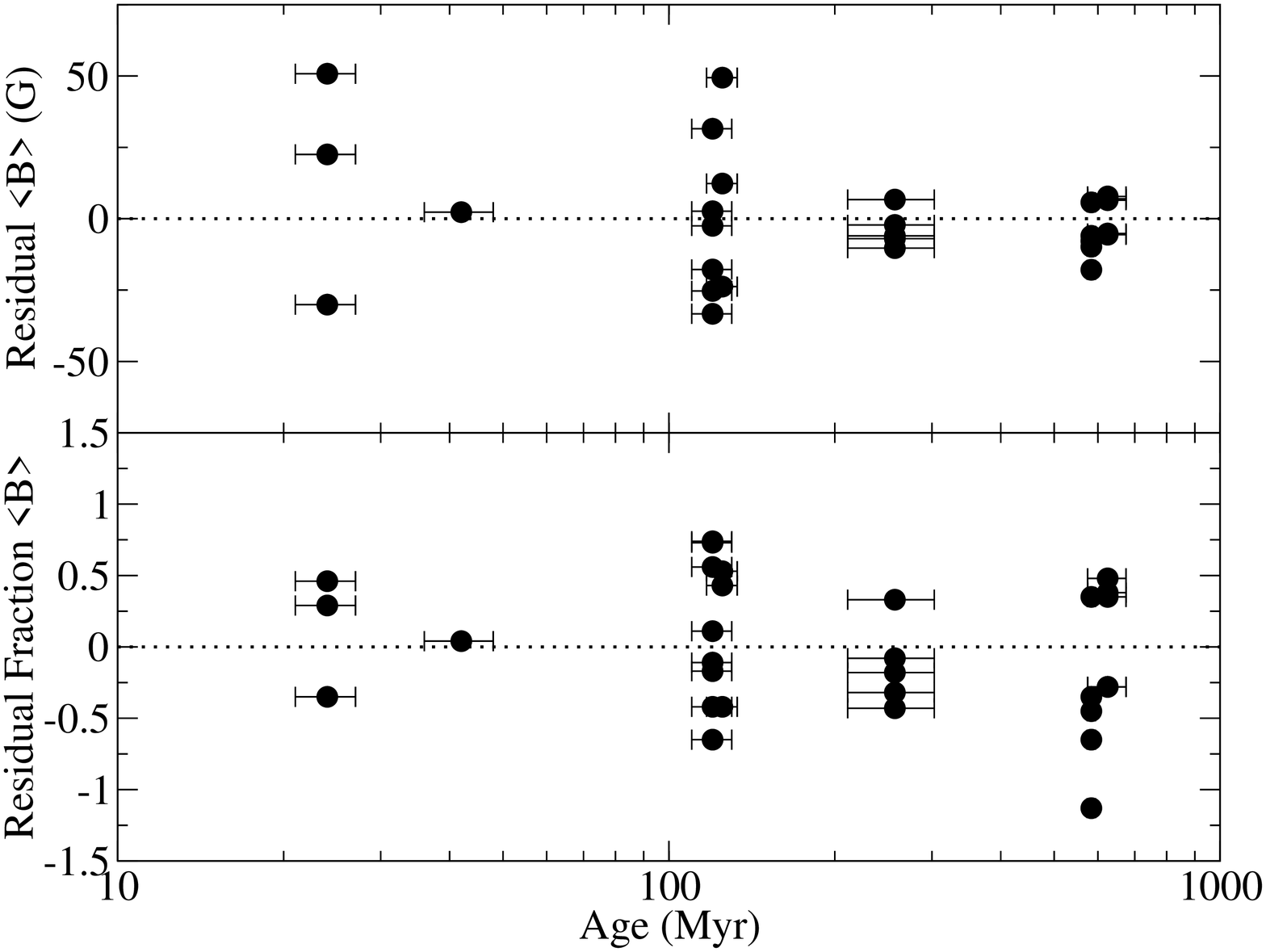}
  \caption{Residual values of the power law fit to \Bmean\ as a function of Rossby number (top) and those residuals divided by the predicted \Bmean\ (bottom).  }
  \label{fig-residualB-age}
\end{figure}

To search for a trend in the mean large-scale magnetic field strength as a function of age, beyond the trend as a function of Rossby number, we calculated the difference between the power law fit in $R_o$ and the observed mean large-scale field values, excluding the two saturated regime stars (BD-072388 and LO Peg).  This is the residuals to the power law fit in Rossby number.  Plotting this residual against age shows a decreased scatter to older ages, illustrated in Fig.\ \ref{fig-residualB-age}.  
Rotational evolution models predict the development of a steep gradient in the internal rotation profile of solar-type stars at the zero-age main sequence, with a rapidly rotating core and a slowly rotating outer convective envelope, which gradually becomes flatter as the star evolves on the early main sequence \citep[e.g.][]{Gallet2013-Bouvier-ang-mom-evol,Gallet2015-Bouvier-ang-mom-evol2}. The decreasing magnetic field scatter observed between 120 and 650 Myr is qualitatively consistent with these predictions, provided the dynamo process is indeed sensitive to the early rotational history of solar-type stars.
However, if we plot this residual as a fraction of the power law values, effectively the fractional residuals of the fit, the scatter appears to be constant as a function of age (Fig.\ \ref{fig-residualB-age}). 
If this scatter is physical, then the process giving rise to it, such as cyclical magnetic variability, appears to operate as a fraction of the magnetic field value.  However, this fractional process does not seem to be age dependent, with the precision currently allowed by our sample.  The possible impact of long term magnetic variability is discussed further in Appendix \ref{Long term magnetic variability}.

Trends in magnetic geometry are more challenging to find.  Multi-epoch studies of stars with ZDI generally show large changes in the magnetic field geometry over a time span of years.  This can be due to stellar magnetic cycles \citep[e.g.][]{Donati2003-sempol-monitoring-cool-active, Mengel2016-tauBoo-mag-update, BoroSaikia2016-61CygA-solar-like}, or apparently more chaotic long term magnetic variability \citep[e.g.][]{Jeffers2011-nonSolar-HD171488-3epochs, Jeffers2014-epsEri-mag-var, BoroSaikia2015-HNPeg}.  This intrinsic variability complicates searches for trends in magnetic geometry.  However, our results continue to support the observation from \citet{Petit2008-sunlike-mag-geom}, Paper I, and \citet{See2016-mag-geom-cycles-PolTor-Rossby}, that very slowly rotating stars have dominantly poloidal fields, while faster rotators have a wider range of poloidal/toroidal ratios.  This transition appears to occur around a Rossby number of 1.0, or a rotation period of 15-20 days, and seems to occur at longer rotation periods than are available in our sample.  However, the transition is not precisely defined, partly because a given star can exhibit a range of poloidal/toroidal ratios, due to its long term magnetic variability.  
We also support the trend from \citet{See2015-toroidal-axisymm} and Paper I that dominantly toroidal magnetic fields are dominantly axisymmetric (i.e. symmetric about the stellar rotation axis), while dominantly poloidal magnetic fields have a wide range of axisymmetries. 

In their study of six solar analogues between 100 and 600 Myr old, \citet{Rosen2016-mag-young-solar-twins} found that the magnetic energy in $l=3$ spherical harmonics was larger than the energy in the $l=2$ harmonics for their two oldest stars ($\sim$600 Myr). We do not find the same trend in our sample.  None of our stars older than 200 Myr have an $l=3$ energy above 50\% of the $l=2$ energy.  The only two stars with this ratio significantly above one are BD-072388 and HII 739 (from paper I).  BD-072388 is the fastest rotator in our sample, while HII 739 is the hottest star in our sample (\teff\ $= 6066 \pm 89$ K).  Our sample is largely cooler than the sample of \citet{Rosen2016-mag-young-solar-twins} (4400-5400 K, vs 5800 K).  So if this effect is strongly dependent on temperature, that could explain the difference.  Or this may simply be a coincidence due to their small sample size.  
The stars TYC 6349-0200-1, HIP 76768, TYC 5164-567-1, HII 296, LO Peg all have ratios of $l=3$ to $l=2$ energy near 1, and these are all fast rotators with some of the smallest Rossby numbers in the sample.  This is consistent with the general trend of stars with smaller Rossby number and faster rotation having more complex ZDI maps.  However, it is still unclear how much of this is driven by changing resolution of the maps and how much of this is real changes in the magnetic field structure.  

\begin{figure*}
  \centering
  \includegraphics[width=5.0in]{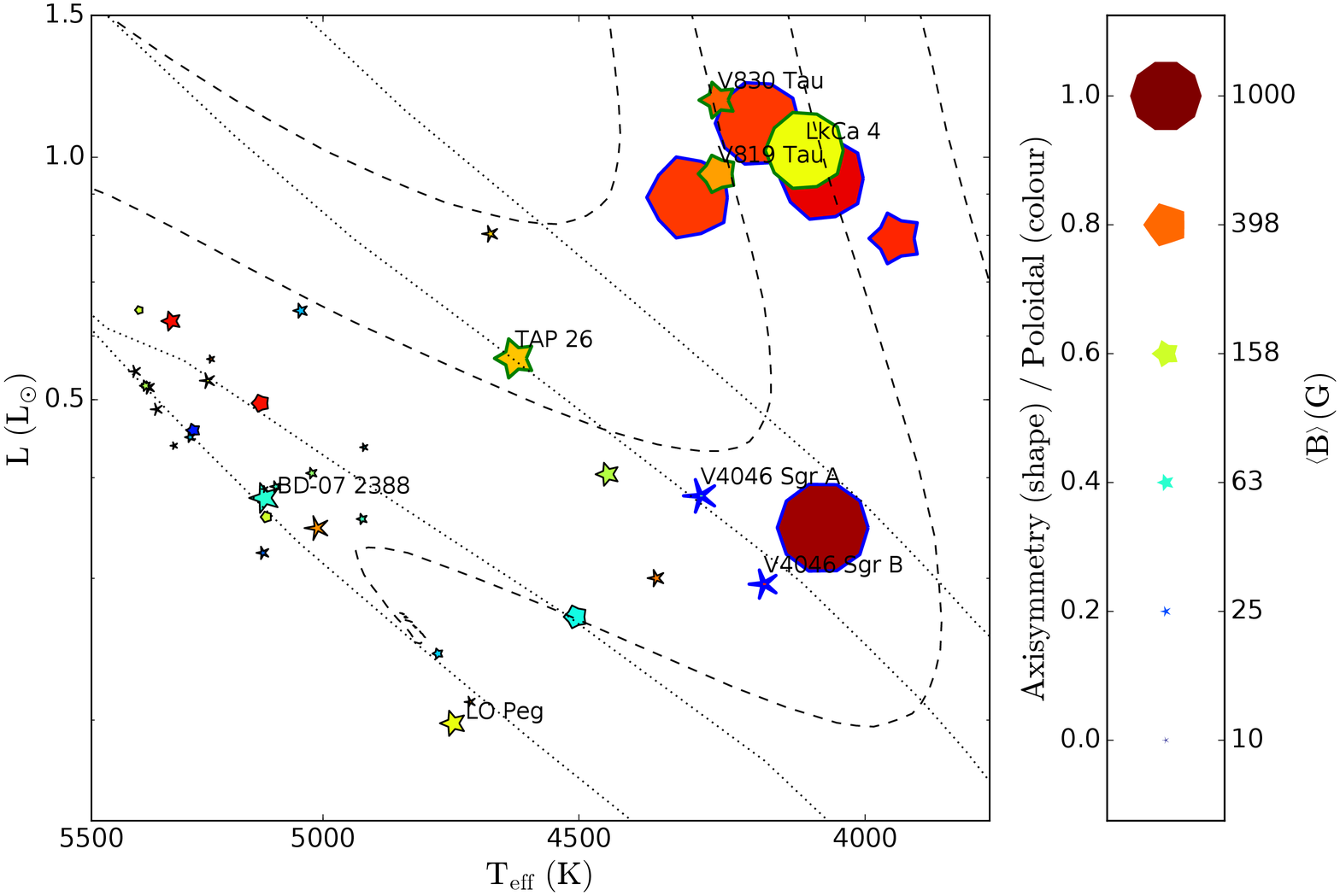}
  \includegraphics[width=5.0in]{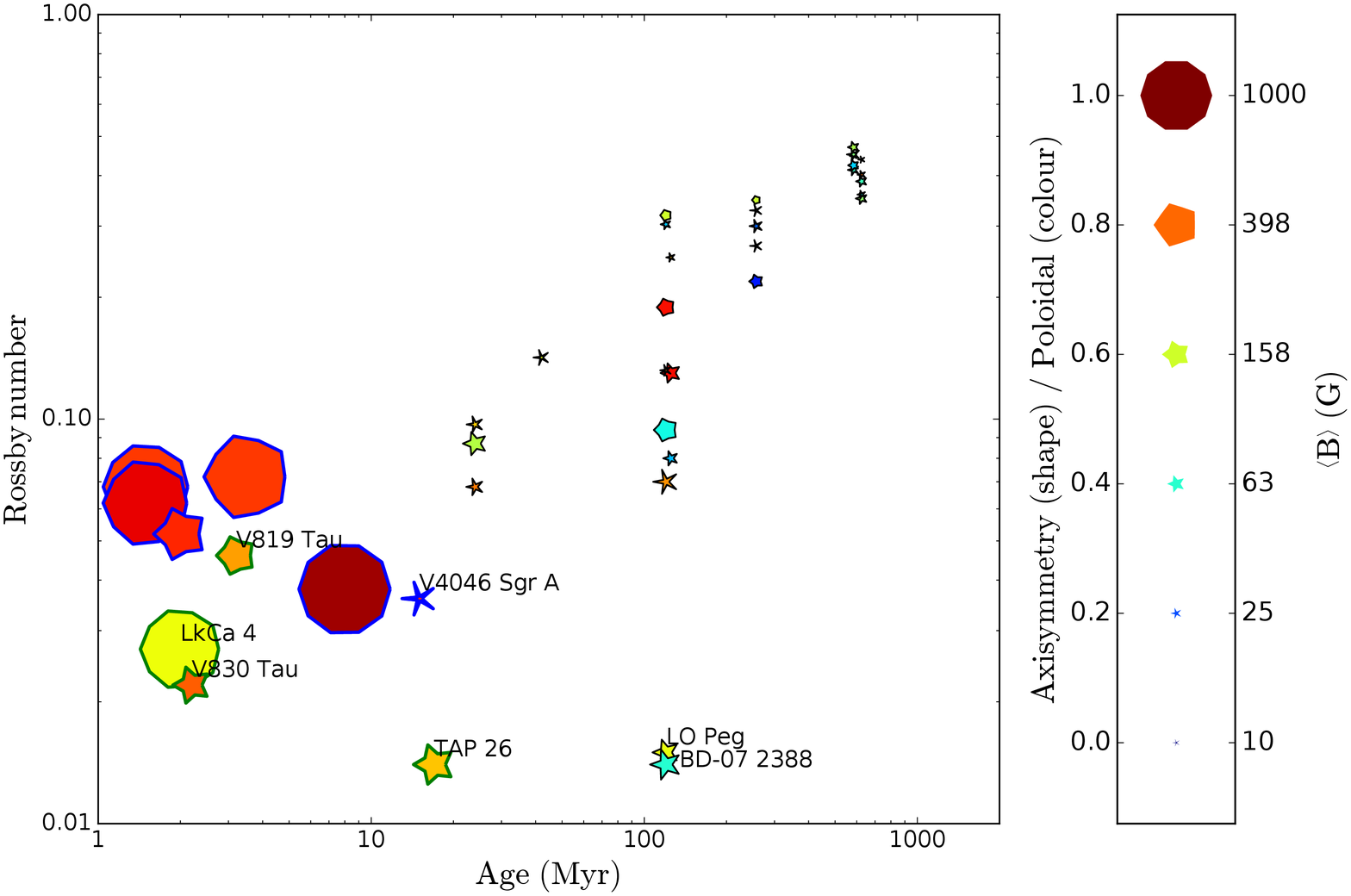}
  \caption{H-R diagram (top) and age-Rossby number plane (bottom), with stars from the Toupies project (thin black outlines), together with some classical T Tauri stars from the MaPP project (thick blue outlines), and some weak-line T Tauri stars from the MaTYSSE project (thick green outlines).  Symbol size corresponds to magnetic field strength, symbol color is how poloidal or toroidal the magnetic field is, and shape is the axisymmetry of the poloidal component of the magnetic field.  In the H-R diagram, evolutionary tracks (dashed lines) are shown for 1.2, 1.0, 0.8 and 0.6 $M_{\odot}$, isochrones (dotted lines) are shown for 10, 20, 50 and 100 Myr.  
    The development of a significant radiative core corresponds to the bend where the stars move from the Hayashi track (largely vertical) onto the Henyey track (more horizontal).  
    Evolutionary tracks and isochrones models are from \citet{Amard2016-eol-traks-rotating}.  }
  \label{fig-cg-hrd-age-ro}
\end{figure*}

Comparing the stars in our sample to younger T Tauri stars is helpful for investigating evolutionary changes in magnetic fields on the pre-main sequence.  In particular, we consider the classical T Tauri stars from the MaPP project (BP Tau, \citealt{Donati2008-BPTau-ZDI}; AA Tau, \citealt{Donati2010MNRAS-AATau-ZDI}; TW Hya, \citealt{Donati2011-TWHya-ZDI}; V4046 Sgr A \& B, \citealt{Donati2011-V4046Sgr-ZDI}; GQ Lup, \citealt{Donati2012-GQLup-ZDI}; and DN Tau, \citealt{Donati2013-DNTau-ZDI}), and weak-line T Tauri stars from the MaTYSSE project (LkCa 4, \citealt{Donati2014-LkCa4-wTTs-mag-planet}; V819 Tau \& V830 Tau, \citealt{Donati2015-V819Tau-WTTs-mag}; and TAP 26, \citealt{Yu2017-TAP26-wTTs-mag-planet}), in roughly the same mass range as our sample.  We considered this in Paper I, but we revisit it here with our expanded sample of stars, and with the expanded sample of weak-line T Tauri stars.  The classical and weak-line T Tauri stars differ in that the classical stars are strongly accreting, while the weak-line stars are accreting at a much lower level.

The classical and weak-line T Tauri stars are shown on an H-R diagram, together with our stars, in Fig.~\ref{fig-cg-hrd-age-ro}.  The classical T Tauri stars show stronger, more poloidal, and more axisymmetric magnetic fields than our sample (apart from the two most evolved T Tauri stars V4046 Sgr A \& B).  This follows the proposal of \citet{Gregory2012-TTauri-B-structure}, that the different magnetic properties are driven by different convective properties, with the T Tauri stars being mostly convective.  This is essentially the same result as in our Paper I, since the new stars we add here are clustered around the ZAMS.  The weak-line T Tauri stars complicate the hypothesis somewhat.  TAP 26 is partly convective and has a similar magnetic field strength and geometry to our later pre-main sequence stars, which is consistent with the magnetic field being driven by structure.  The star LkCa 4 is mostly or fully convective and has a consistent strength and axisymmetry to the classical T Tauri stars, although it may be less poloidal.  However, the two stars V819 Tau and V830 Tau seem to have intermediate magnetic properties between our sample and the classical T Tauri stars, with magnetic field strengths closer to our sample, and magnetic geometries that are mostly poloidal but more complex and non-axisymmetric, and seem to fall in the mostly or fully convective regime of the H-R diagram.  Thus it remains unclear how well the weak-line T Tauri stars fall into this scenario, but observations of more stars are needed to further test this idea. 

\subsection{Differential rotation}

\begin{figure*}
  \centering
  \includegraphics[width=3.3in]{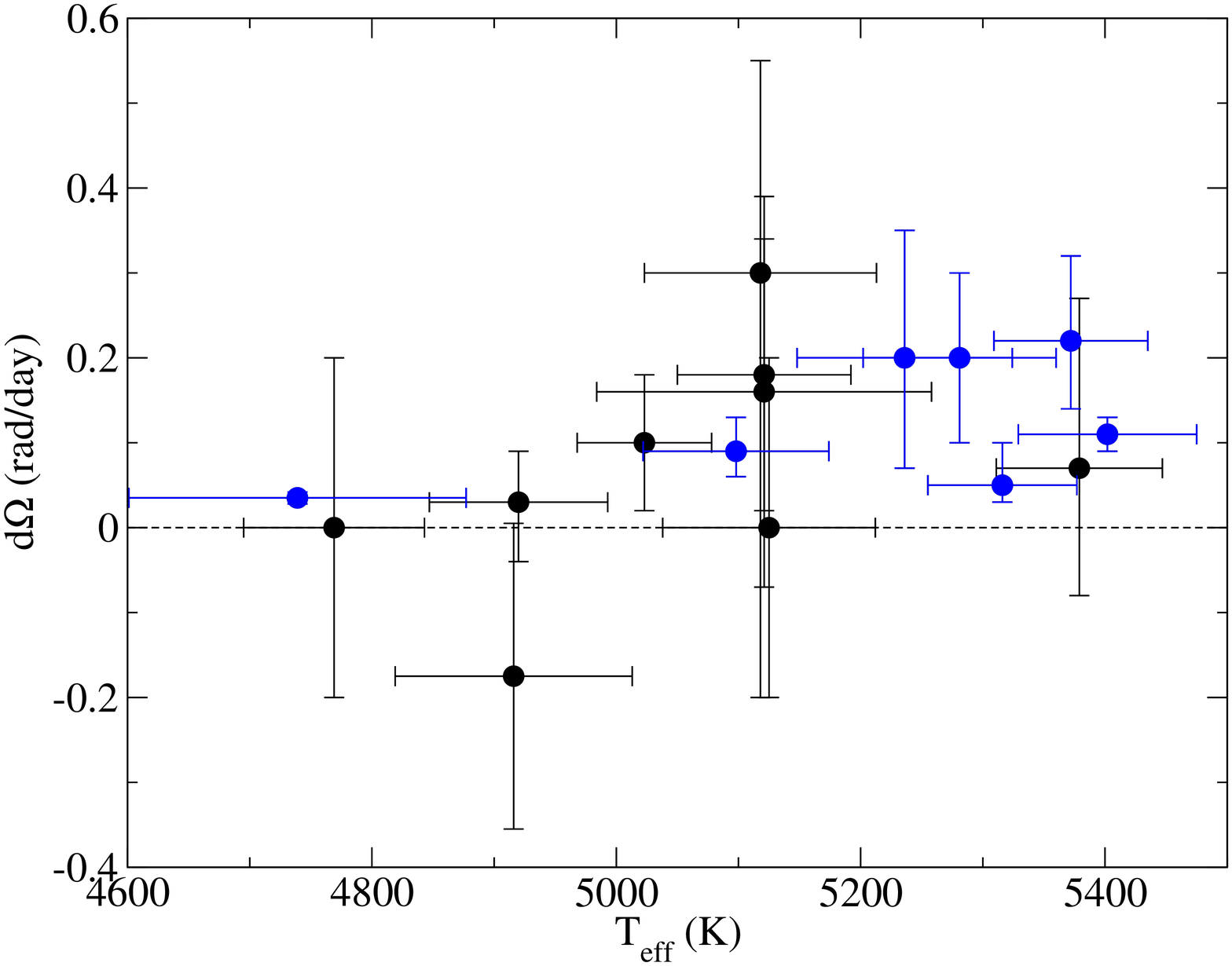}
  \includegraphics[width=3.3in]{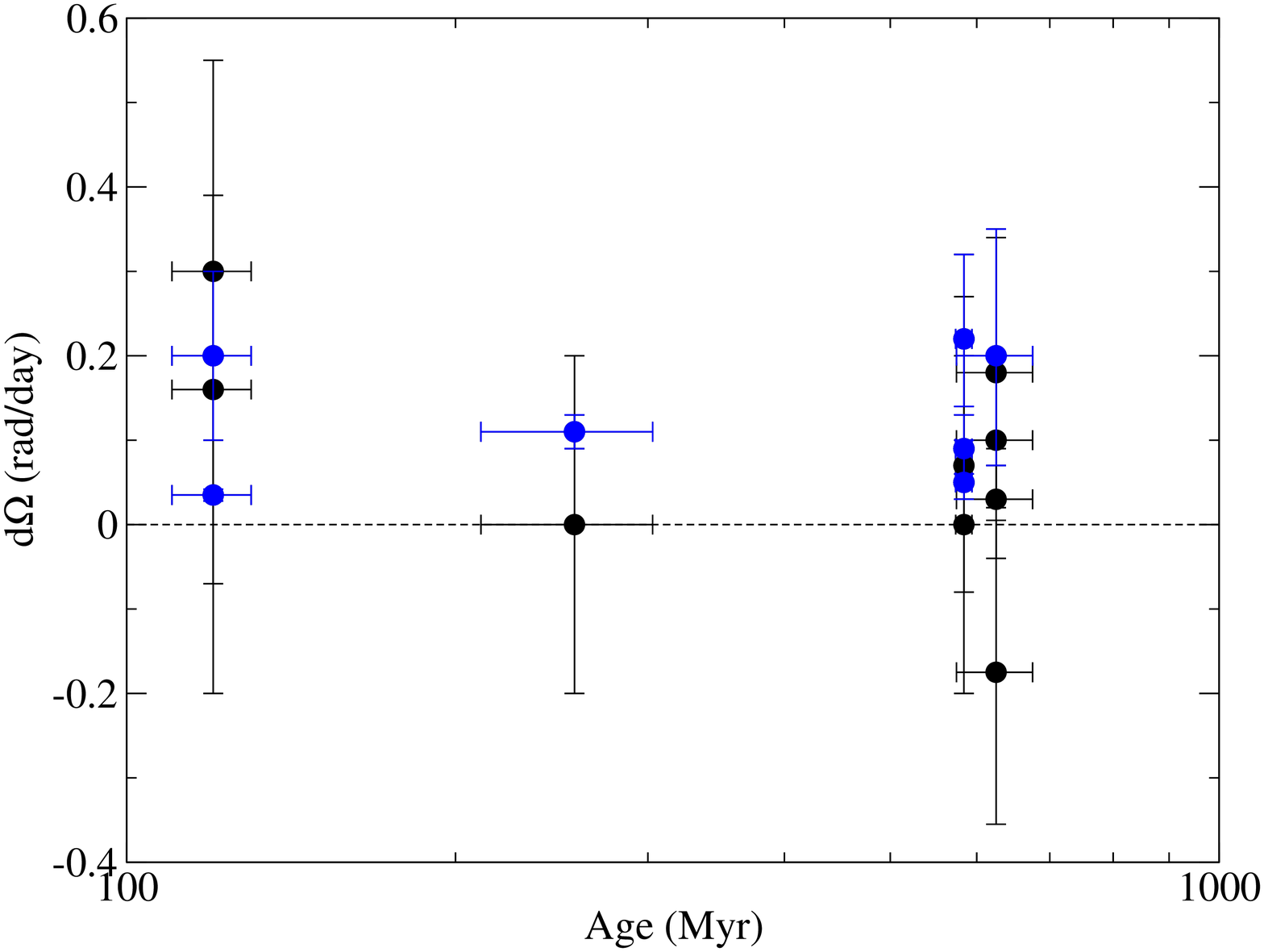}
  \caption{Differential rotation rate (\dOmega) as a function of effective temperature (left) and age (right).  Stars with significantly non-zero differential rotation are in blue. }
  \label{fig-dOmega-trends}
\end{figure*}

We have searched for trends in latitudinal differential rotation using the values derived this paper, and the literature value for LO Peg from \citet{Barnes2005-LOPeg-DI}.  
We see no clear trend in latitudinal differential rotation with rotation period, although there are a large uncertainties on our \dOmega\ values, and most stars are non-detections.  There is a trend towards decreasing values of the ratio ${\rm d}\Omega / \Omega_{\rm eq}$ for the faster rotators, although that is largely driven by BD-072388 and LO Peg.  The large $\Omega_{\rm eq}$ of BD-072388 and LO Peg implies small values for the ratio, and smaller limits on the ratio provided by our uncertainties.  

We find a weak trend towards increasing \dOmega\ with \teff, illustrated in Fig.\ \ref{fig-dOmega-trends}.  The range of \teff\ in our sample is small, thus the tend is not strong.  However, the hotter stars have larger \dOmega, while the cooler stars have a small value (LO Peg) or are non-detections typically with smaller limits.  A similar trend, with a similar degree of confidence, appears in ${\rm d}\Omega / \Omega_{\rm eq}$.
If we consider convective turnover time rather than \teff, we get a similar quality trend, with \dOmega\ decreasing with increasing convective turnover time.  Indeed this correlation may be slightly better, but the larger uncertainties on convective turnover time makes this unclear.  We can speculate that larger convective turnover times, and larger convective cells, redistribute angular momentum more efficiently, leading to less differential rotation.  
This trend in \dOmega\ with \teff\ is qualitatively similar to the trend reported by Barnes et al. (MNRAS, in press, doi:10.1093/mnras/stx1482),  
who considered a range of literature \dOmega\ values for stars from 3000 to 7000 K.  A few other overviews of literature \dOmega\ values have found similar trends \citep[e.g.][]{CollierCameron2007-diff-rot-review}.

We find no clear trend in \dOmega\ with age, illustrated in Fig.\ \ref{fig-dOmega-trends}.  There seems to be a comparable range of values in the sample around 120 Myr as there is in the sample around 600 Myr.  However, the age sampling of our \dOmega\ values is sparse, and does not extend to the youngest portion of our sample.  One of the motivations for investigating \dOmega\ is the large radial internal differential rotation predicted by rotational evolution models.  If the internal radial differential rotation changes importantly between 120 and 600 Myr, it does not seem to be reflected in surface latitudinal differential rotation.  However, if the surface latitudinal differential rotation is primarily controlled by the convective properties of the stellar envelope, this would not be surprising. 

We find no trend in the mean magnetic field \Bmean\ with \dOmega.  This is in strong contrast to the trends with rotation period and Rossby number.  The \dOmega\ values carry large uncertainties and we lack stars in the saturated regime, however the latitudinal surface differential rotation we measure does not seem to be important for the generation of large-scale magnetic fields.  If differential rotation is important for the dynamo generation of magnetic fields in these stars, it must not be related to the latitudinal surface differential rotation \dOmega.

In magnetic geometry, we find no clear trend in the fraction of toroidal magnetic field with \dOmega.
This would seem to argue against the toroidal field being generated by latitudinal differential rotation shearing poloidal field.  However, we caution that there are large uncertainties on \dOmega, and it may be that \dOmega\ is harder to to measure in strongly toroidal stars, thus no strong conclusions can be drawn.
There is no trend in axisymmetry of the magnetic field and \dOmega.  However, the uncertainties on \dOmega\ are noticeably larger for strongly axisymmetric fields, particularly when the total axisymmetry reaches $\sim$70\%.  This is because measuring \dOmega\ requires detectable non-axisymmetric features in the magnetic field at different latitudes, thus when the non-axisymmetric features becomes weak, our ability to measure \dOmega\ becomes weak.

\section{Conclusions}

We have derived detailed magnetic maps for 15 young solar-like stars and characterized their large-scale magnetic field strength and geometry.  We also derived fundamental physical parameters for these stars.   The stars were selected from members of four stellar associations with ages from 120 to 625 Myr.  We find a narrow range of \teff\ for the stars, from 4769 to 5402 K, and most of the stars have rotation periods between 5.9 and 10.6 days, except for one very fast rotator at 0.326 days.  This extends the sample from Paper I to older, slower rotating stars. 

We find that the average large-scale magnetic field decreases with increasing age across our sample, although there is a large scatter around 120 Myr.  The average large-scale magnetic field also decreases with rotation period and Rossby number within our sample, with Rossby number providing the tighter correlation.  At very low Rossby number we see further tentative evidence for saturation of the large-scale magnetic field, with a second apparently saturated star BD-072388.  This star has a similar rotation rate and similar magnetic properties to LO Peg, both of which are similar to the literature values for AB Dor.  This helps further support the hypothesis of saturation of large-scale magnetic fields due to increasing rotation rate, rather increasing convective turnover time.

Among stars older than $\sim$20 Myr, the evolution of the large-scale magnetic field strength can be explained sufficiently well by changing Rossby number.  Once the trend in Rossby number has been subtracted, there is no clear residual trend in age.  However, comparing to T Tauri stars in the same mass range, there are clear differences that cannot be explained by Rossby number.  The oldest T Tauri stars fall close to our proposed saturation value, however the younger objects have much stronger magnetic fields.  This is likely a consequence of changing internal structure, as proposed by \citet{Gregory2012-TTauri-B-structure}, since the youngest T Tauri stars are largely convective.  However the possible impact of accretion and star-disk interactions cannot be completely ruled out from the current initial studies of weak-line T Tauri stars.

This paper has largely strengthened the conclusions of our Paper I.  In the next paper in this series, we will focus on younger stars, and further probe the saturation of the large-scale magnetic field at rapid rotation.

\section*{Acknowledgments} 
This study was supported by the grant ANR 2011 Blanc SIMI5-6 020 01 ``Toupies: Towards understanding the spin evolution of stars'' ({http://ipag.osug.fr/Anr\_Toupies/}). 

This work has made use of data from the European Space Agency (ESA) mission {\it Gaia} (\url{http://www.cosmos.esa.int/gaia}), processed by the {\it Gaia} Data Processing and Analysis Consortium (DPAC, \url{http://www.cosmos.esa.int/web/gaia/dpac/consortium}). Funding for the DPAC has been provided by national institutions, in particular the institutions participating in the {\it Gaia} Multilateral Agreement.
\bibliographystyle{mnras}
\bibliography{massivebib.bib}{}

\appendix
\section{Individual Targets}
\label{Individual Targets}

\begin{figure*}
  \centering
  \includegraphics[width=2.8in]{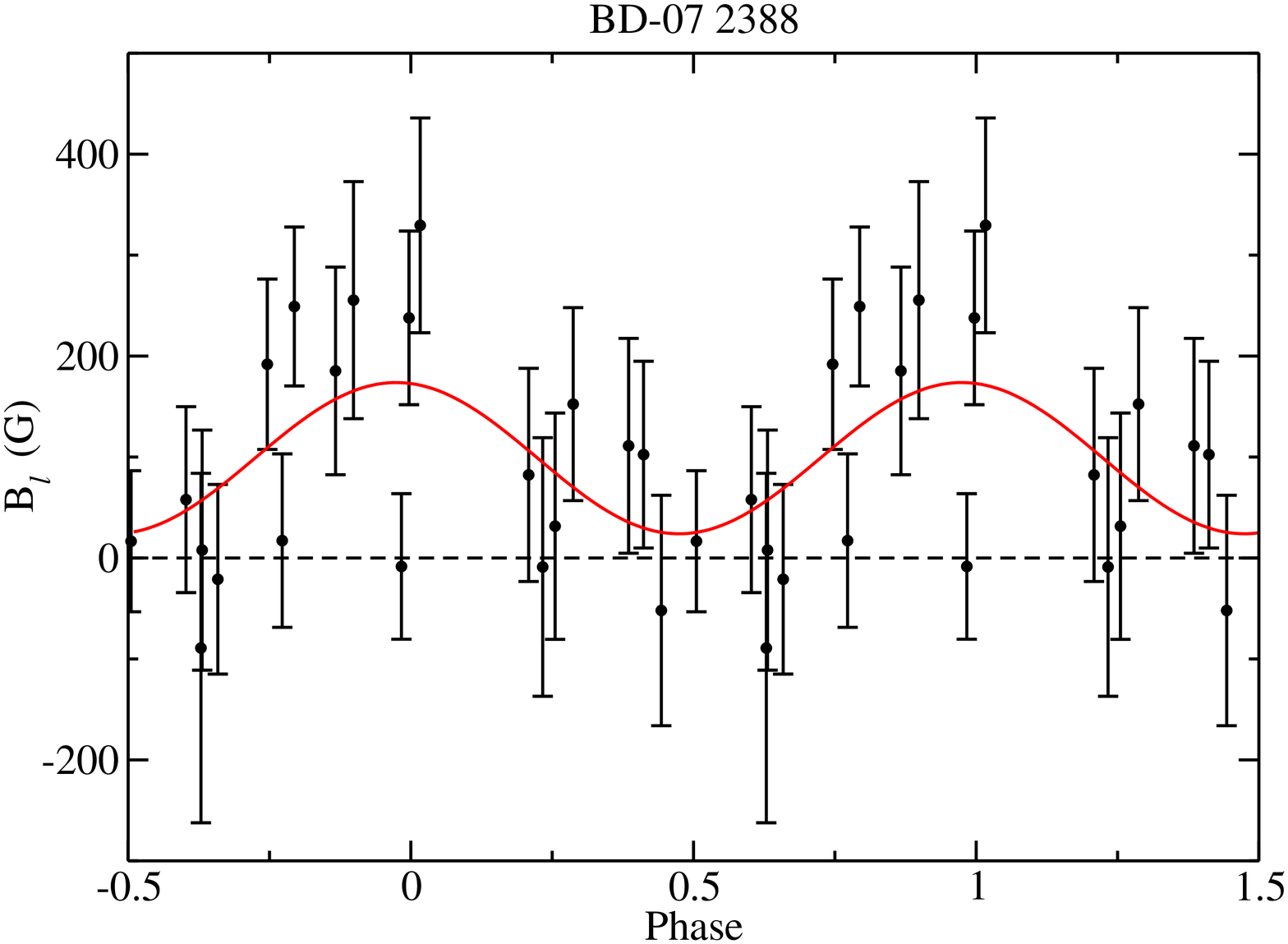}
  \includegraphics[width=2.8in]{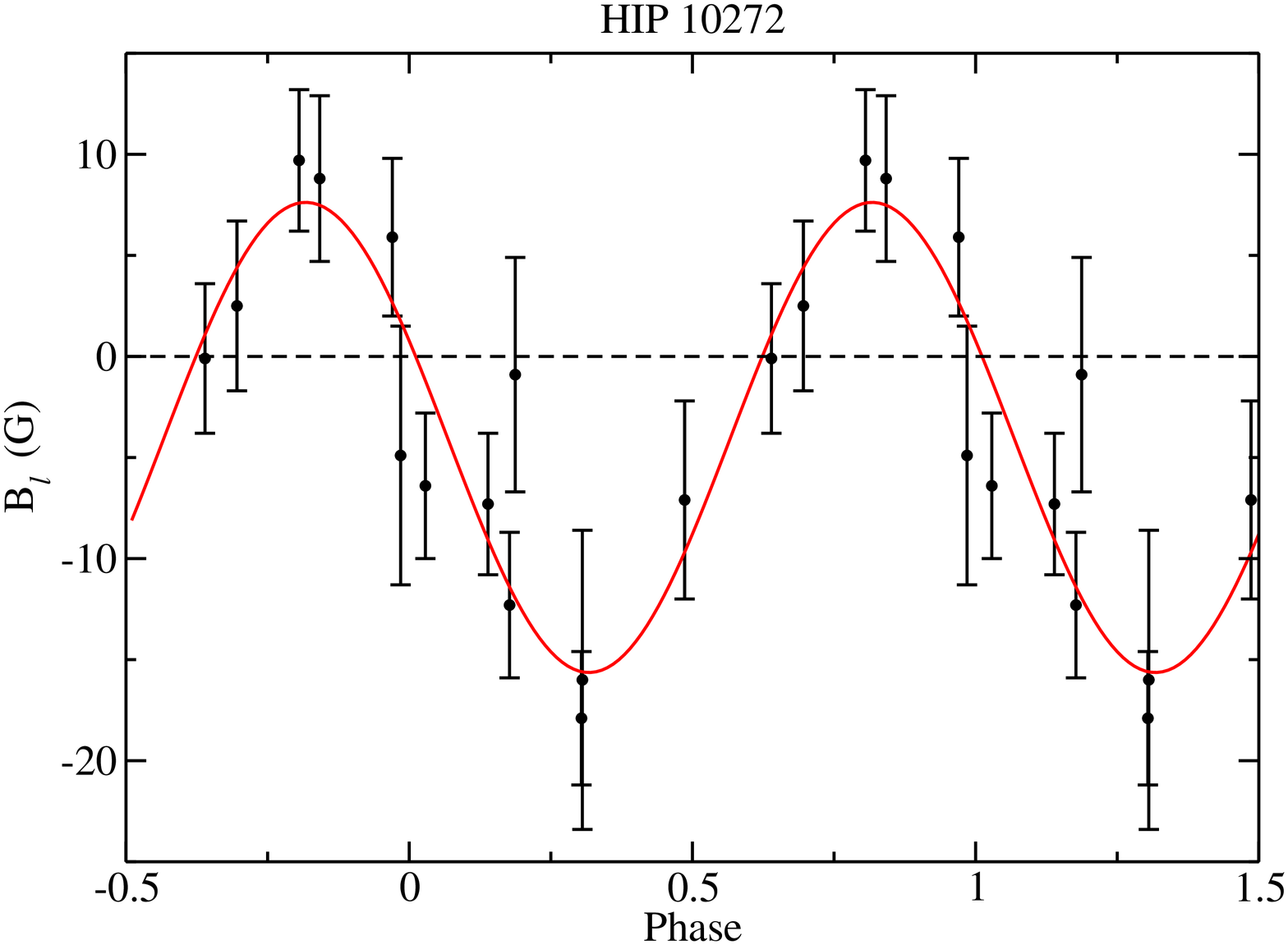}
  \includegraphics[width=2.8in]{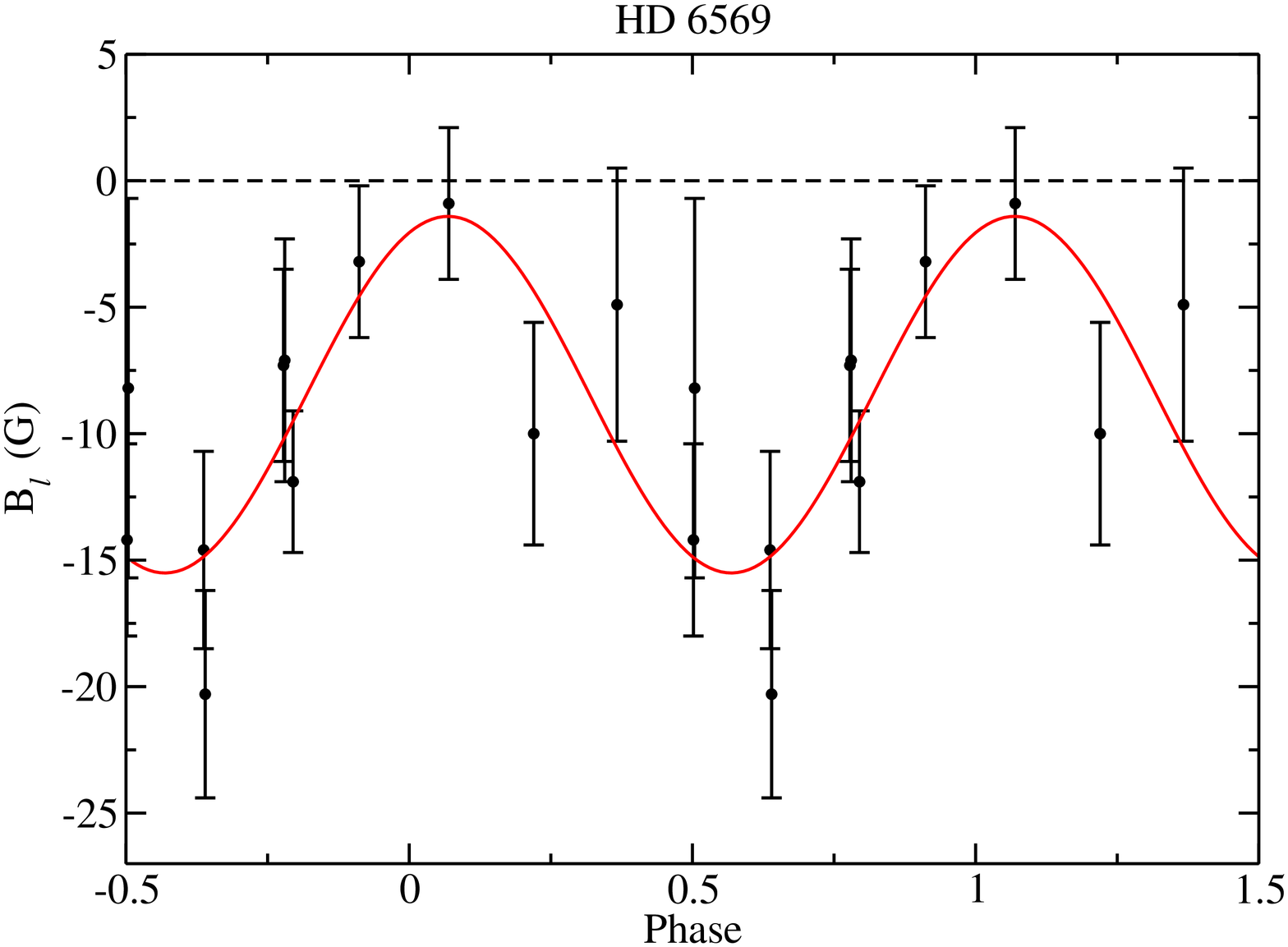}
  \includegraphics[width=2.8in]{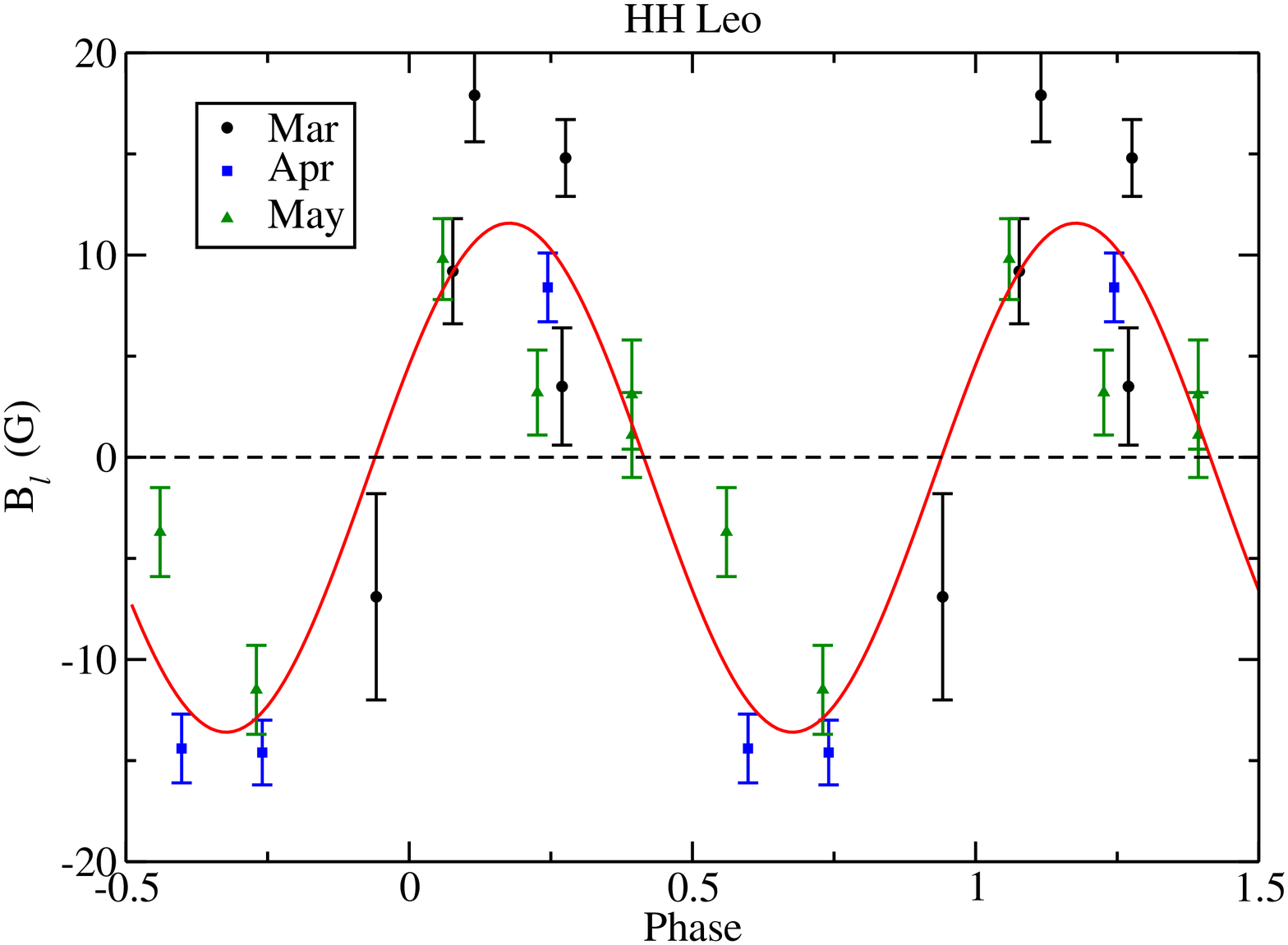}
  \includegraphics[width=2.8in]{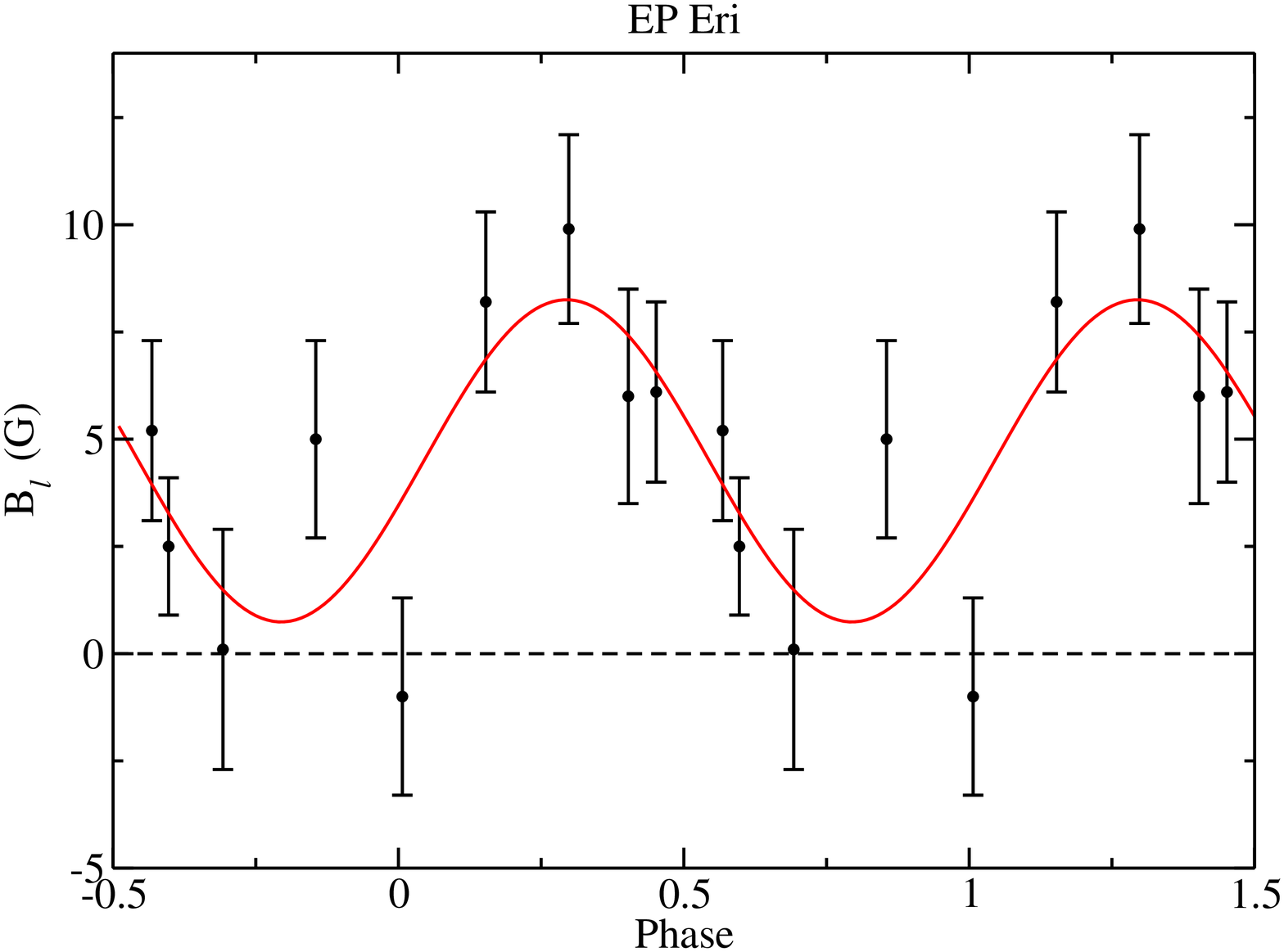}
  \includegraphics[width=2.8in]{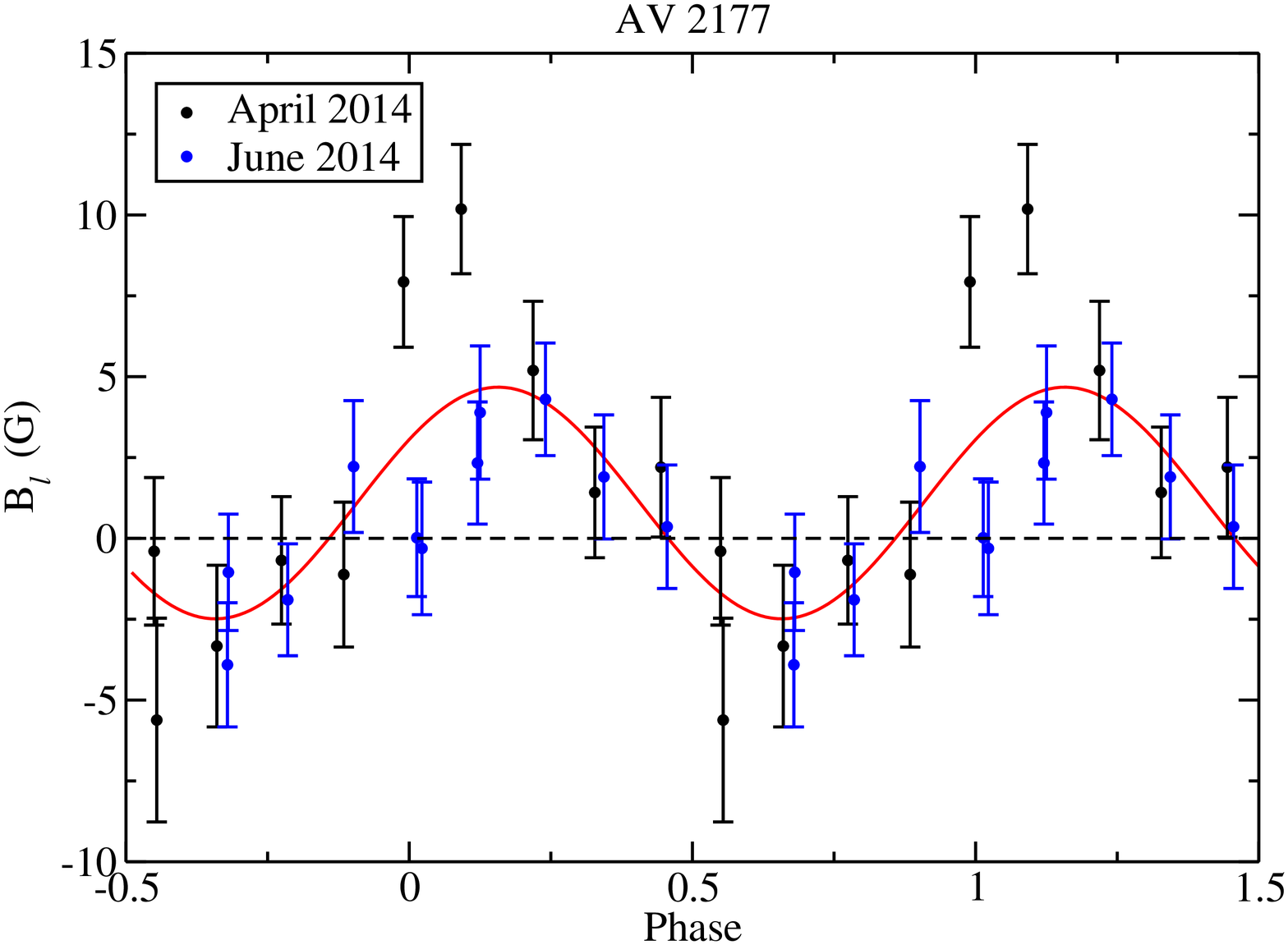}
  \includegraphics[width=2.8in]{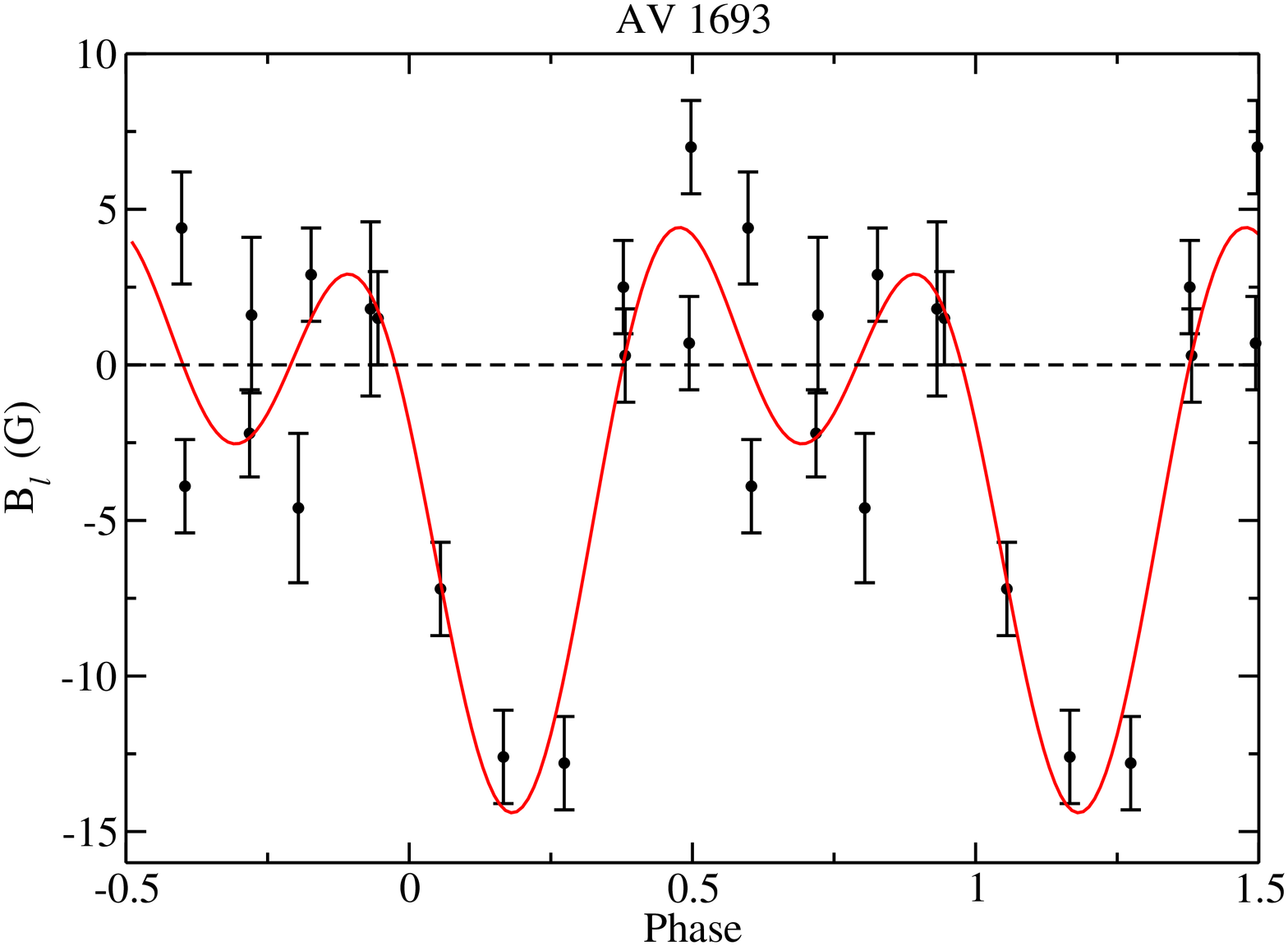}
  \includegraphics[width=2.8in]{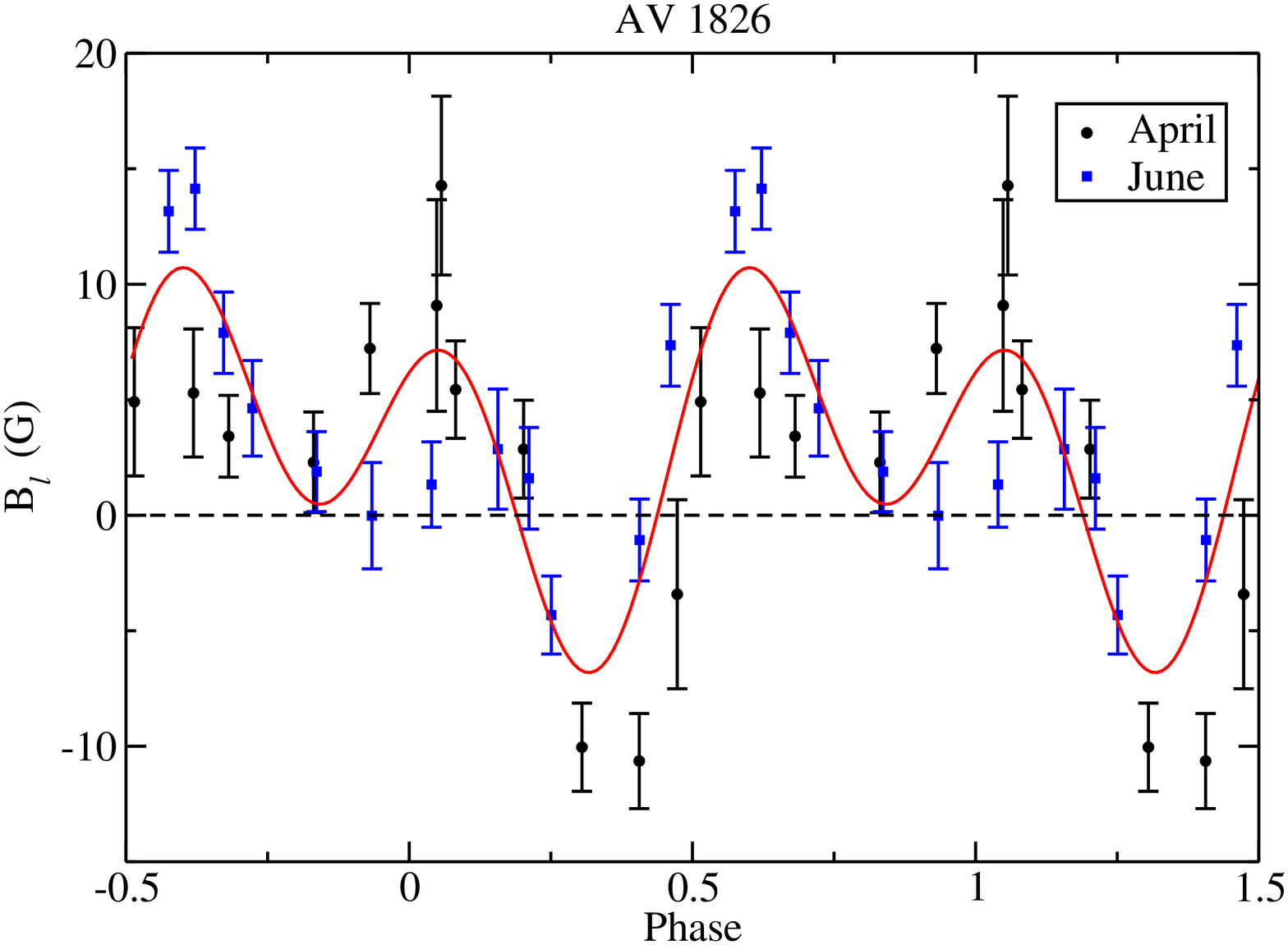}
  \caption{Longitudinal magnetic fields measured for stars in our sample, phased with the rotation periods.  The solid line is the fit through the observations.   }
  \label{fig-bz}
\end{figure*}

\begin{figure*}
  \centering
  \includegraphics[width=2.8in]{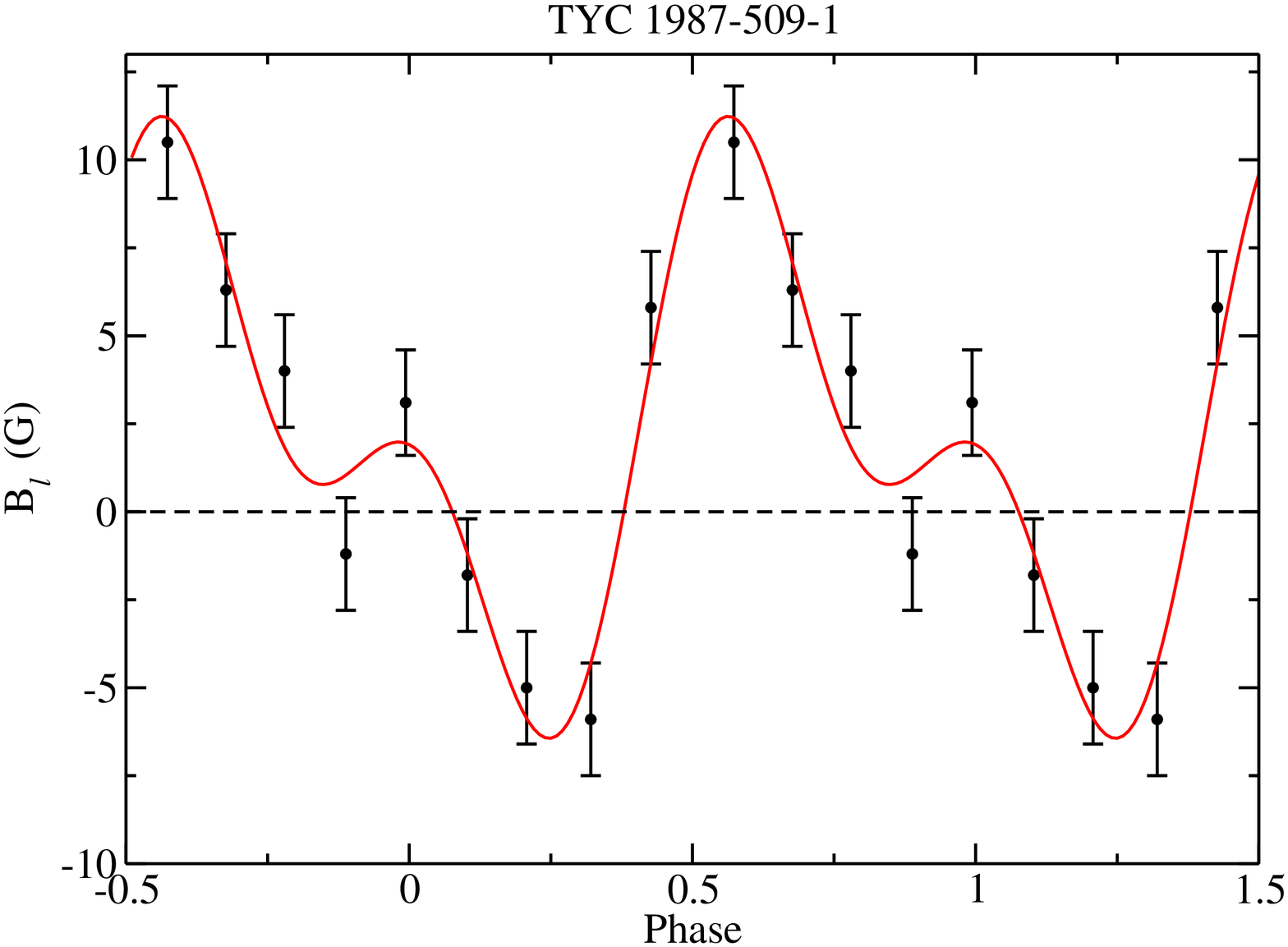}
  \includegraphics[width=2.8in]{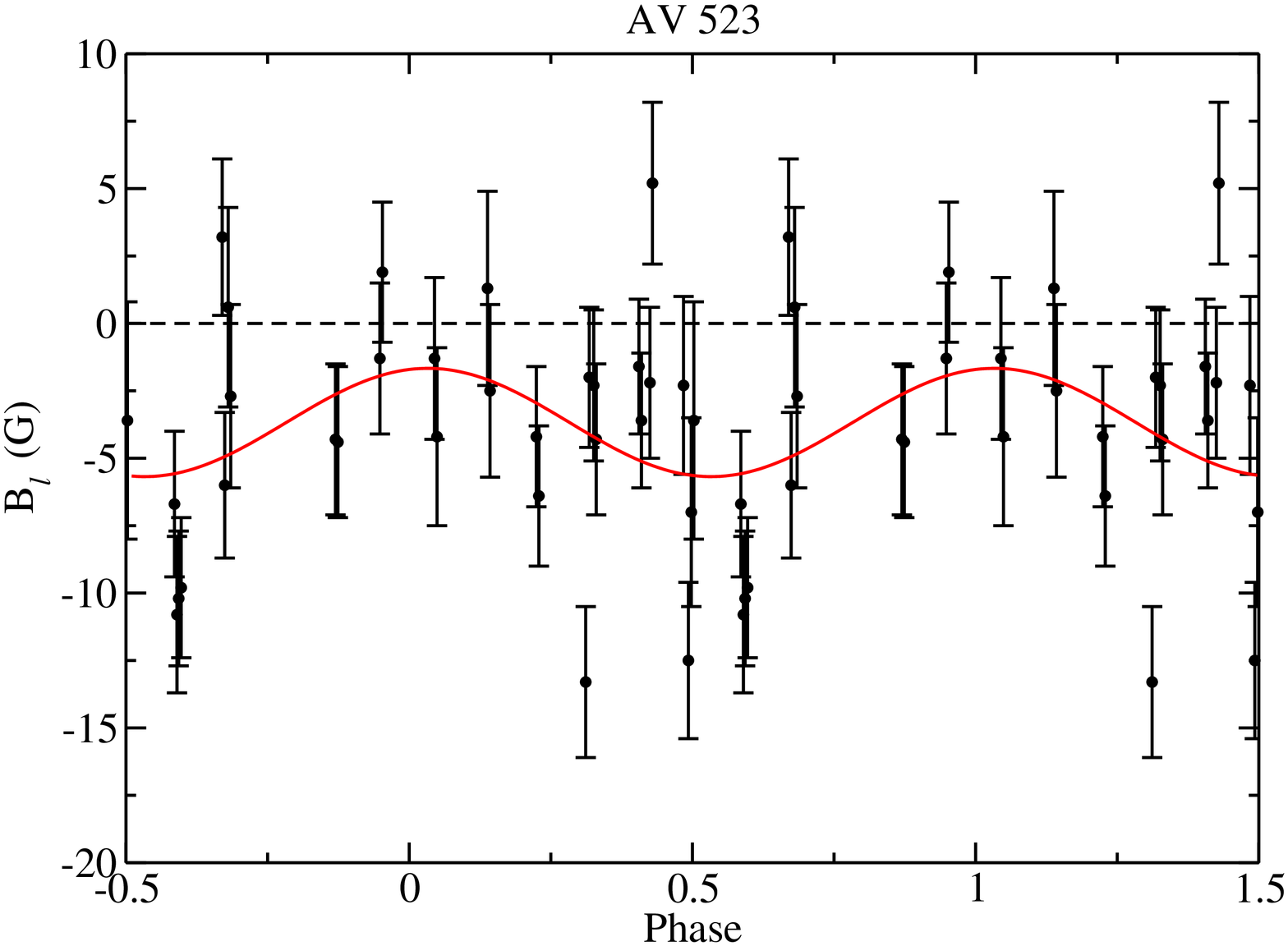}
  \includegraphics[width=2.8in]{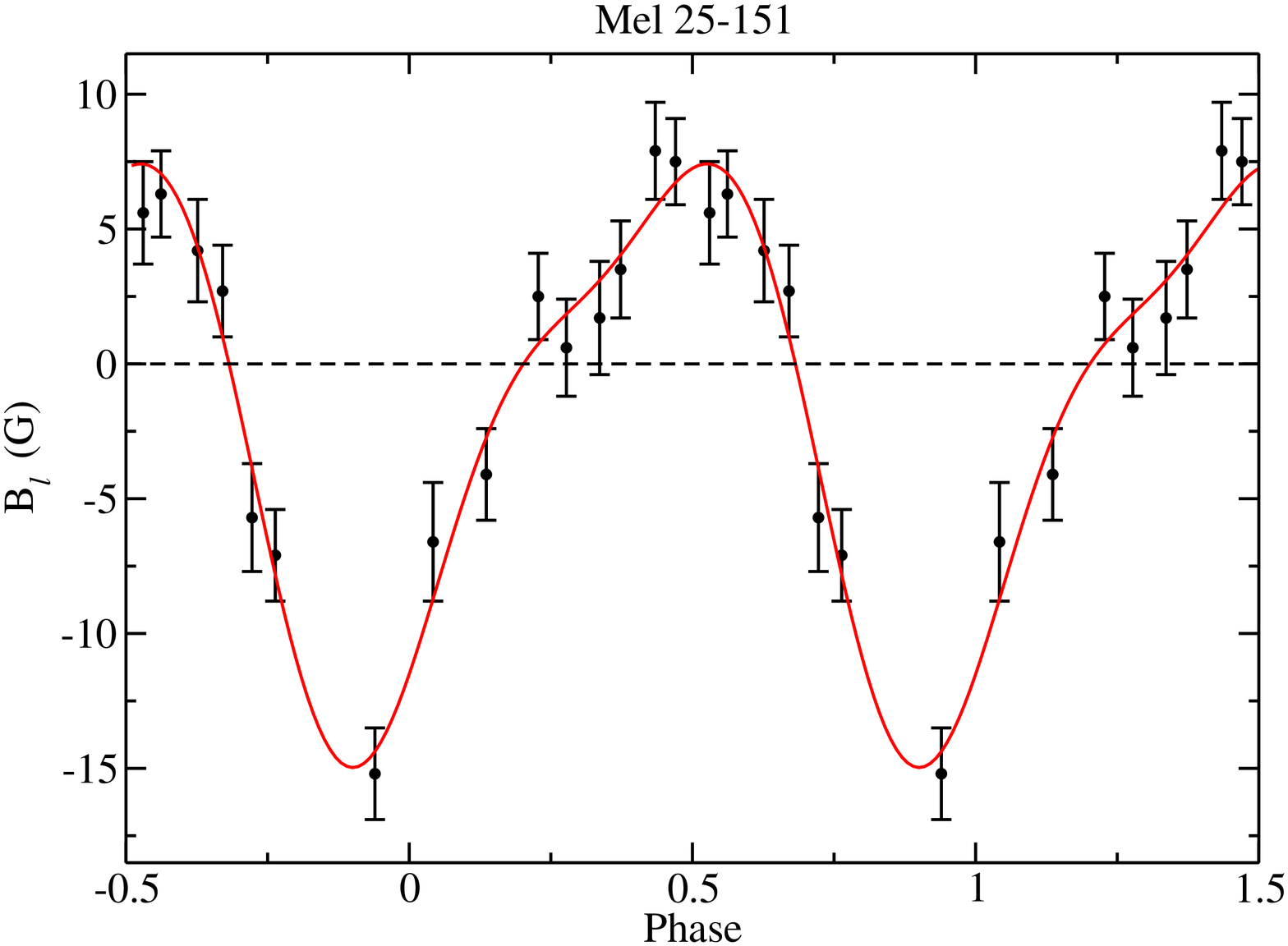}
  \includegraphics[width=2.8in]{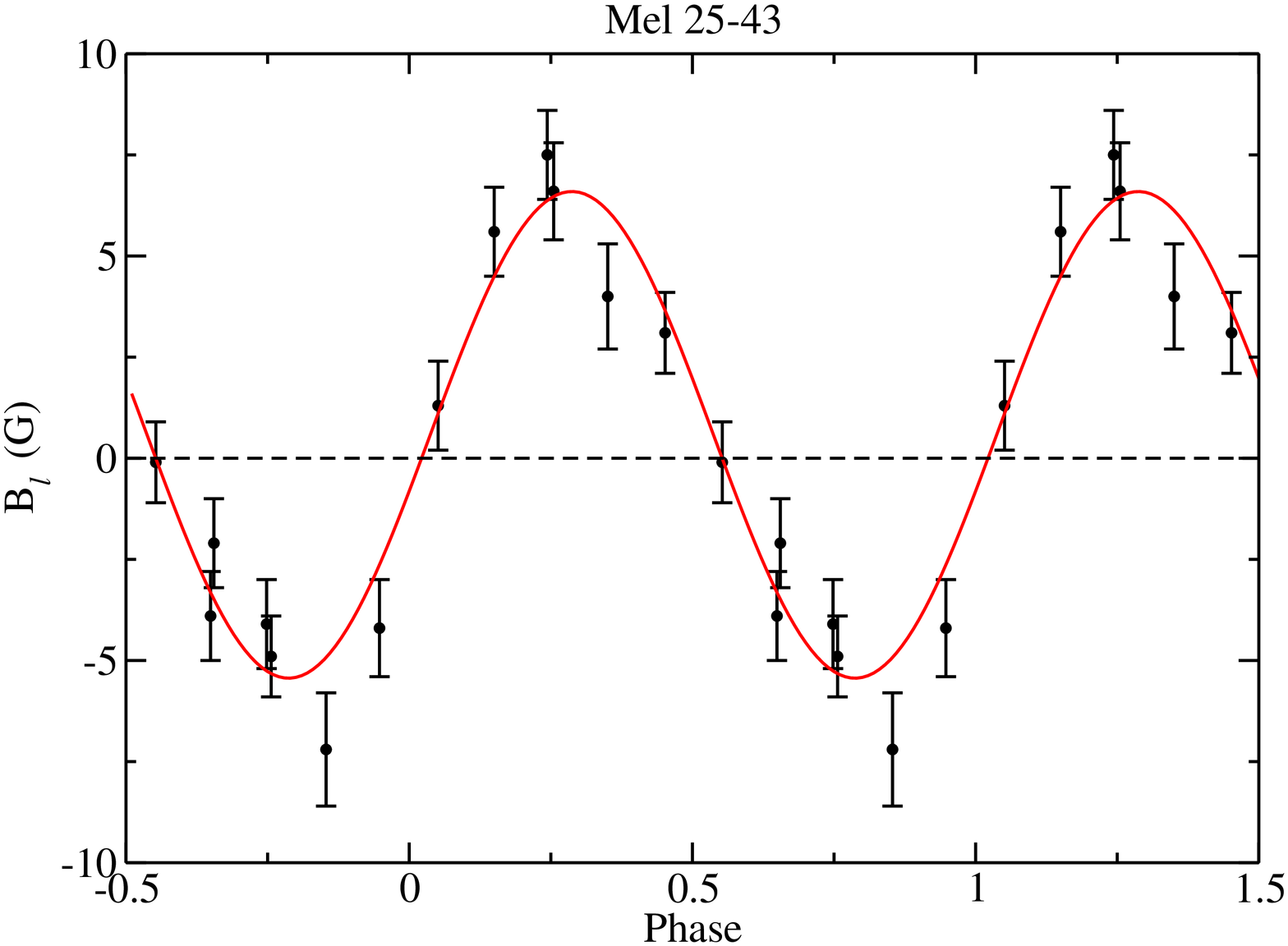}
  \includegraphics[width=2.8in]{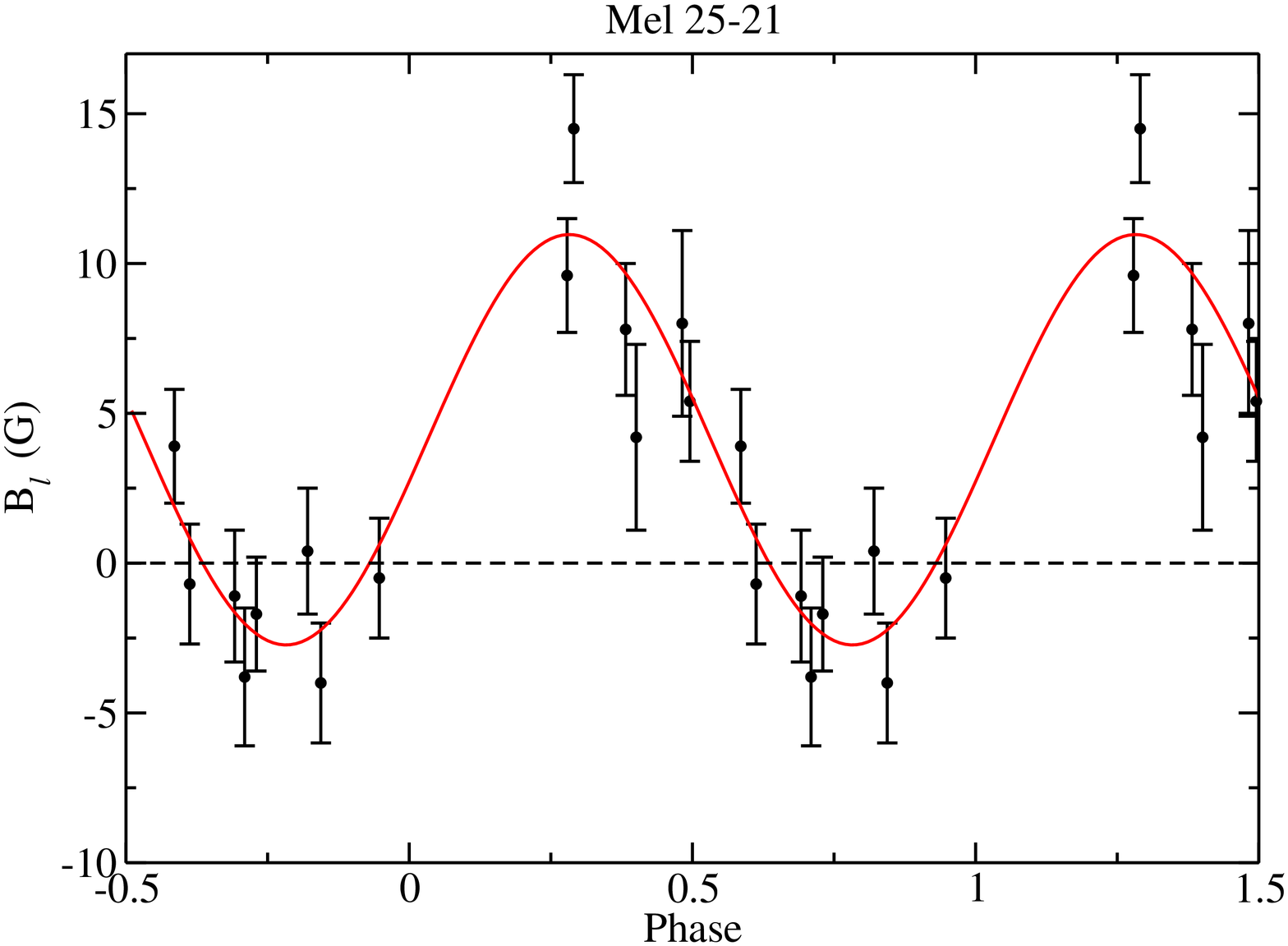}
  \includegraphics[width=2.8in]{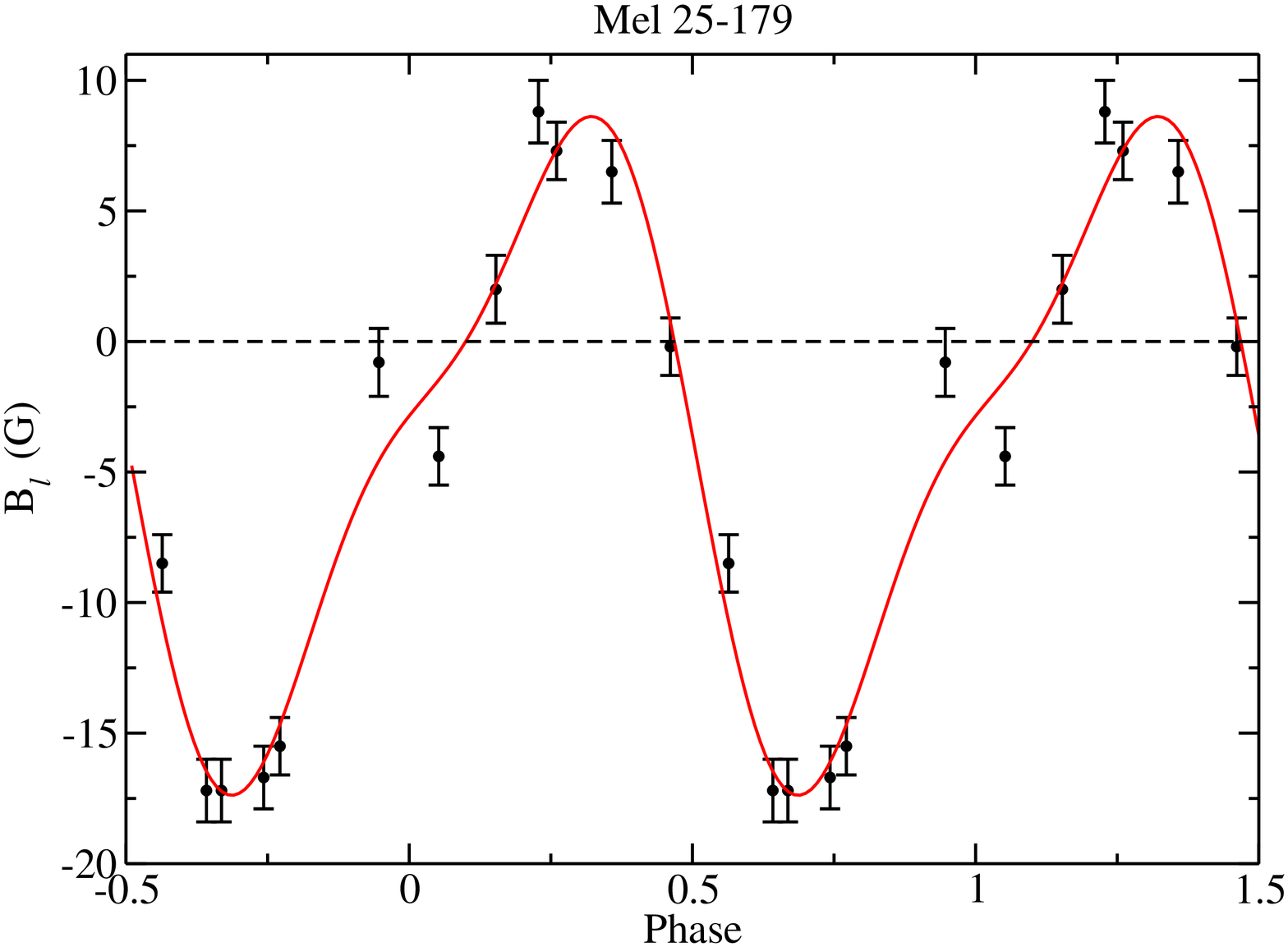}
  \includegraphics[width=2.8in]{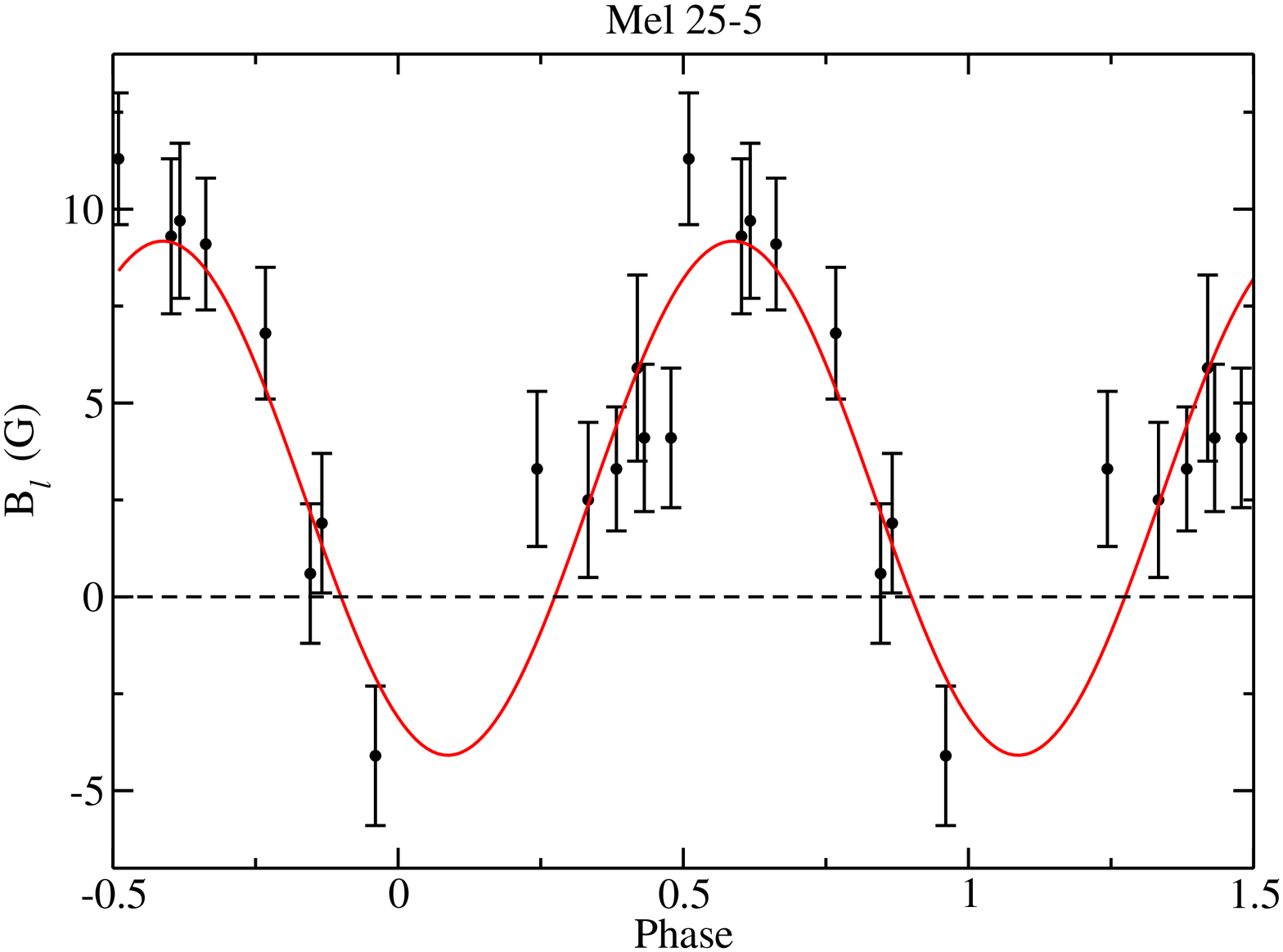}
  \includegraphics[width=2.8in]{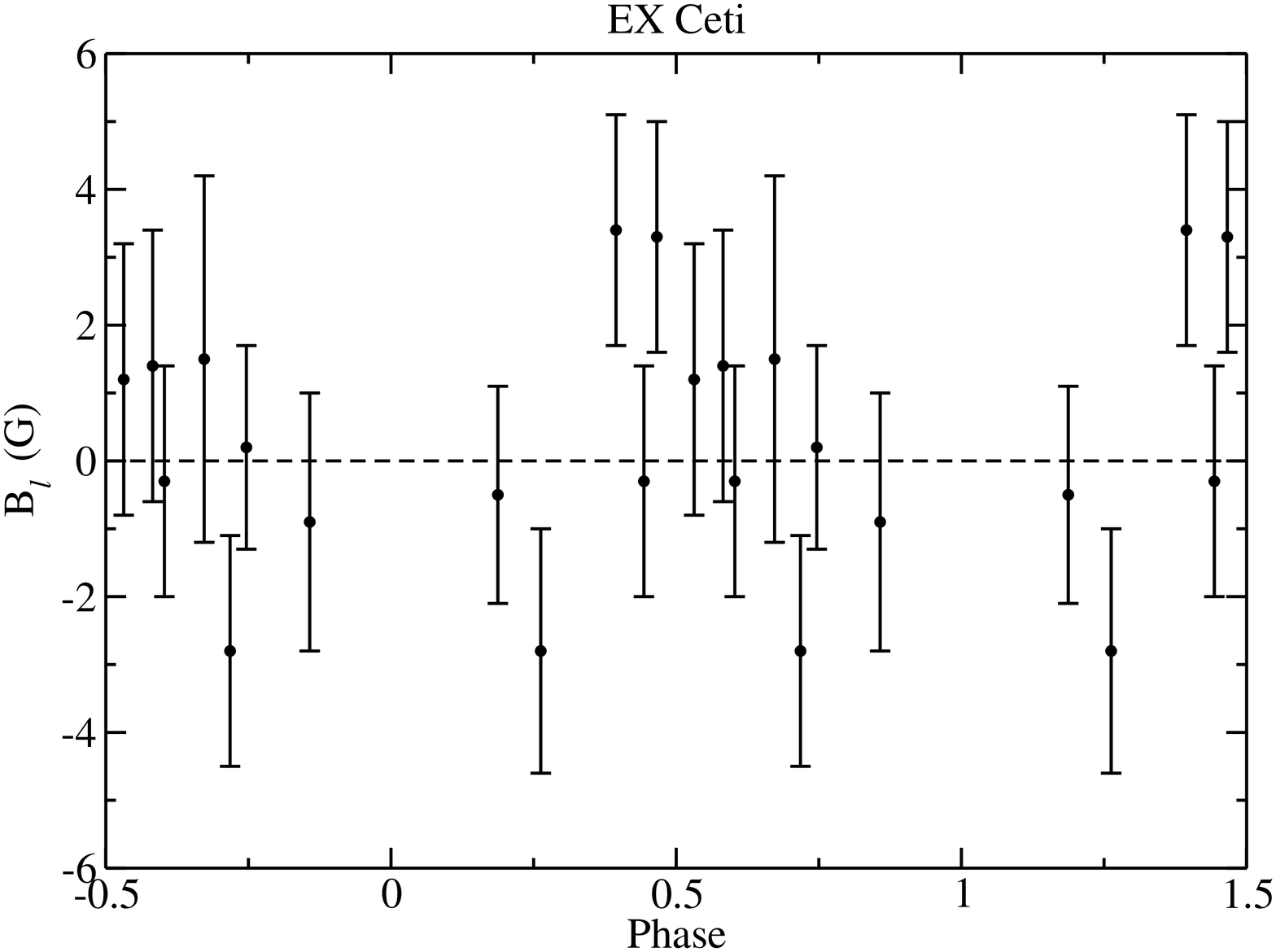}
  \caption{Longitudinal magnetic fields measured for stars in our sample, phased with the rotation period, as in Fig.\ \ref{fig-bz}.  Note that in EX~Cet we do not detect any magnetic field, but it is included here for comparison.  }
  \label{fig-bz2}
\end{figure*}

\begin{figure*}
  \centering
 \includegraphics[width=2.0in]{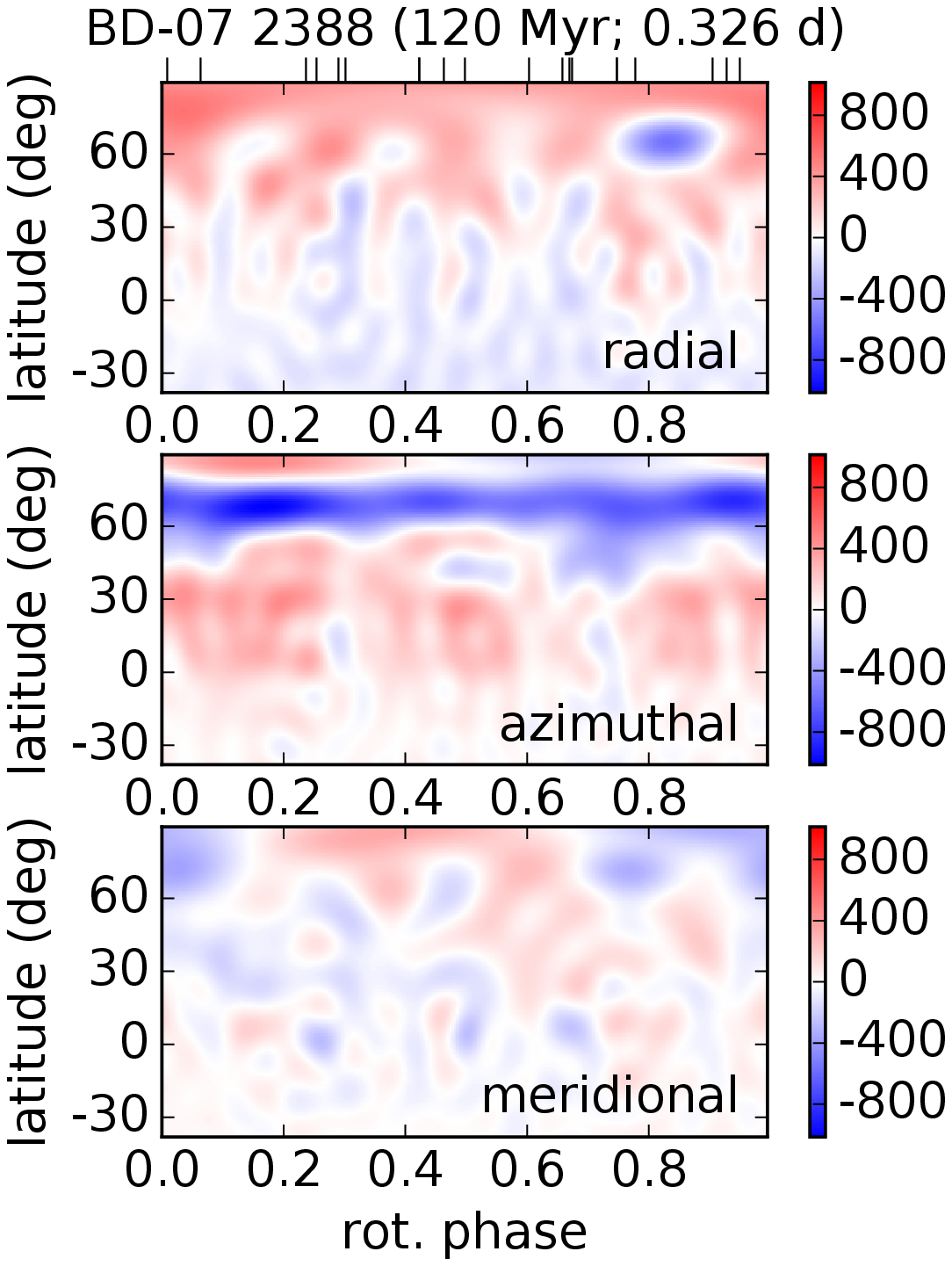}
 \includegraphics[width=2.0in]{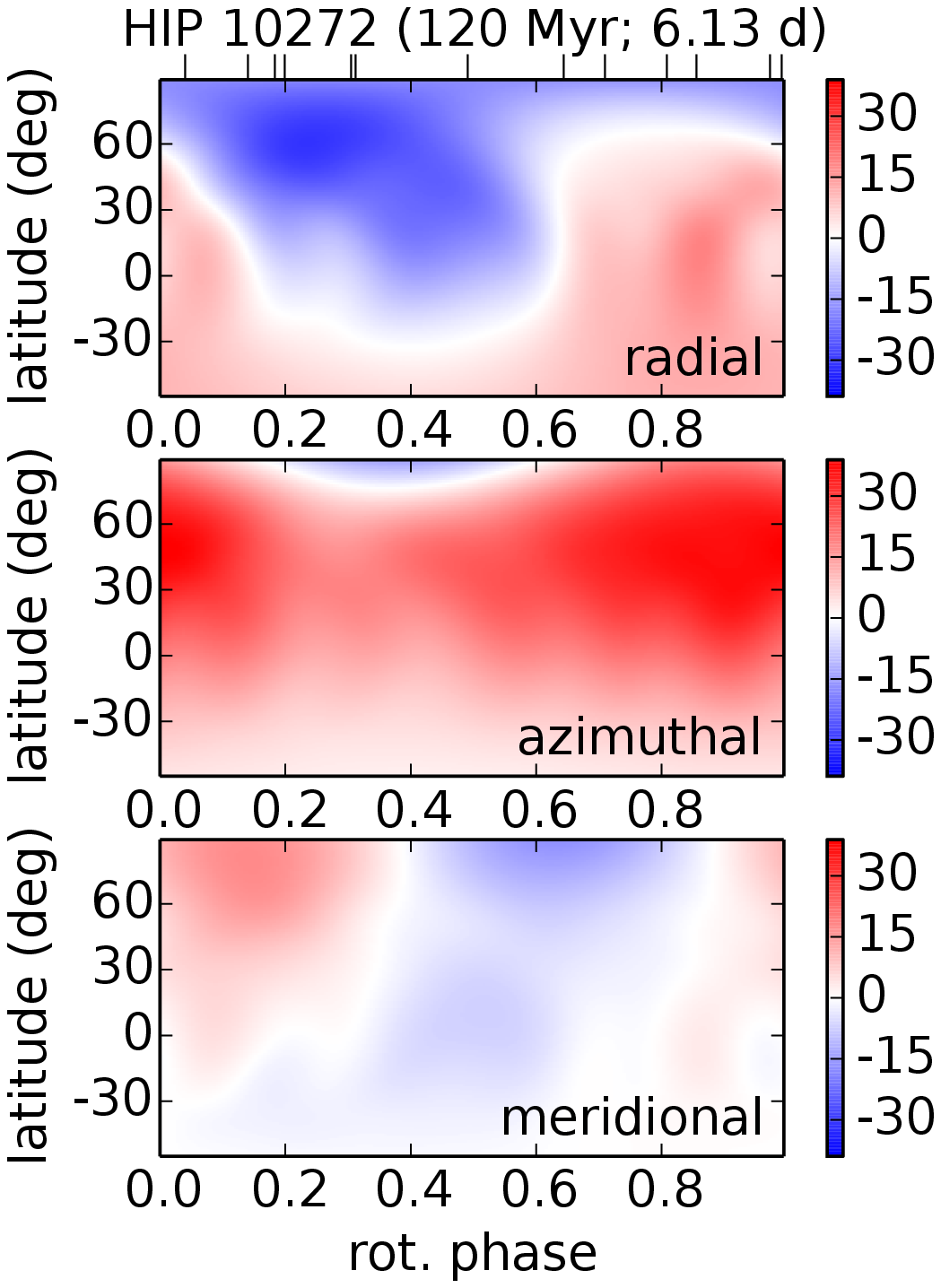}
 \includegraphics[width=2.0in]{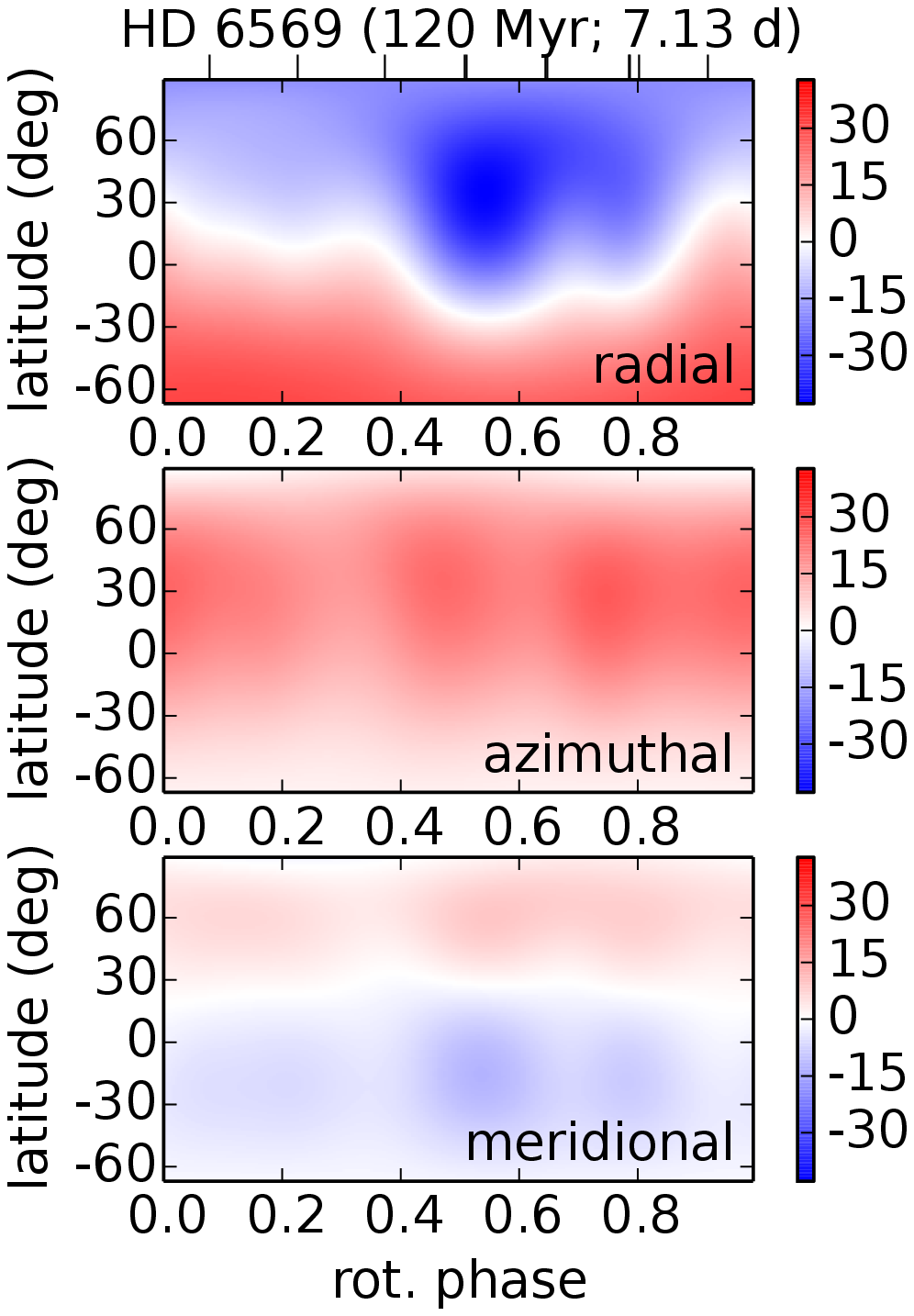}
 \includegraphics[width=2.0in]{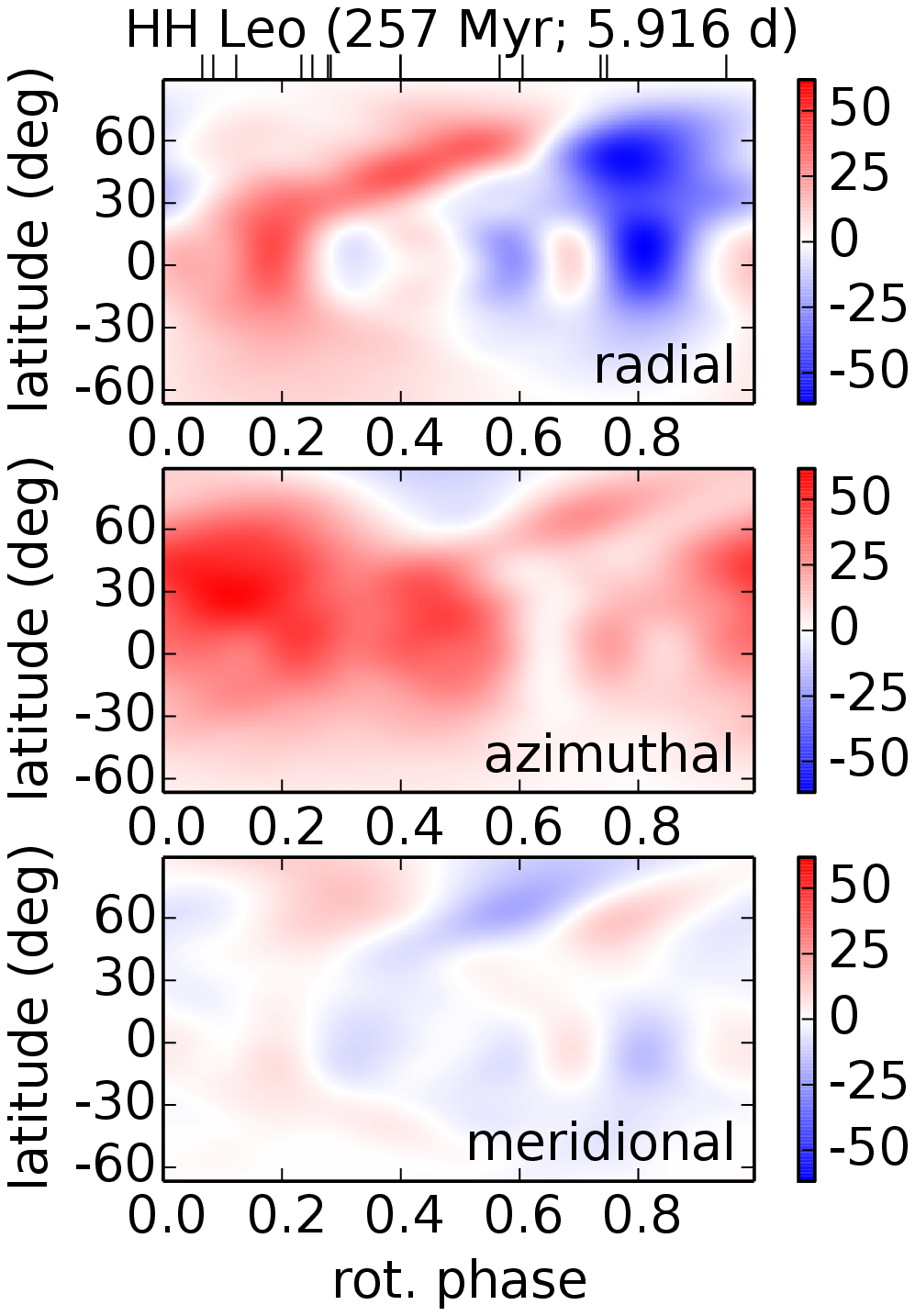}
 \includegraphics[width=2.0in]{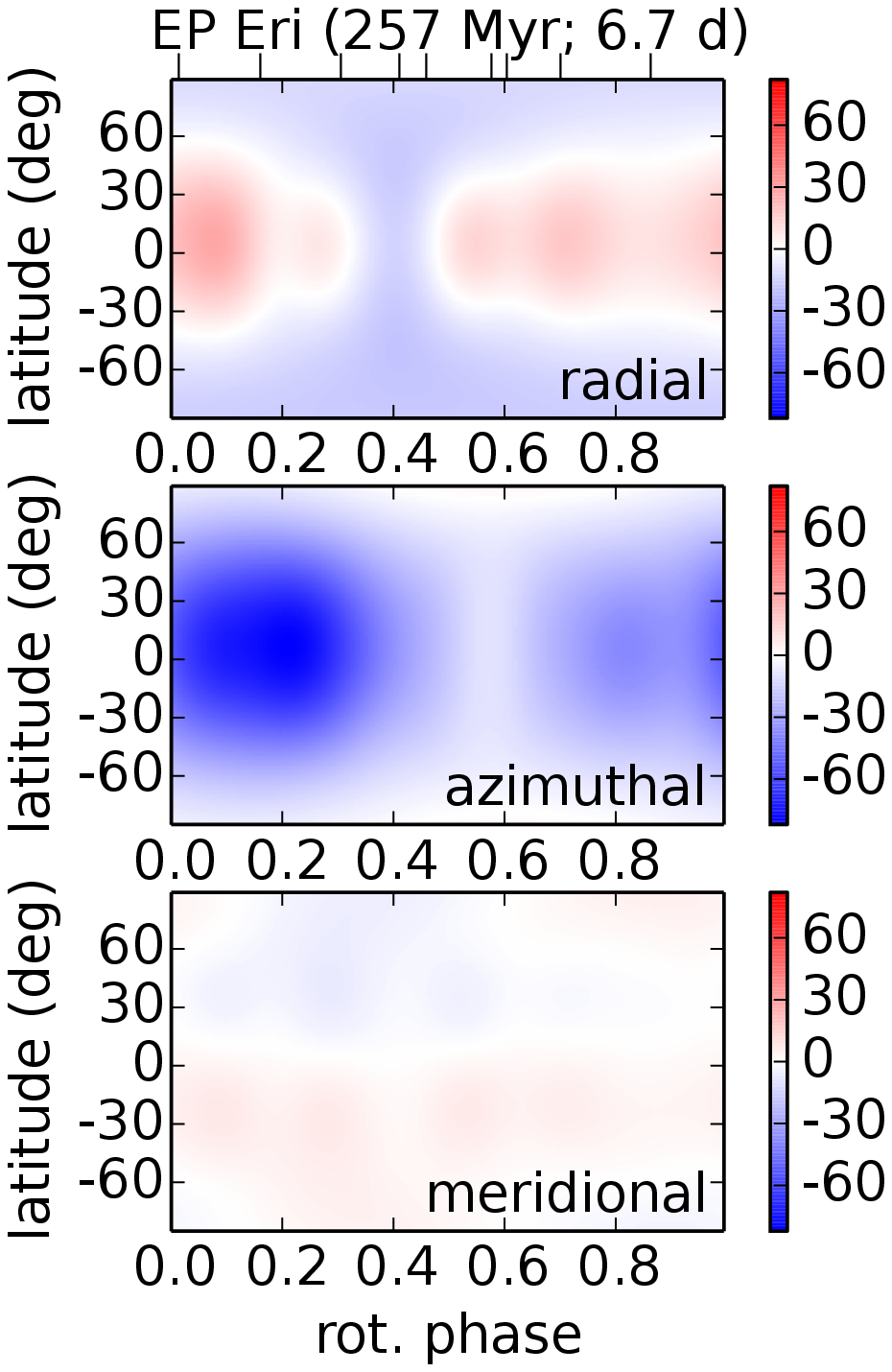}
 \includegraphics[width=2.0in]{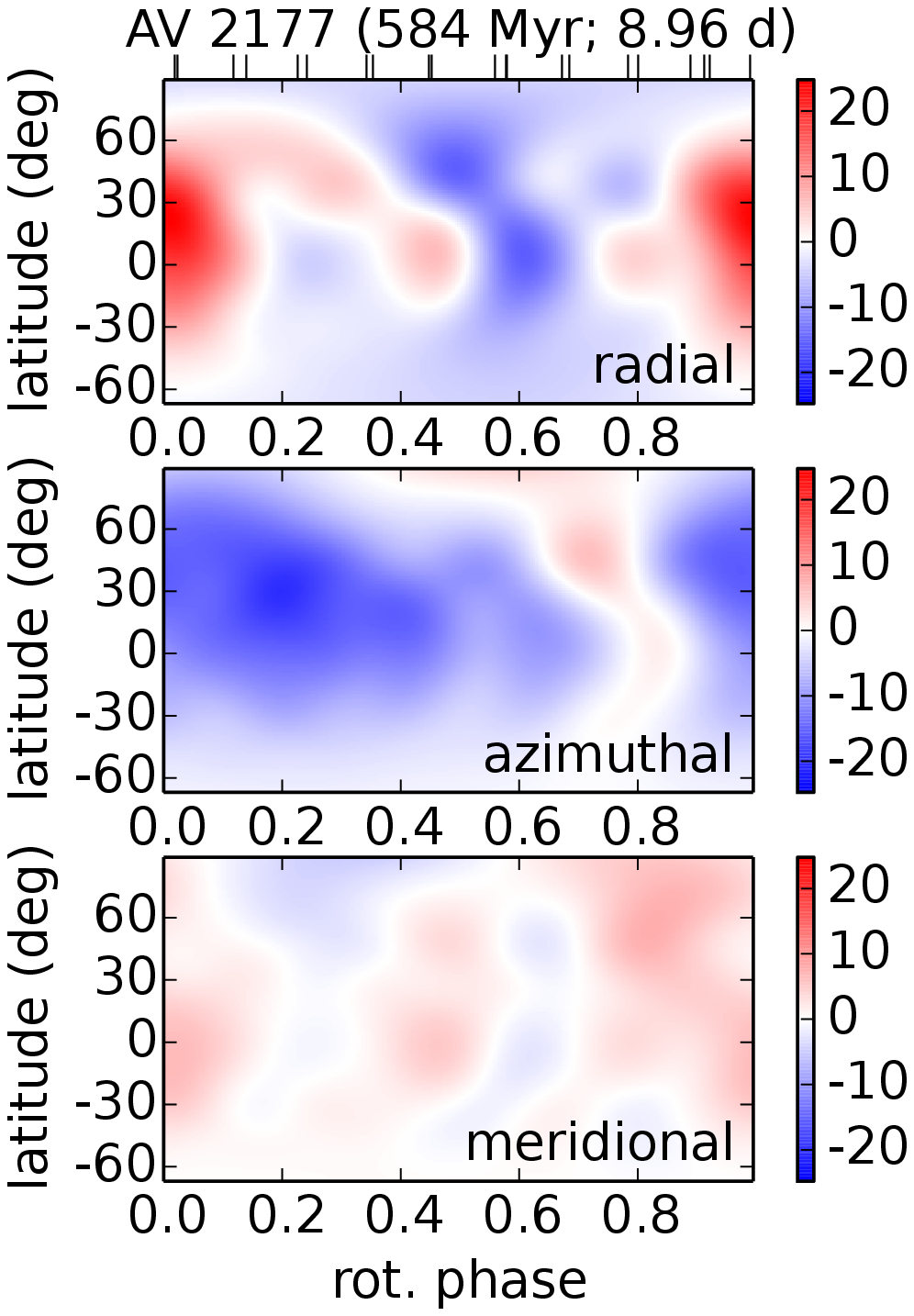}
 \includegraphics[width=2.0in]{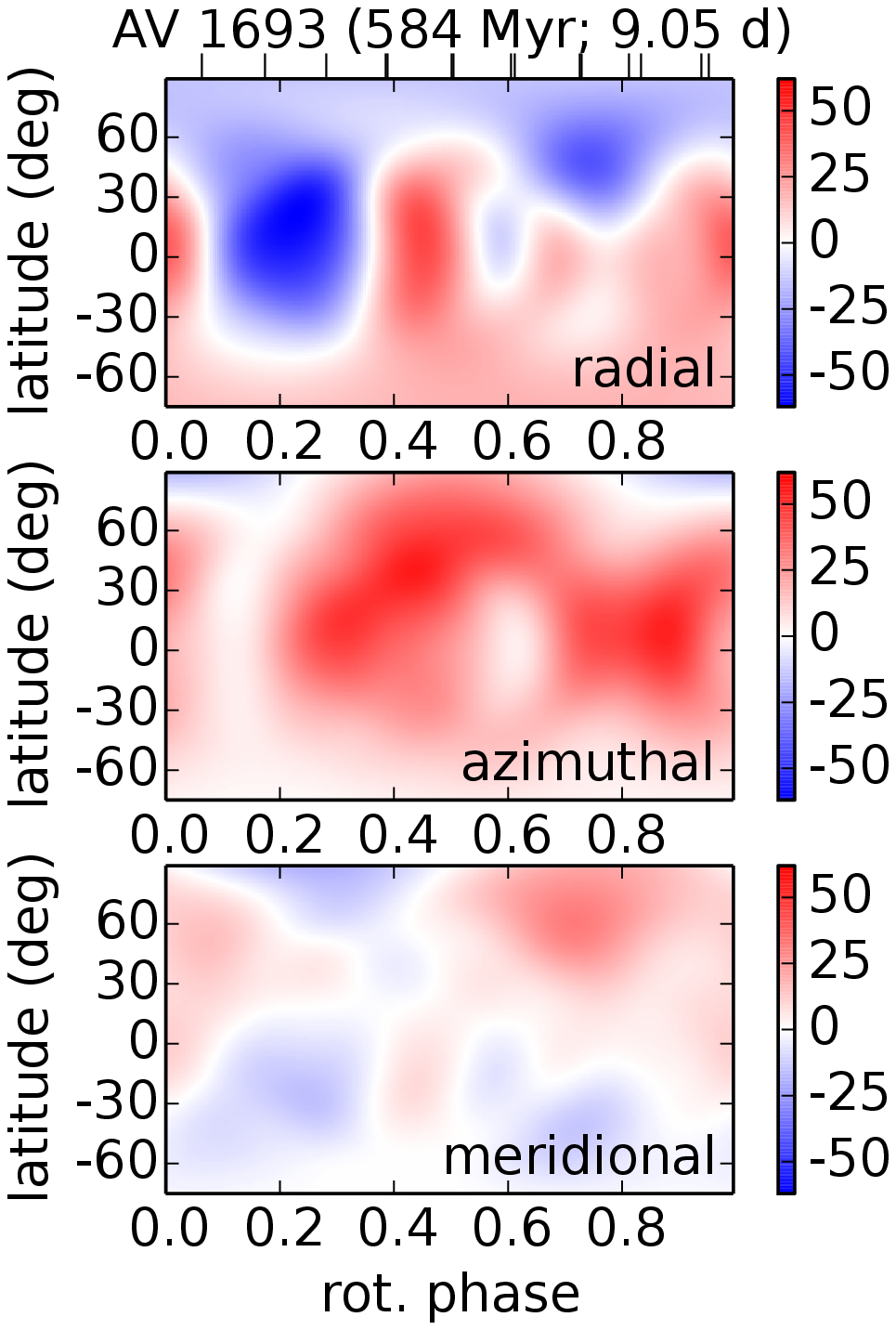}
 \includegraphics[width=2.0in]{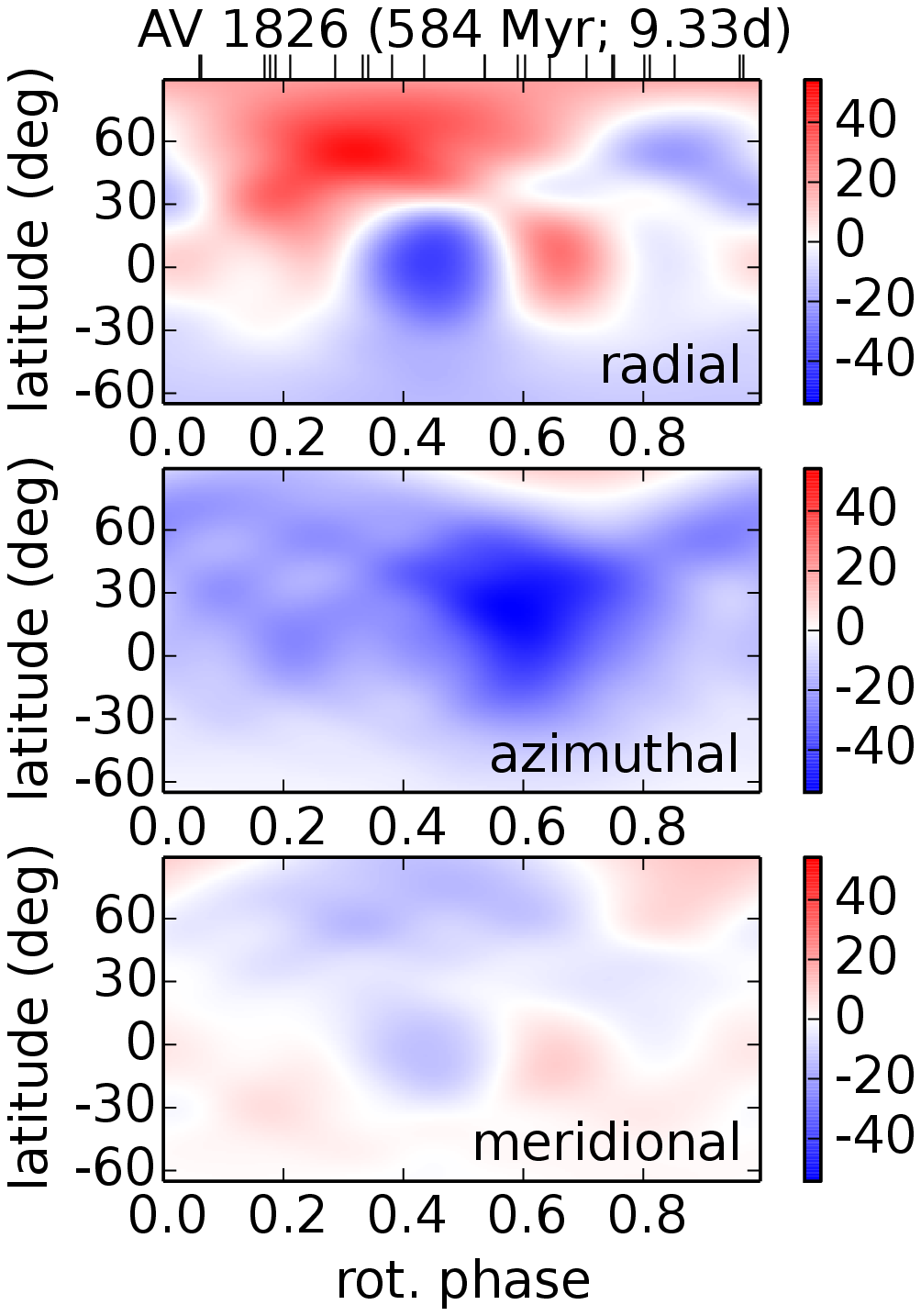}
 \includegraphics[width=2.0in]{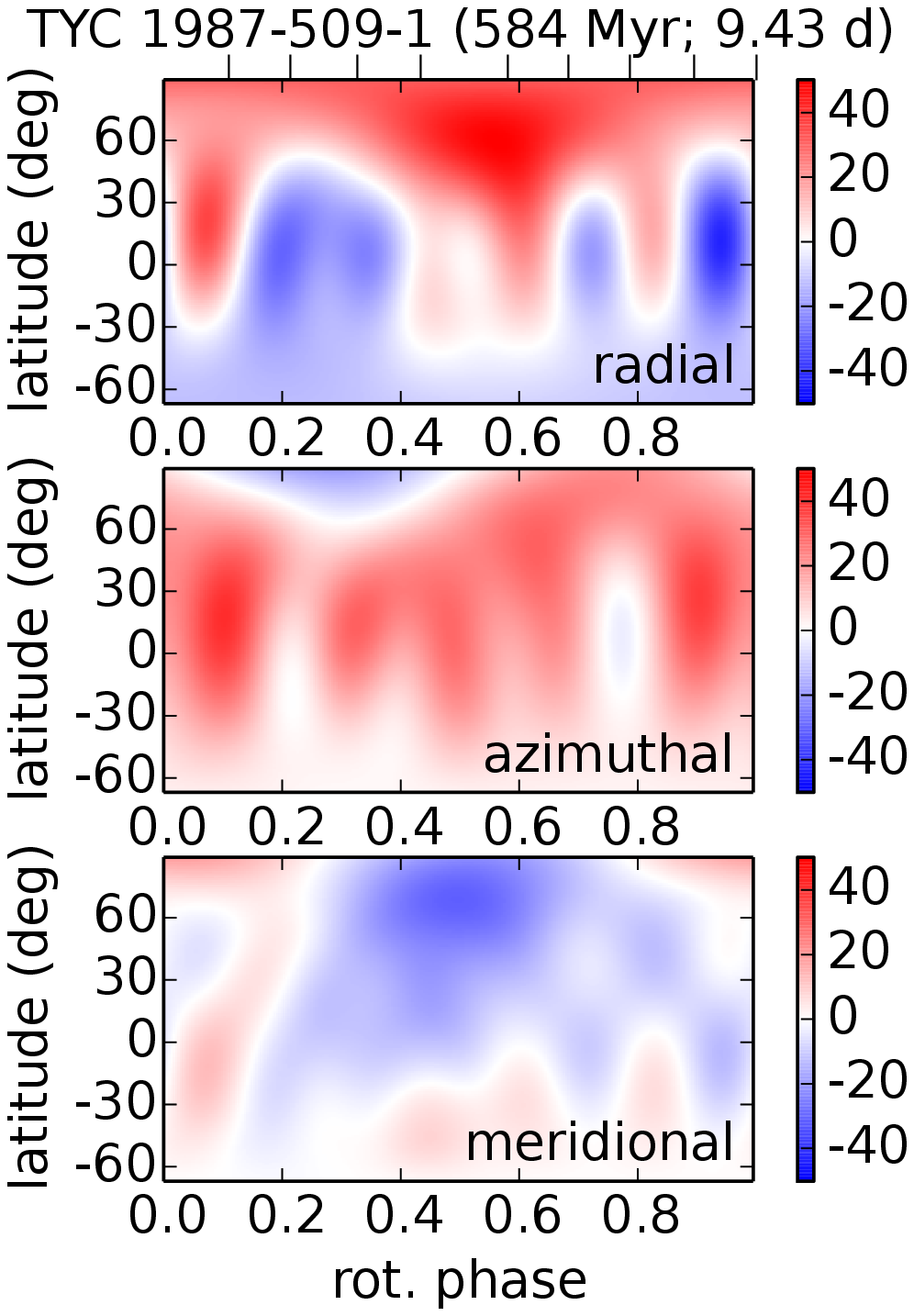}
  \caption{Maps of the derived magnetic fields for the stars in this study.  
  Plotted are the radial (top), azimuthal (middle), and meridional (bottom) components of the magnetic fields.  Sub-figures are labeled by the name of the star, followed by its age and rotation period.  Tick marks at the top of the figure indicate phases at which observations were obtained.  }
  \label{fig-zdi-maps}
\end{figure*}

\begin{figure*}
  \centering
 \includegraphics[width=2.0in]{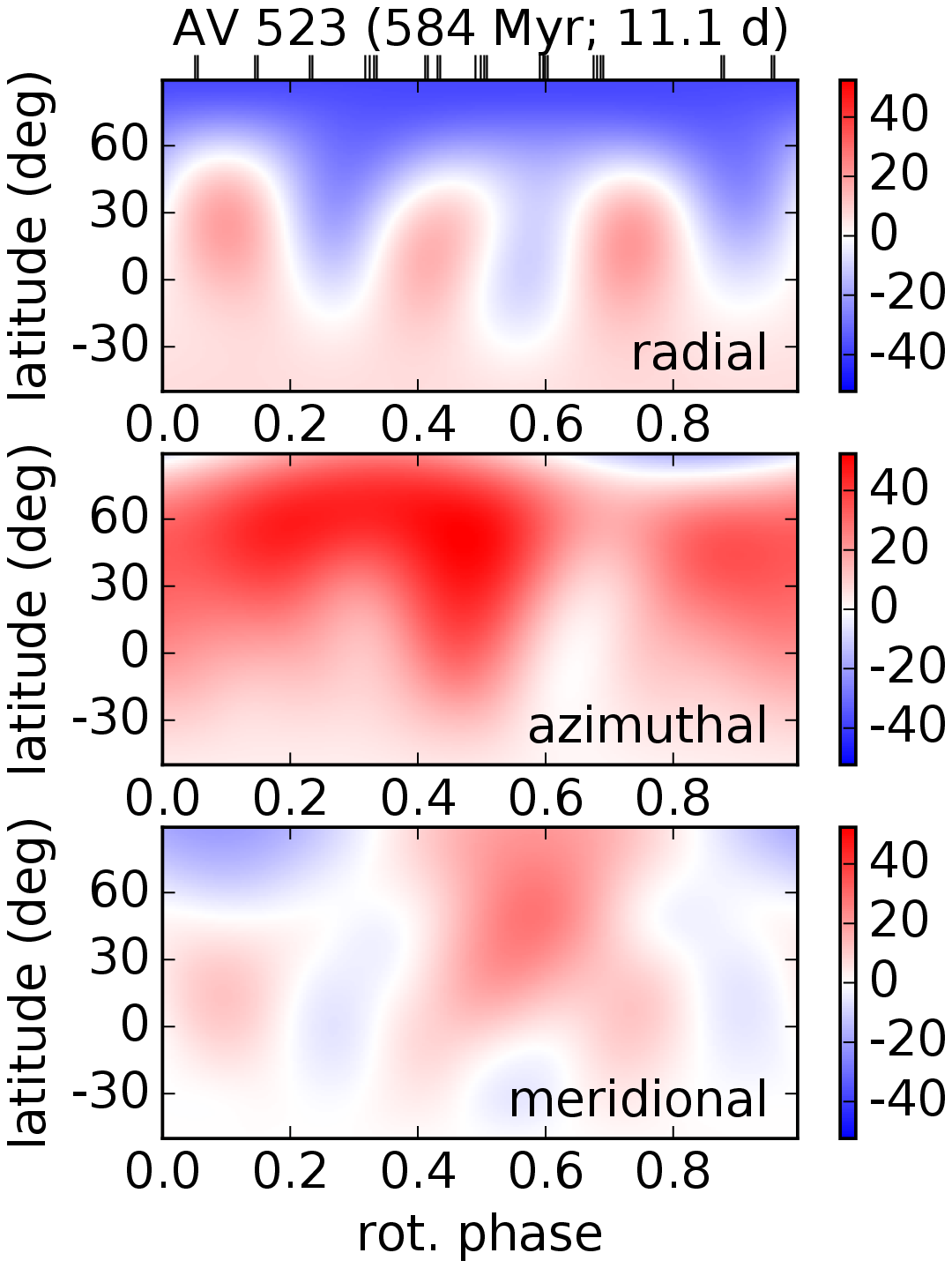}
 \includegraphics[width=2.0in]{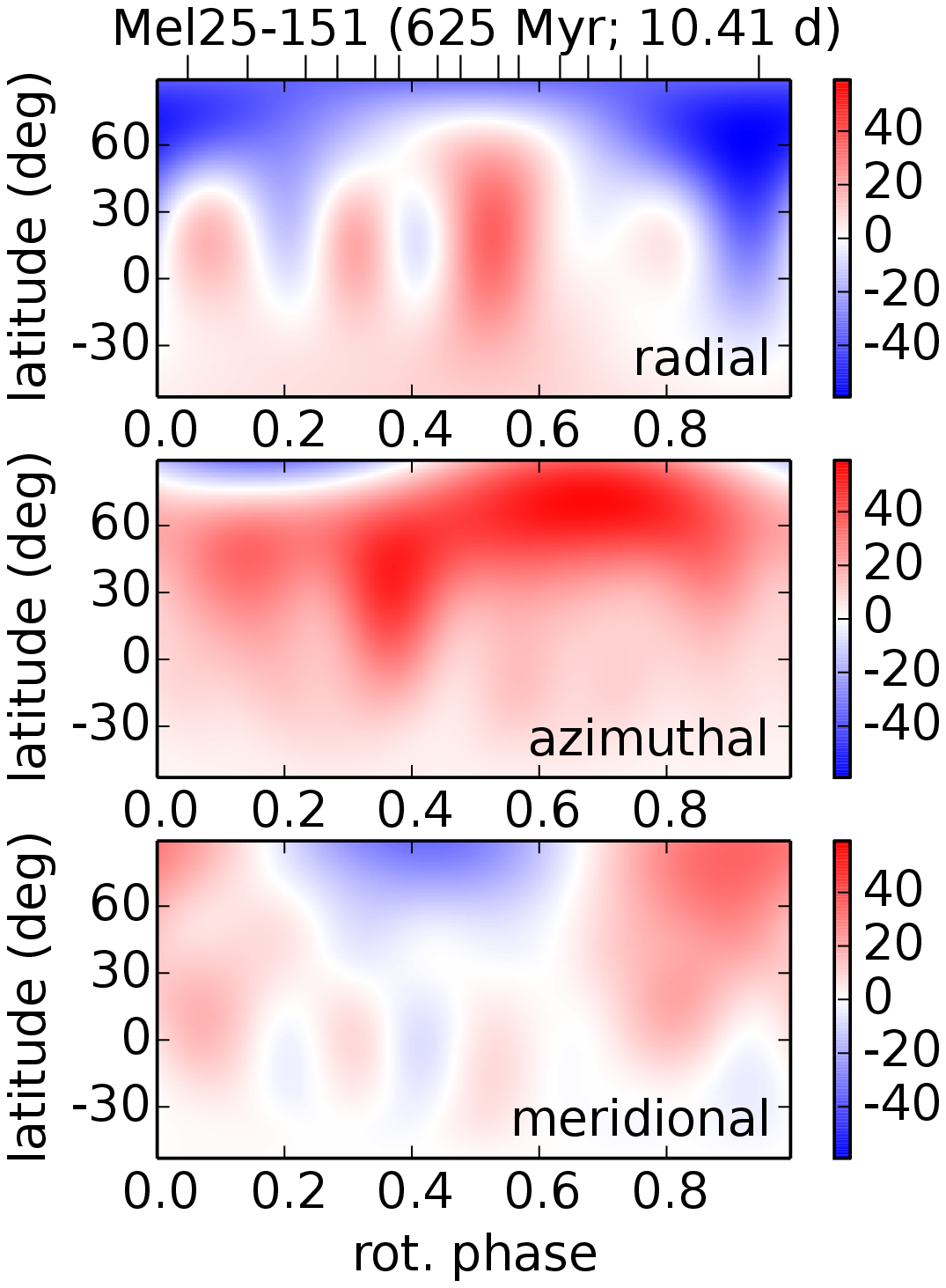}
 \includegraphics[width=2.0in]{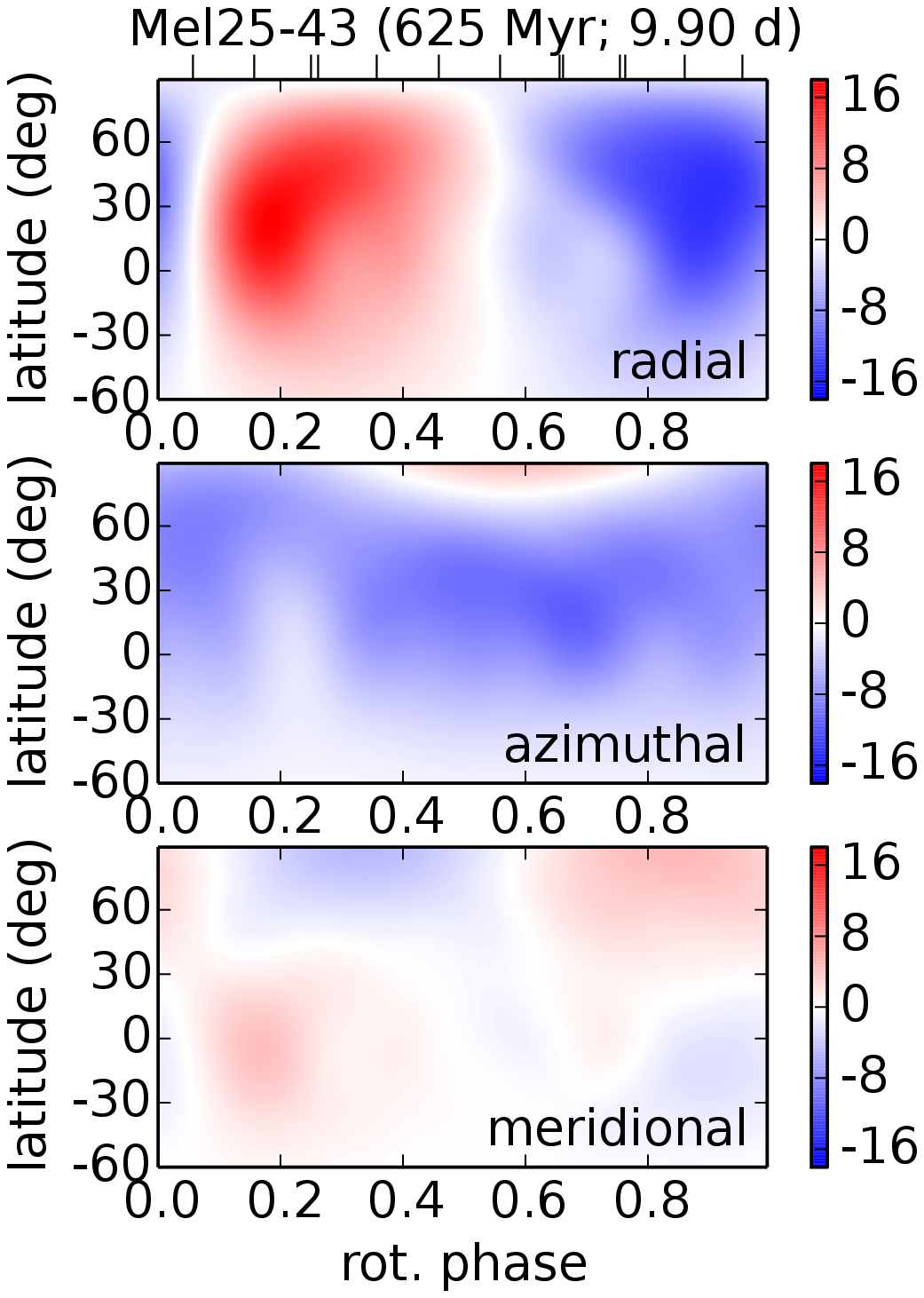}
 \includegraphics[width=2.0in]{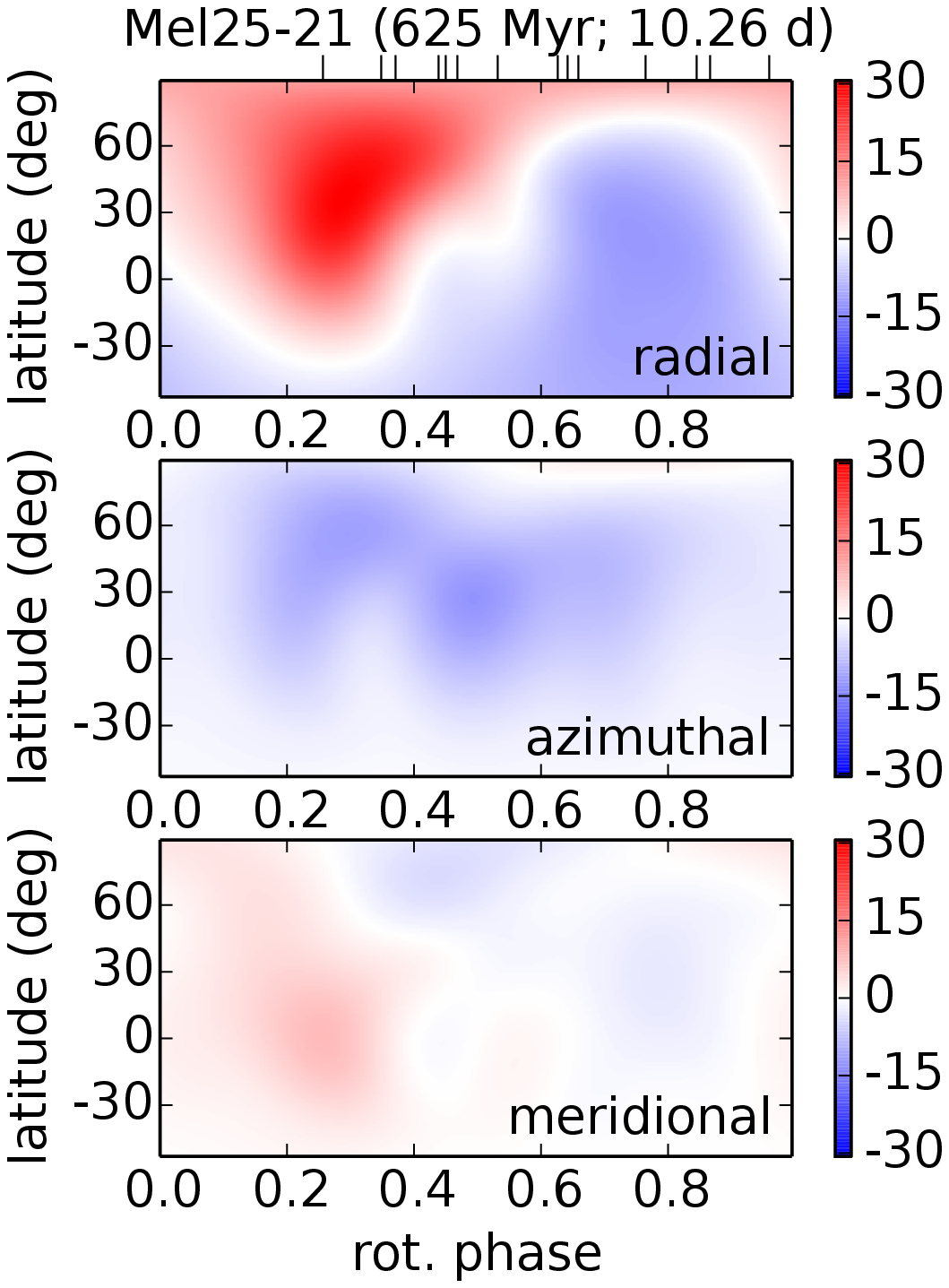}
 \includegraphics[width=2.0in]{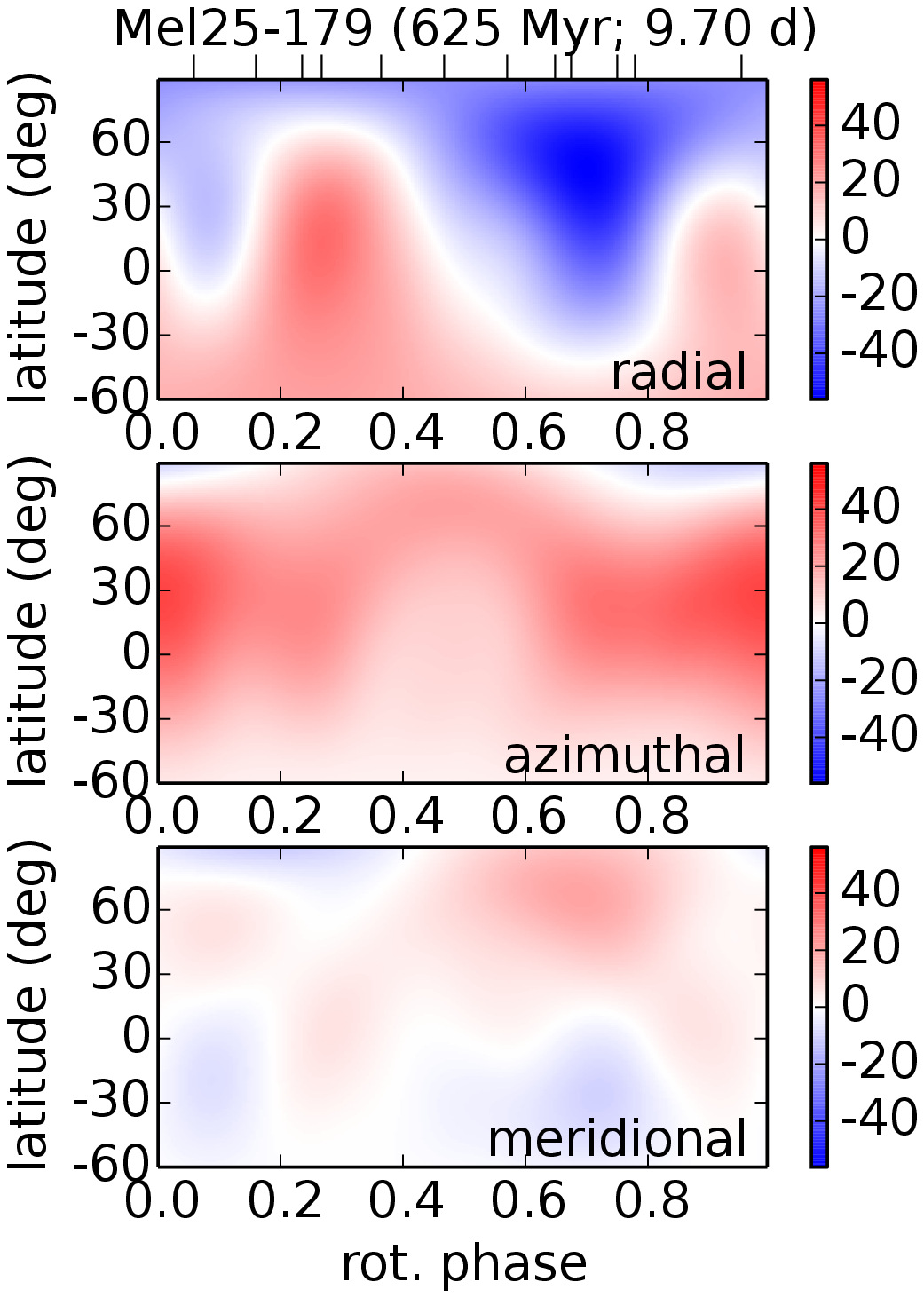}
 \includegraphics[width=2.0in]{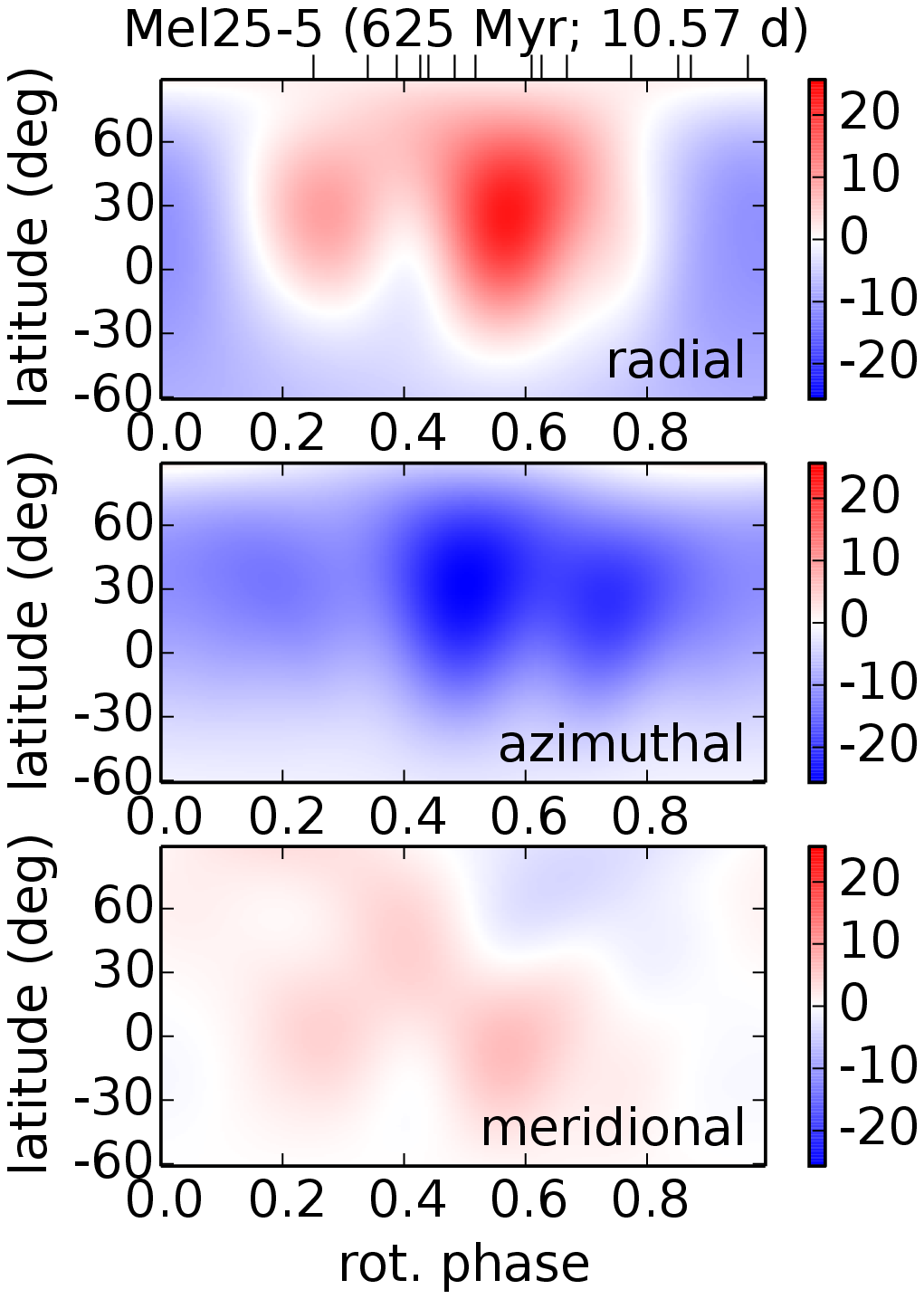} 
  \caption{Maps of the derived magnetic fields for the stars in this study, as in Fig. \ref{fig-zdi-maps}  }
  \label{fig-zdi-maps2}
\end{figure*}

\subsection{BD-07 2388}
BD-07~2388 (TYC 5426-4-1) is a member of the AB Dor association \citep{Torres2008-youngNearbyAssoc}.  \citet{Kiraga2012-photo-periods} find a photometric rotation period of $P = 0.32595$ days.  They do not quote an uncertainty, but note that it was based on 550 observations. 

\citet{Elliott2015-binaries-young-assoc} find that the star is a binary with a $0.11 \pm 0.01$ arcsec separation, and a difference in $K$ magnitudes of $-0.619$ ($K$ primary $= 7.404 \pm 0.030$; $K$ secondary $= 8.023 \pm 0.031$).  Since this is unresolved by 2MASS, we use these magnitudes, and the bolometric calibration for $K$ magnitudes from \citet{Pecaut2013-PMS-BC-withJ} to compute bolometric magnitudes.  Despite the relatively small magnitude difference in $K$, there are no clearly detectable lines of the secondary in our spectra or LSD profiles.  While the temperature of the secondary is unknown, this implies that the secondary is more than one magnitude fainter in $V$.  Based on the intrinsic colors of \citet{Pecaut2013-PMS-BC-withJ}, and the observed $V$ magnitude of the system of \citet{Kiraga2012-photo-periods} (9.323), this implies a difference of 1.8 magnitudes in $V$.  Thus in our analysis we treat the star as spectroscopically single.  

The distance to this star is poorly constrained, since there is no Hipparcos parallax, no Gaia parallax yet, and the star is part of the AB Dor moving group, but not a cluster with a small spacial extent.  \citet{Torres2008-youngNearbyAssoc} derive a dynamical distance to the star (the distance the star should have to be co-moving with the association, based on its proper motion).  Using this distance we derive a luminosity of $1.4 \pm 0.6$ $L_\odot$ and a radius of $1.5 \pm 0.3$ \Rsun.  However, these parameters place the star $\sim 2\sigma$ above the association isochrone.  It is possible the proper motion measurements of the star were influenced by the unrecognized secondary, producing this mild inconsistency.  Thus we prefer to fix the star to the association isochrone and use this constraint to determine the luminosity, radius, and mass.  

Our period search using longitudinal field values produced ambiguous results.  The S/N of the longitudinal field values is poor, due to the complex magnetic field in the star, the modest S/N of the observations, and the very high \vs\ that distributes the signal in $V$ over many spectral pixels.  Periods near 0.325 days are a minimum in $\chi^2$ (reduced $\chi^2$, $\chi_{\nu}^2 = 0.98$), but so are periods at 0.388 days ($\chi_{\nu}^2 = 0.88$), 0.280 days ($\chi_{\nu}^2 = 1.05$), and an apparent alias at 0.64 days ($\chi_{\nu}^2 = 0.95$).  The 0.325 days period is the best for a second order fit ($\chi_{\nu}^2 = 0.70$, compared to 0.85, 0.87, and 0.82, respectively), however 0.388 days is slightly better for a first order fit, and the difference between periods in $\chi^2$ is not large.  
Our period search from radial velocity measurements is a little less ambiguous.  Periods around 0.326 days are best for first, second and third order fits.  However secondary minima at 0.485 days and 0.244 days cannot confidently be ruled out.  
The period from ZDI is less ambiguous, however it still produces two values for the period, $0.3265 \pm 0.0010$ days or $0.489 \pm 0.001$ days, with equal entropy values.  Although we note that the 0.326 days period produces a simpler phasing of the $V$ profiles by eye.  Since the 0.32595 day period of \citet{Kiraga2012-photo-periods} is the only one present in all our estimates, and often marginally the best period, we conclude that this is indeed the true rotation period of the star.

The high \vs\ of BD-07~2388 leads to larger uncertainties on the parameters derived from spectroscopic analysis.  As a test, we performed the full spectrum fit for all the observations individually, then took the average and standard deviation of the results.  The average derived parameters were very close to our single spectrum analysis, in this case by less than one standard deviation.  More over, the standard deviation of values from individual windows was typically twice the standard deviation of values for one window but different observations.  This confirms that, even in this extreme case, the uncertainties on the physical parameters are dominated by systematic uncertainties (both errors in atomic data, and in this case the heavily spotted photosphere) not random photon noise.

Given the uncertainty in the stellar radius and the danger of a systematic error, discussed above, we estimate the inclination of the star from the maximum entropy ZDI solution.  The best inclination from ZDI was $38 \pm 13 $ degrees. 

The relatively low S/N of the Stokes $V$ spectra make a precise determination of differential rotation difficult, despite having observations over several rotation cycles.  Our best fit model is ${\rm d}\Omega = 0.16\pm 0.23$ rad/day and a period of $0.325 ^{+0.02}_{-0.03}$ days, which is not significantly different from zero.  Thus we do not detect differential rotation in this star.

\subsection{HIP 10272}
HIP 10272 (HD 13482, BD+23 296) is a member of the AB Dor association \citep{Zuckerman2004-ABDor-members, Torres2008-youngNearbyAssoc, McCarthy2014-ABDor-detailed}.  It has a rotation period of $6.13 \pm 0.03$ days from \citet{Messina2010-RACE-OC-periods}, based on photometry.  The star has a K4 companion at 1.8'' distant \citep{Torres2008-youngNearbyAssoc}.  With Narval's 1.6'' diameter pinhole, the secondary should have been outside the pinhole in all observations, however with poor seeing some light from the secondary may have been collected.  We see no evidence for the secondary in our spectra, thus we conclude that we successfully observed a single star. 

The two stars appear to have been unresolved in the 2MASS catalog, thus our luminosity estimate based on this is overestimated.  The luminosity from this value falls well above the association isochrone, and is inconsistent with our spectroscopic \lgg.  \citet{McCarthy2014-ABDor-detailed} note the difficulty in finding a luminosity for the star that is consistent with the AB Dor association isochrone.  They propose that either the Tycho-2 photometry that they used contains light from an unresolved star, or that the distance from \citet{van_Leeuwen2007-Hipparcos_book} is overestimated by $\sim$10 pc.  If we use the Hipparcos photometry for the components of the system, and the $V$ bolometric correction of \citet{Pecaut2013-PMS-BC-withJ}, we find a luminosity that is closer to the association isochrone, but still inconsistent by more than $3\sigma$.  Given the ambiguities in the luminosity from photometry, we assume the star is on the association isochrone (effectively the ZAMS), and derive a luminosity, radius, and mass from that.  

Our search for a period based on $B_l$ measurements produced $6.2 \pm 0.6$ days ($\chi_{\nu}^2 = 0.9$).  The apparent radial velocity variability in the star is very small, and it produces two ambiguous periods ($6.0^{+0.7}_{-0.5}$ and $4.2 \pm 0.2$ days), but it is consistent with the periods from $B_l$ and the literature.
From the maximum entropy ZDI solution, we find a period of $6.4$ days, and from the ZDI minimum $\chi^2$ solution we find $6.4 \pm 0.2$.
Therefore we adopt the $6.13 \pm 0.03$ day period of \citet{Messina2010-RACE-OC-periods}, as it is consistent with our values and more precise.  

Given the potential systematic uncertainties in the luminosity and radius for the star, we prefer to estimate the inclination of the rotation axis from the maximum entropy ZDI solution.  From this we find $i = 55 \pm 20^\circ$, which is rather uncertain but consistent with our period, radius and \vs\ values. 

We estimated differential rotation by searching for the minimum $\chi^2$ ZDI solution, and found a solution of ${\rm d}\Omega = 0.2\pm 0.1$ rad/day, $\Omega_{\rm eq} = 1.08 \pm 0.06$ rad/day ($P = 5.8 \pm 0.3$ days).  The minimum in $\chi^2$ is mostly well defined, however it does have a weak (lower probability) tail towards lower differential rotation and longer periods.  Due to the relatively weak signal in $V$ relative to the noise, we consider this differential rotation measurement somewhat tentative.  The impact of differential rotation on the map of this star is small, it pushes slightly more magnetic energy into higher order and less axisymmetric spherical harmonics, but only by a couple percent of the total energy.

\subsection{HD 6569}
HD 6569 (HIP 5191, TYC 5275-1735-1, BD-15 200) is a member of the AB Dor association \citep{Zuckerman2004-ABDor-members, Torres2008-youngNearbyAssoc, McCarthy2014-ABDor-detailed}.
The star has a rotation period of $7.13 \pm 0.5$ days from \citet{Messina2010-RACE-OC-periods}, based on photometry.  

Our period search from longitudinal magnetic field values yielded a broad but distinct and unique minimum of $6.6^{+1.4}_{-1.9}$ days, from a first order fit ($\chi_{\nu}^2 = 0.8$).  
The large uncertainties are due to the small dataset, spanning only 11 days, and the weakness of the $B_{l}$ values relative to the noise.  We find no significant radial velocity variability in the star, and attempts construct a periodogram from the $v_r$ data allow almost any period.  
From a ZDI search we find a period of $6.8 \pm 0.2$ days, which is again rather uncertain but unique.  
Therefore we confirm the $7.13 \pm 0.05$ days rotation period of \citet{Messina2010-RACE-OC-periods}, but are unable to improve on it.  

We attempted to determine differential rotation for this star, but did not find a clearly unique solution. 
The best solution is for ${\rm d}\Omega = 0.30^{+0.25}_{-0.5}$ rad/day, $\Omega_{\rm eq} = 0.97^{+0.05}_{-0.07}$ rad/day ($P = 6.45^{+0.5}_{-0.3} $ days), but this is not significantly superior to a solution with no differential rotation.  Therefore we cannot reliably constrain differential rotation in the star.

\subsection{HH Leo}
HH Leo (HD 96064, HIP 54155, TYC 4924-1114-1, BD-03 3040) is a member of the Her-Lyr association \citep{Eisenbeiss2013-HerLyr-age}.  It is a hierarchical triple system, with the B and C components forming a close binary at $\sim$11'' from the A component \citep{Eisenbeiss2013-HerLyr-age}.  Our observations focused on the brighter and more massive A component. with Narval's 1.6'' diameter pinhole the B and C components fell well outside the pinhole, and we see no evidence of these components in our spectra.  

\citet{Cutispoto1999-rot-periods} found a rotation period of $6.9 \pm 0.3$ days, based on photometric observations spanning $\sim$13 nights.  
(This value appears to have been used by \citet{Eisenbeiss2013-HerLyr-age}, but without a clear citation).

Given the low precision of the rotation period from \citet{Cutispoto1999-rot-periods}, we attempted to refine this value.  
The period search based on $B_l$ measurements yielded a minimum at $6.01 \pm 0.06$ days, but for a reduced $\chi_{\nu}^2$ of $\sim 5$ for both first and second order fits.  With a second order fit there is a probable alias at 12.0 days ($\chi_{\nu}^2 = 2.6$), third order fits produce several ambiguous minima but none with $\chi_{\nu}^2$ below 3.  This poor $\chi^2$ is likely due to differential rotation modifying the magnetic field structure during the observations.  Restricting the data to only the 6 observations from May 2015, we find a best period of $6.2 \pm 0.4$ days with a reduced $\chi_{\nu}^2$ of $\sim 3.5$, for a first order fit (higher order fits could not be reliably performed due to the small data set).  This is much more uncertain due to the smaller dataset, but produces a more reasonable $\chi^2$.
The radial velocity variability is very weak and leads to an inconclusive periodogram, but it does have minima at both 6.0 and 6.8 days.  
From a ZDI period search, without differential rotation, we find $P = 5.95 \pm 0.02$, with an alias at $\sim$12 days.  The 12 day alias is not allowed by our radius and \vs, which limits the period to be $< 7.0$ days (at $1\sigma$).
While our period disagrees with \citet{Cutispoto1999-rot-periods} at $\sim 3\sigma$, none of our periodograms produce an acceptable period near 6.9 days (apart from a marginal minimum for $v_r$).  Indeed periods in the 5.9-6.1 day range are the closest acceptable values in our data to 6.9 days.

Due to the long time span of our observations, differential rotation must be included in the ZDI model to produce an acceptable $\chi^2$.
From a differential rotation search we find ${\rm d}\Omega = 0.11\pm 0.02$ rad/day, $\Omega_{\rm eq} = 1.062 \pm 0.003$ rad/day ($P = 5.915 \pm 0.017$ days).  We adopt this rotation period, as it is likely the most accurate period from our data, and more precise than the litterateur value.  This allows for a reduced $\chi^2$ of 1.2, which suggest that the large-scale magnetic field has not evolved strongly over 3 months, since the observations can be fit with a simple differential rotation law.  However since the best achieved reduces $\chi^2$ is slightly larger than for most other stars in the sample (usually values closer to 1.0 are achievable) some weak, marginal evolution of the large-scale magnetic field may have occurred.

\subsection{EP Eri}
EP Eri (HD 17925, HIP 13402, TYC 5292-897-1, BD-13 544) is a member of the Her-Lyr association \citep{Eisenbeiss2013-HerLyr-age}.  
A rotation period of 6.76 days was found by \citet{Donahue1996-periods-Sindex}, which is the mean of 10 periods (from  6.56 to 7.20 d) based on chromospheric S index variability.  \citet{Noyes1984-CaHK-Rossby} found a period of 6.6 days, from S index variability.  \citet{Cutispoto1992-early-phot-periods} found a photometric (UBVRI) period of 6.5 days, and \citet{Messina2001-periods-act-diss} report a period of 6.57.  
\citet{Eisenbeiss2013-HerLyr-age} quote a rotation period of 6.725 days, but do not provide a clear source for this value.
The most reliable value appears to be 6.76 or 6.56 from \citet{Donahue1996-periods-Sindex}, so we take the period to be $6.76 \pm 0.20$ days. 

From the variability in the longitudinal magnetic field, we find a rotation period of $6.4 \pm 1.0$ days ($\chi_{\nu}^2 = 1.5$).  
Radial velocity variability produces a period of $6.8 \pm 0.5$ days.  
These periods are all consistent with the values of \citet{Donahue1996-periods-Sindex}, but imprecise since we have few observations obtained over a short time frame. The ZDI period search gives $6.9 \pm 0.15$ days.  
The inclination angle, from the radius period and \vs\ is $85^{+5}_{-30}$, while the maximum entropy inclination from ZDI is $75 \pm 15^\circ$, which is consistent. 

With 9 observations, covering less than 1.5 rotation cycles, we cannot well constrain surface differential rotation in the star.  We only find ${\rm d}\Omega < 0.3$ rad/day and the associated $\Omega_{\rm eq} < 0.95$ rad/day ($P > 6.6 $ days).  Therefore we assume differential rotation is negligible over the short time-span of our observations.  

For EP Eri we struggle to fit the star well with a solar metallicity, but there is not a well established cluster metallicity.  Thus we include metallicity as a free parameter in both spectral analyses.  We find a marginally enhanced metallicity of [Fe/H] = $0.08 \pm 0.05$ dex.

\subsection{EX Cet}
EX Cet (HD 10008, HIP 7576, TYC 4687-325-1, BD-07 268) is a member of the Her-Lyr association \citep{Eisenbeiss2013-HerLyr-age}.  
\citet{Strassmeier2000-periods-preDI} found a photometric rotation period for the star, $P = 7.15$ days.  They used 35 observations over 78 days, with an amplitude of 0.015 in the Str\"omgren y-band.  

We do not detect a magnetic field in this star.  We have 13 observations, none of which provide even marginal detections in Stokes $V$ by the LSD detection criteria.  One of the observations is of slight lower S/N, but the other 12 are of similarly good quality.  From these we measure longitudinal magnetic fields consistent with zero, with a mean uncertainty on $B_l$ of 1.8 G.  Thus the longitudinal magnetic field of the star was consistently below 5.4 G, as a $3\sigma$ upper limit.  This is surprising, as the other stars in our sample have a peak $B_l$ of 7 G at the lowest, and typically peak above 10 G.  Thus if the magnetic field of any other star in the sample was observed with this precision it would have been detected.  

Since we do not detect the magnetic field of the star, we are unable to constrain the rotation period from $B_l$ (all periods yield $\chi_{\nu}^2 < 1.5 $), or from ZDI.  The apparent radial velocity variability is very marginal, and can be fit with a constant $v_r$ at a $\chi_{\nu}^{2} = 1.2$, thus we are unable to constrain the rotation period from this either.  

Based on our measured \vs\ and derived radius, the rotation 7.15 days period of \citet{Strassmeier2000-periods-preDI} implies an inclination of $28^\circ$.  For randomly oriented inclination axes, the probability of observing an inclination $i$ goes as $\sin i$.  That makes this inclination, and hence the 7.15 days period a bit unlikely, but not enough to reject this rotation period and assume a larger value.

\subsection{AV 2177}
AV 2177 (Cl* Melotte 111 AV 2177, BD+26 2362, TYC 1990-108-1) is a member of the Coma Berenices open cluster \citep{Kraus2007-ComaBer-members, CollierCameron2009-ComaBer-periods}. \citet{CollierCameron2009-ComaBer-periods} find photometric rotation periods of 8.38 days in 2004 and 8.47 days in 2007.  

We find a clear systematic velocity drift in our observations, strongly suggesting that this is an SB1 system.  Our observations do not sample the orbit well, and no good constraints can be made.  However, we see a velocity amplitude of at least 2.5 \kms, with a $\sim$0.1 \kms\ change between nights.
The $v_r$ values increase from the 9th of April until the 24th of April, drops in the gap between our observations to a more negative value than seen in April, and then increase from the 6th of June until the 19th of June when it reaches a value similar to 9th of April.  This suggests we have seen something close to an orbital cycle, and that the orbital period is less than 3 months.
This velocity variability was subtracted out of all subsequent analysis.  

From the $B_l$ values of the full dataset, we find somewhat ambiguous periods of 7.9 days ($\chi_{\nu}^2 = 1.5$) and 9.1 days ($\chi_{\nu}^2 = 1.4$).  Thus we turn to ZDI to refine these periods.

We obtained two datasets of this star separated by approximately 2 months, one between the 9th and 24th of April (2014) and the other between the 6th and 19th of June (2014).  Both datasets are relatively small, with 10 and 11 observations respectively.  Additionally, the magnetic field in AV 2177 is weak, and so the amplitude of the signal in our observed profiles is small.  This makes ZDI maps from the individual datasets relatively uncertain.  Nevertheless we performed ZDI on the two datasets individually, to check for large changes in the magnetic field of the star.  We find the magnetic geometry is largely consistent, however there is a spot of negative radial field (roughly on the opposite side of the star from the positive spot) seen in June that was not detected in April.  While this may be intrinsic evolution of the magnetic field, it may also have simply been undetected in the lower S/N observations in April.   The spot of positive radial field is consistent between the two epochs, and the toroidal band of azimuthal field is consistent between the two epochs.  The apparent gap in the band of azimuthal field appears in both maps, at the same rotation phase.  

In order to improve our sensitivity, and to search for differential rotation, we also performed ZDI using both the April and June data sets.  This assumes that the change in the magnetic field is only due to differential rotation.  However since this time period represents less than 10 stellar rotations, this approximation is likely reasonable (for the large-scale magnetic field).
From this we find evidence for marginal differential rotation of $\Delta\Omega = 0.05^{+0.05}_{-0.02}$ rad/day, with a best equatorial rotation rate of $\Omega_{eq} = 0.700^{+0.009}_{-0.006}$ rad/day (P $= 8.98^{+0.08}_{-0.12}$ d) at $1\sigma$.
This model produces a statistically acceptable fit to the observations (at $\chi_\nu^2 = 0.95$) while still being well regularized.  Thus there is no clear evidence for intrinsic evolution of the magnetic field in AV 2177.  Consequently we use this map including the April and June data for all subsequent analysis.

\subsection{AV 1693}
AV 1693 (Cl* Melotte 111 AV 1693, BD+24 2462, TYC 1989-361-1) is a member of the Coma Berenices open cluster \citep{Kraus2007-ComaBer-members, CollierCameron2009-ComaBer-periods}. \citet{CollierCameron2009-ComaBer-periods} find a photometric rotation period of 9.05 days, based on 2183 observations from SuperWASP.  

Our period search from longitudinal magnetic fields produces a period of $9.0 \pm 0.4$ days, although a second order fit is needed (first order best $\chi_{\nu}^2 = 10.2$, second order $\chi_{\nu}^2 = 3.6$).  The period search from radial velocities produces an ambiguous pair of periods at $9\pm1$ days and $4.5\pm0.5$ days, with a second order fit (a first order fit prefers the $\sim4.5$ day period).  However the radial velocity amplitude is not much larger than the uncertainty, thus this is not a very reliable period measure.
The period search from ZDI found $9.11 \pm 0.15$ days.
Our periods are fully consistent with the period of \citet{CollierCameron2009-ComaBer-periods}, thus we adopt their likely more precise value, but use a conservative uncertainty, for a value of $9.05 \pm 0.10$ days.

With 15 observations over 17 nights, almost two full rotation cycles, we have the possibility of measuring differential rotation. 
Using the ZDI based search, we find optimal values of ${\rm d}\Omega = 0.22^{+0.10}_{-0.08} $ rad/day, and $\Omega_{eq} = 0.728^{+0.013}_{-0.011} 0.$ rad/day ($P_{eq} =  8.63^{+0.13}_{-0.15}$ days) at $1\sigma$.
This differential rotating model also improves the best reduced $\chi^2$ from 1.5 to 1.2, thus the star has significant differential rotation.

\subsection{AV 1826}
AV 1826 (Melotte 111 AV 1826, BD+27 2139, TYC 1991-1235-1) is a member of the Coma Berenices open cluster \citep{Kraus2007-ComaBer-members, Mermilliod2008-ComaBer-members, CollierCameron2009-ComaBer-periods}. 
\citet{CollierCameron2009-ComaBer-periods} found photometric rotation periods of 9.26 days in 2007 and 4.79 days in 2004, and concluded that the shorter period is an alias of the longer period, due to an unfortunate spot distribution.  \citet{Terrien2014-ComaBer-photo} found a photometric period of $9.483 \pm 0.028$ days, which is approximately consistent with the value of \citet{CollierCameron2009-ComaBer-periods}.  
Both values are roughly consistent with our period measurements from the longitudinal field and from ZDI, therefore we adopt the value of $9.483 \pm 0.028$ days, since it is based on more observations obtained over a longer time period.  

Our observations of this star were obtained in two separate runs, separated by approximately 2 months, from the 9th to the 24th of April (2014) then from the 5th to the 19th of June (2014).  Both datasets are relatively small, with 12 observations in each month.
Periodograms based on the longitudinal magnetic field measurements from the April run yield best periods around 9 days (first order $\chi_{\nu}^2 = 3.8$, second order $\chi_{\nu}^2 = 1.3$), as do the measurements from the June run (first order $\chi_{\nu}^2 = 4.2$, second order $\chi_{\nu}^2 = 1.0$), however both data sets require a second order fit to produce an acceptable $\chi^2$.  Using the full dataset produces a best period of 9.4 days, favoring the period from \citet{Terrien2014-ComaBer-photo} (first order $\chi_{\nu}^2 = 4.5$, second order $\chi_{\nu}^2 = 3.6$).  There appears to be moderate but significant differences between the April and June $B_l$ curves (Fig.\ \ref{fig-bz}).  This could be simply due to differential rotation on the star, or due to the appearance or disappearance of large-scale magnetic structures, and ZDI maps are necessary to investigate this.  

There appears to be a small drift in the $v_r$ values between the April and June observations, of $\sim$0.2 \kms.  Most likely this is due to binarity, making the star an SB1, but this is not entirely clear.  We find no evidence for lines of a secondary star in our spectra or LSD profiles.  

We performed ZDI first on the two datasets independently.  The resulting maps have similarities, both containing strong toroidal components, and a strong positive radial spot at high latitudes.  However, there are some significant differences between the maps, particularly an offset in phase between the high latitude radial spot and the maximum in the lower latitude toroidal loop, which suggests differential rotation may be significant.
We performed a search for differential rotation, using the full dataset.  We find a primary $\chi^2$ minimum at $P = 9.34 \pm 0.08$ days ($\Omega = 0.673 \pm 0.006$ rad/day) and $\Delta\Omega = 0.09^{+0.04}_{-0.03}$ rad/day at $1\sigma$.
However, a secondary minimum exists at $P = 8.73 \pm 0.12$ days ($\Omega = 0.719$ rad/day), and $\Delta\Omega = 0.12 \pm 0.04$.  The primary minimum provides the better $\chi^2$, and has the only period consistent with the photometric value, thus we adopt $P = 9.34 \pm 0.08$ days and $\Delta\Omega = 0.09^{+0.04}_{-0.03}$ as the correct solution.  Our 9.34 day period is slightly shorter than the photometric 9.48 day period, which is consistent with a deferentially rotating star, since the photometric value assumed solid body rotation and the spots were likely not located exactly at the stellar equator, yielding a slightly slower rotation period.

With this period we can fit the combined set of $V$ LSD profiles to a reduced $\chi^2$ of 1.0, by only allowing for differential rotation (Fig.~\ref{fig-sample-diff-rot-search}). Thus we conclude that changes in the surface magnetic structure over this two month time period are mostly due to searing from differential rotation, rather than the emergence or disappearance of magnetic flux on large scales.

\subsection{TYC 1987-509-1}
TYC 1987-509-1 (BD+29 2215, 1SWASP J114837.70+281630.5) is a member of the Coma Berenices open cluster \citep{CollierCameron2009-ComaBer-periods}.  The star has a photometric rotation period of 9.43 days from \citet{CollierCameron2009-ComaBer-periods}, based on 1138 observations from SuperWASP.

Our observations for this star were obtained over only a 9 day period.  Initially observations were planned for a 15 day time-span, but poor weather and scheduling conflicts at the telescope limited this to only 9 days.  Thus our observations provide sufficient phase coverage of the star, and can rule out rotation periods shorter than $\sim$9 days, but cannot firmly confirm the 9.43 day period of \citet{CollierCameron2009-ComaBer-periods}.  From a period search using $B_l$ the rotation period must be longer than 8.2 days, and a second order sinusoid is needed to provide an acceptable fit to the data (first order $\chi_{\nu}^2 = 5.0$, second order $\chi_{\nu}^2 = 1.6$).  The radial velocity variability is compatible with the 9.43 day period, but the variability is only marginally significant, so the constraint on the period is weak.  From the period search using ZDI, we find a limit on the period of $>8.3$ days.  No reliable differential rotation values could be determined, as we do not have repeated observations of the same rotation phase.  

Based on radius, period, and \vs, the inclination is only constrained to be $>82^\circ$ at $1\sigma$ and $>63^\circ$ at $2\sigma$.  To further constrain this inclination, we searched for the inclination which provides the maximum entropy solution from ZDI.  This produced $i = 67 \pm 5^\circ$, which is consistent with the constraint from radius, period, and \vs, and thus we adopt it.  

In a differential rotation search, we do no find a useful constraint.  The minimum $\chi^2$ (or maximum entropy) is at ${\rm d}\Omega \sim 0$ (${\rm d}\Omega = 0.07^{+0.20}_{-0.15}$ rad/day, $P = 8.4 \pm 0.5 $ at $1\sigma$), however since the uncertainties are very large the result is not useful.
Thus we assume no differential rotation in our final model.

\subsection{AV 523}
AV 523 (Cl* Melotte 111 AV 523, TYC 1988-6-1, 1SWASP J121253.23+261501.3) is a member of the Coma Ber cluster \citep{Kraus2007-ComaBer-members, CollierCameron2009-ComaBer-periods}.  \citet{CollierCameron2009-ComaBer-periods} measured a photometric rotation period for the star, but found some ambiguity between a 5.44 days and 10.9 days period, though they favor the longer period.  

From our period search based on longitudinal field measurements, we find a $\chi^2$ minimum near $10 \pm 1$ days.  However this is a broad minimum with a relatively high $\chi^2$ ($\chi_{\nu}^2 = 2.1$).  There is significantly more signal in the LSD profiles, since the magnetic field is relatively axisymmetric and toroidal.  The radial velocity variability is not sufficiently significant to constrain the rotation period.  

Using ZDI to search for a period, we find a unique maximum in entropy (for a fixed $\chi^2$) of 11.1 days.  Similarly we find a unique minimum in $\chi^2$ (for a fixed entropy) at $11.1 \pm 0.2$ days, with the uncertainty taken from the contour in $\chi^2$.  This is consistent with the value of \citet{CollierCameron2009-ComaBer-periods}.  Given the possible ambiguity in the period from \citet{CollierCameron2009-ComaBer-periods}, we adopt our rotation period from ZDI as the best value.  

Since the star is a particularly slow rotator, the constraints period and \vs\, provide on the inclination are relatively poor: $i > 54^\circ$.  Thus we adopt the inclination from ZDI.  From both entropy and $\chi^2$ we find a best inclination of $\sim 50^\circ$.  From the $\chi^2$ contour we derive a formal uncertainty $i = 50 \pm 7^\circ$.  

The differential rotation search did not produce a definite result.  This is not surprising, since our observations span less than 1.5 rotation cycles, with few repeated phased.
Values of ${\rm d}\Omega$ from  -0.2 to 0.2 rad/day are allowed within $1\sigma$.
Thus we do not measure differential rotation for the star, and in our analysis assume it is zero.

\subsection{Mel25-151}

Mel25-151 (Cl Melotte 25 151, HD 240629, V1362 Ori, HIP 23701, TYC 110-1206-1, BD+06 829) is a member of the Hyades \citep{Perryman1998-Hyades-age-dist, Delorme2011-periods-Hyades-Praesepe}.  \citet{Koen2002-var-from-hipp} reported variability in 78 Hipparcos photometric observations with a frequency of 0.29974 $d^{-1}$, but with a significance only a bit above their detection threshold.  This is inconsistent with our period estimates below, thus given the high uncertainty we reject this as the rotation period.   \citet{Delorme2011-periods-Hyades-Praesepe} found a rotation period of 10.41 days from SuperWASP photometry.  They do not report an uncertainty, so we assume it to be $\pm 0.1$, although this may be an overestimate.  

Our radial velocity measurements indicate that the star is an SB1, with a decrease in $v_r$ of $\sim$0.75 \kms\ over 15 days.  However, we see no evidence for the secondary in our spectra, or in our LSD profiles.  The velocity variability is consistent with a decrease in $v_r$ of 0.05 \kms\ per day, and does not yield a useful rotation period estimate.
\citet{Bender2008-Hyades-binaries-sb2} detect the secondary faintly in the infrared, report an orbital period of 629 days.  \citet{Patience1998-hyades-binaries} found a third component to the system by speckle imaging, 1.45 magnitudes fainter in $K$ and at a 0.85'' separation, which is too faint to appear in our observations. 
Only one component could be seen in our spectra, and the velocity shifts were corrected for in the rest of our analysis.  

From our $B_l$ measurements, we find a period of $10.1 \pm 0.7$ days.  While the uncertainty on the value is large due to the relatively short timespan of our observations, this period represents a clear unique $\chi^2$ minimum in the periodogram.
We use a second order fit here (first order $\chi_{\nu}^2 = 1.7$, second order $\chi_{\nu}^2 = 0.7$) in order to fit the extrema of the $B_l$, although in this case the second order term is less necessary than in most cases. 
From the ZDI rotation period search, assuming no differential rotation, we find a period of $10.06 \pm 0.10$ days. 
We do not find any clear differential rotation value, due to having few observations at repeated phases, and spanning only $\sim$1.5 rotation cycles.
We can limit it to ${\rm d}\Omega = 0.03^{+0.06}_{-0.07}$ rad/day and an associated period range of $P = 9.9^{+0.5}_{-0.4}$ days, at $1\sigma$, with a very strong covariance between rotation frequency and differential rotation.  
Thus in our final model we assume no differential rotation.

\subsection{Mel25-43}
Mel25-43 (Cl Melotte 25 43, V988 Tau, HD 284414, HIP 20482, TYC 1272-912-1, BD+19 708) is a member of the Hyades \citep{Perryman1998-Hyades-age-dist, Delorme2011-periods-Hyades-Praesepe}.  A rotation period of 9.90 days was reported by \citet{Delorme2011-periods-Hyades-Praesepe}, based on SuperWASP photometry.  There is no uncertainty reported for this period, so we assume a value of $\pm 0.1$, although this may be an overestimate. 

From $B_l$, we find an broad but unambiguous minimum in the periodogram of $10.0 \pm 1.1$ days ($\chi_{\nu}^2 = 1.2$).
From a ZDI period search we find a similar minimum in $\chi^2$ of $10.1 \pm 0.4$ days.
These values are consistent with \citet{Delorme2011-periods-Hyades-Praesepe}, but less precise due to our smaller dataset, therefore we adopt their rotation period.  

From $v_r$ we find a decreasing trend over the course of our observations, strongly suggesting that the star is an SB1, but we find no clearly identifiable rotational modulation on top of this variability.  We find $v_r$ running from 38.1 to 37.5 \kms\ over the 15 days of our observations. Indeed \citep{Perryman1998-Hyades-age-dist} note the star as a spectroscopic binary.  The original 1997 Hipparcos reduction found the star to be part of a binary with a 590 days period, with a semi-major axis of $7.6 \pm 2.4$'' (a Gaia solution is not yet available).  
\citet{Bender2008-Hyades-binaries-sb2} manage to spectroscopically detect the secondary of the system in the infrared.  Based on the infrared observations of the secondary, and a larger dataset of optical observations for the primary, they find a mass ratio (secondary/primary) of $0.66 \pm 0.05$,  with $M_{1}\sin^2 i = 0.249 \pm 0.046 M_{\odot}$ and $M_{2}\sin^2 i = 0.165 \pm 0.017 M_{\odot}$.  
We find no evidence for lines from the secondary in our optical spectra of the star, or in our LSD profiles, and thus treat the star as single in our subsequent analysis.  

Due to the uncertain contribution from the secondary in the system, the luminosity of the primary is somewhat uncertain.  The star has a fairly precise parallax from Hipparcos and $J$ magnitude from 2MASS, however with these values the star falls slightly above the Hyades isochrone on the HR diagram.  Based on the mass ratio of \citet{Bender2008-Hyades-binaries-sb2}, the secondary may not be enough to fully explain this excess luminosity, however no direct measurement of the secondary's luminosity exist.  Consequently, we prefer to determine the mass of the star by assuming it falls on the cluster isochrone.  For the inclination of the rotation axis, we use the maximum entropy ZDI solution, rather than relying on the uncertain radius, coupled with \vs\ and the rotation period.  

The search for differential rotation, from ZDI with $\chi^2$, found an uncertain, marginal value of  ${\rm d}\Omega = 0.18^{+0.16}_{-0.16}$ rad/day and $\Omega_{\rm eq} = 0.68^{+0.05}_{-0.05}$ rad/day ($P = 9.3^{+0.8}_{-0.6}$ days), at $1\sigma$.  Since we have few phases with observations from different rotation cycles, this value is rather uncertain.  In our final magnetic map we do adopt this differential rotation value, but do not consider it a clear detection of differential rotation.

\subsection{Mel25-21}
Mel25-21, (V984 Tau, HD 284253, HIP 19934, TYC 1276-86-1, BD+21 612) is a member of the Hyades \citep{Perryman1998-Hyades-age-dist, Delorme2011-periods-Hyades-Praesepe}. \citet{Delorme2011-periods-Hyades-Praesepe} find a rotation period of 10.26 days, from SuperWASP photometry.  Since they do not report an uncertainty, in our analysis we conservatively assume 0.1 days uncertainty on their value, although this may be an overestimate.

Our period determination is rather uncertain, since the observations were obtained over 13 days, and the rotation period is near ten days.  From $B_l$ we find a period of $9.6 \pm 1.3$ days, with a broad but unambiguous $\chi^2$ minimum ($\chi_{\nu}^2 = 1.2$), which is consistent with \citet{Delorme2011-periods-Hyades-Praesepe}.  There is no significant radial velocity variability in the observations, and thus our radial velocity measurements are compatible with the period of \citet{Delorme2011-periods-Hyades-Praesepe}, but do not significantly constrain the rotation period.  
From ZDI we find a best period of $9.73 \pm 0.20$ days, assuming no differential rotation, which is again rather uncertain but roughly consistent with \citet{Delorme2011-periods-Hyades-Praesepe}.  

A differential rotation search from ZDI yielded marginal results.  The best values were ${\rm d}\Omega = 0.2^{+0.15}_{-0.13}$ rad/day and $\Omega_{\rm eq} = 0.68 \pm 0.03$ rad/day ($P = 9.15 \pm 0.45$ days).  But this is highly uncertain, and within $2\sigma$ of no differential rotation.  Given the few rotation phases with repeated observations, such a large uncertainty is not surprising.  While this is not a confident detection of differential rotation, we do adopt this value for our final magnetic map.

\subsection{Mel25-179}
Mel25-179 (Cl Melotte 25 179, HD 285830, HIP 20827, TYC 680-104-1, BD+14 699) is a member of the Hyades cluster \citep{Perryman1998-Hyades-age-dist, Delorme2011-periods-Hyades-Praesepe}.  A rotation period of 9.70 days was found by \citet{Delorme2011-periods-Hyades-Praesepe}.  Since they gave no uncertainty on the period, we conservatively assume $\pm 0.1$ days, although this may be an overestimate.   \citet{Perryman1998-Hyades-age-dist} found no evidence for multiplicity.  

From our radial velocity measurements we find a general decreasing trend, from 39.78 \kms\ to 39.60 \kms, over the 15 days of our observations.  We find no clear rotational modulation in addition to this variation.  This trend suggests the star may be an SB1.  We see no evidence for lines of a secondary in the spectrum or in our LSD profiles.  Thus we conclude that a secondary star is not contributing significantly to our observations.  \citet{Patience1998-hyades-binaries} found the star to be a binary, based on speckle observations.  They found a K-band magnitude difference of $5.5 \pm 0.4$, separation of $0.91 \pm 0.02$'', estimated mass of 0.1\Msun.  With such a large magnitude difference, the star reported by \citet{Patience1998-hyades-binaries} almost certainly does not contribute to our observations.

From our period search based on $B_l$ we find $P = 10.2 \pm 0.5$ days, which is a broad but clearly unambiguous minimum in $\chi^2$.
This minimum appears clearly in a first order fit ($\chi_{\nu}^2 = 6.8$), but adding a second order term greatly improves the fit quality ($\chi_{\nu}^2 = 3.0$), thus we adopt the second order fit. 
The period search from a ZDI, using no differential rotation, found $10.21 \pm 0.08$ days, which is consistent with \citet{Delorme2011-periods-Hyades-Praesepe}.
Since the \vs\ of the star is small, the fractional uncertainty is large, and thus the constraint on inclination from \vs\ and period is weak.  Instead we performed an inclination search using ZDI, and found $56 \pm 4$ degrees, in good agreement with the value based on period, radius, and \vs.

The search for differential rotation with ZDI produced a small value that is marginally consistent with no differential rotation, with significant uncertainties: ${\rm d}\Omega = 0.10\pm 0.08$ rad/day and $\Omega_{\rm eq} = 0.638^{+0.017}_{-0.020} $ rad/day ($P = 9.85^{+0.32}_{-0.25}$ days).  Our observations span only roughly 1.5 rotation cycles, and have only three phases with repeated observations.  Thus despite the relatively high S/N of the Stokes $V$ signal, there is still a large uncertainty on the derived differential rotation, and our value is not a definite detection.

\subsection{Mel25-5}
Mel25-5 (Cl Melotte 25 5, HIP 16908, TYC 1247-684-1, BD+20 598) is a member of the Hyades \citep{Perryman1998-Hyades-age-dist, Delorme2011-periods-Hyades-Praesepe}.  The star has a $10.57$ day rotation period from \citet{Delorme2011-periods-Hyades-Praesepe}, based on photometry from SuperWASP.  Since \citet{Delorme2011-periods-Hyades-Praesepe} provide no uncertainty, we assume a $\pm 0.10$ days uncertainty in our analysis, although this likely overestimates the uncertainty on their value.

Our period search from $B_{l}$ finds a best period of $10.0^{+1.4}_{-1.2}$ days ($\chi_{\nu}^2 = 1.2$), with an apparent alias at 4.6 days ($\chi_{\nu}^2 = 1.9$).  Our observations only span 13 days, which leads to an imprecise period.  We do not detect significant radial velocity variability, and thus cannot constrain the rotation period with it. However the small amount of marginal variability is compatible with a 10.6 day period.  From a ZDI search with no differential rotation, we find an imprecise but unique best period of $10.04^{+0.17}_{-0.18}$.

The differential rotation search for the star excludes large values of ${\rm d}\Omega$, but does not confidently detect a non-zero value.  The best values are ${\rm d}\Omega = -0.17\pm 0.18$ rad/day and $\Omega_{\rm eq} = 0.58 \pm 0.04$ rad/day ($P = 10.8 \pm 0.8$ days), which is consistent with no differential rotation and the period of \citet{Delorme2011-periods-Hyades-Praesepe}.  Since our observations only span 13 days, and the rotation period is 10.57 days, we have few observations at the same phase and different rotation cycles, which leads to a very weak constraint on differential rotation.  Thus we adopt zero differential rotation in our final map.

\section{Zeeman Doppler Imaging code}
\label{ZDIpy}

In this project we developed a new independent implementation of a Zeeman Doppler Imaging code.  The goal of this was to match the functionality of the \citet{Donati2006-tauSco} code, and produce identical results to the code used in Paper I, but be easier for the authors to use and extend in the future.  The ZDI code inverts a time series of Stokes $V$ profiles to reconstruct the large-scale vector magnetic field of a star.  The magnetic field is expressed as a combination of spherical harmonics as in \citet{Donati2006-tauSco}.  The maximum entropy fitting routine of \citet{Skilling1984-max-entropy-regularisation} is used, to provide a regularized fit to data that both minimizes $\chi^2$ and maximizes entropy.  

\subsection{The model}
The forward model, which generates model line profiles from stellar model, proceeds as follows.  The model star is assumed to be spherical, and the surface is tiled with surface elements of approximately equal area.  The elements are organized in latitudinal rings, with the number of elements in the ring going as the sine of the co-latitude.  Areas of the elements are calculated exactly for the portion of the sphere they subtend.  The poles of this spherical coordinate system are aligned with the rotation axis of the star.

The stellar magnetic field is expressed using a set of spherical harmonics, following \citet{Donati2006-tauSco} (see also \citealt{Vidotto2016-vector-field-spher-harm} for more details). Specifically:
\begin{equation}
B_{r}(\theta,\phi) = \sum_{l=0}^{L} \sum_{m=0}^{l} \mathrm{Re} [\alpha_{l m} Y_{l m}(\theta,\phi) ] 
\end{equation}
\begin{equation}
B_{\theta}(\theta,\phi) = -\sum_{l=0}^{L} \sum_{m=0}^{l} \mathrm{Re}[\beta_{l m} Z_{l m}(\theta,\phi) +  \gamma_{l m} X_{l m}(\theta,\phi)] 
\end{equation}
\begin{equation}
B_{\phi}(\theta,\phi) = -\sum_{l=0}^{L} \sum_{m=0}^{l} \mathrm{Re}[\beta_{l m} X_{l m}(\theta,\phi) - \gamma_{l m}  Z_{l m}(\theta,\phi)]
\end{equation}
where:
\begin{equation}
Y_{l m} = c_{lm} P_{l m}(\cos \theta) e^{i m \phi} 
\end{equation}
\begin{equation}
X_{l m}(\theta,\phi) = \frac{c_{lm}}{l+1} \frac{i m}{\sin \theta} P_{l m}(\cos \theta) e^{i m \phi} 
\end{equation}
\begin{equation}
Z_{l m}(\theta,\phi) = \frac{c_{lm}}{l+1} \frac{\partial P_{l m}(\cos \theta)}{\partial \theta} e^{i m \phi}
\end{equation}
and
\begin{equation}
c_{lm} = \sqrt{\frac{2l+1}{4\pi} \frac{(l-m)!}{(l+m)!}}.
\end{equation}
Here $P_{l m}(\cos \theta)$ is the associated Legendre polynomial (sometimes written as $P_{l m}(\theta)$ for simplicity), of degree $l$ and order $m$.  The co-latitude is $\theta$, and the longitude is $\phi$.  The sums over spherical harmonics are carried out here over positive $m$, and over degree $l$ up to a limiting sufficiently large value $L$.  The complex valued coefficients $\alpha_{l m}$, $\beta_{l m}$, and $\gamma_{l m}$ describe the magnetic field, with $\alpha_{l m}$ corresponding to the radial poloidal field, $\beta_{l m}$ the tangential poloidal field, and $\gamma_{l m}$ the toroidal field.  The magnetic field is saved using these coefficients, and when being fit these coefficients are the free parameters.  

The code allows for brightness spots on the star, however in this paper we assume a homogeneous surface brightness.  Since the code is designed to work on single lines, or more typically LSD profiles, there is a strong degeneracy in the model between spot temperature and spot filling factor.  Rather than assume a spot temperature and allow filling factor to be a free parameter, we prefer to use a total pixel surface brightness.  The reader can then interpret this as a filling factor with their preferred spot temperature if they wish.  This is implemented by assigning every surface element on the model star a relative brightness (by default 1.0).  These can then be fit by modeling a series of Stokes $I$ line profiles, as in regular Doppler Imaging.  However, in this paper we assume a homogeneous surface brightness for all stars.  Most stars in this sample show very little variability in Stokes $I$, and thus brightness maps cannot be reliably derived.  For this reason, and to be consistent with Paper I, this feature was not used.

The local line models, effectively the emergent flux at one point on the stellar surface, are calculated using a Gaussian as an approximation for the Stokes $I$ line profile.  The code also has the option of using a Voigt profile (the convolution of a Gaussian and Lorentzian) as a more realistic approximation for the Stokes $I$ line.  These Voigt profiles are calculated using the approximation of \citet{Humlicek1982-comp-voigt}, which is computationally efficient and more than sufficiently accurate for our purposes.  However, since adopting a Voigt profile does not have a large impact on the magnetic maps for our sample, we use a Gaussian local profile in order to be consistent with Paper I.

The local line models for Stokes $V$ are calculated using the weak field approximation \citep[e.g.][]{Landi-Landolfi2004-polarization-lines-book}
\begin{equation}
 V(\lambda) =  g_{eff}  \frac{\lambda_{0}^{2} e}{4 \pi m_{e} c} B_{l} \frac{\mathrm{d}I}{\mathrm{d}\lambda}, 
\end{equation}
where $g_{eff}$ is the effective Land\'e factor for the line, $\lambda_{0}$ is the rest wavelength of the line, $e$ is the electron charge, $m_{e}$ is the electron mass and $c$ is the speed of light.  This approximation is valid when Zeeman splitting is smaller than the intrinsic line width, up to magnetic fields of approximately 1 kG.  $B_{l}$ is the line of sight component of the magnetic field, and must be calculated from the magnetic vector for each surface element.  This is typically adequate for magnetic fields below $\sim$1 kG, as is the case in our study, unless extremely high precision is needed.  As a future step we plan to include an Unno-Rachkovsky approximation for the line \citep{Unno1956-Zeeman-pol, Rachkovsky1962-mag-pol-sol}, should line profiles outside of the weak field regime be needed, although this will likely be significantly more computationally intensive.  

The full disk integrated line models are calculated by summing local line models from all visible surface elements, scaled by the projected area of the surface element, its brightness, and the limb darkening.  These are then normalized by the sum of the continuum levels, which are also scaled by the projected area, brightness, and limb darkening.  A linear limb darkening law \citep[e.g.][]{Gray2005-Photospheres} in the form: 
\begin{equation}
  I_{c}/I_{c}^{0} = 1 - \eta + \eta \cos(\omega)
\end{equation}
is used, where $\eta$ is the limb darkening coefficient, $I_{c}/I_{c}^{0}$ is the brightness relative to disk center, and $\omega$ is the angle from disk center. The local line profiles are Doppler shifted according to the line-of-sight projection of the rotational velocity of the surface element.  The set of wavelength points used for calculating the local profiles takes this Doppler shift into account, so no interpolation is needed at this stage.  The wavelength grid used throughout the calculation is same as the wavelength grid of the observed spectra, so no interpolation is needed when comparing to the observations.  

Differential rotation is implemented using a linear law in the form:
\begin{equation}
  \Omega(\theta) = \Omega_{\rm eq} + {\rm d} \Omega \sin^2 \theta ,
\end{equation}
where $\Omega(\theta)$ is the rotation frequency at co-latitude $\theta$ and ${\rm d}\Omega$ controls the degree of differential rotation, as mentioned in Sect.\ \ref{Rotation period and differential rotation}.  This is used to calculate the modified rotation phase of the surface elements at each co-latitude.  This is also applied to the rotational velocities of the surface elements, typically leading to slightly lower velocities near the pole, and a slightly modified rotationally broadened line profile.  

Once the disk integrated line profile is calculated, it is convolved with a Gaussian instrumental profile, corresponding to the resolution of the spectrograph.  This provides a full synthetic model observation that can be compared to the real observation.  

\subsection{Inversion}

With a solution to the forward problem in hand, which generates a set of model line profiles for a given magnetic field, we now turn to the inverse problem, fitting the magnetic field given a set of observed line profiles.  This is formally an `ill posed problem', in that there are multiple magnetic field configurations that can fit a given set of observations equally well.  Thus simple $\chi^2$ minimization is not enough to produce a robust unique solution.  We adopt an additional regularization parameter to help ensure uniqueness of the solution, in our case entropy.  This a standard solution to this problem that DI and ZDI codes face.  We adopt the maximum entropy fitting routine of \citet{Skilling1984-max-entropy-regularisation}, following \citet{Donati1989-early-ZDI-test}, \citet{Donati1997-ZDI-tests}, \citet{Hussain2001-DOTS-descript}, and \citet{Donati2006-tauSco}.  Many fitting routines using maximum entropy regularization and attempt to maximize the quantity $Q = S - \lambda \chi^2$, where $S$ is the entropy, $\chi^2$ is a goodness-of-fit statistic, and $\lambda$ is a parameter tuned by the user to control the degree of regularization.  The \citet{Skilling1984-max-entropy-regularisation} routine instead tries to maximize $S$ subject to the inequality constraint $\chi^2 \leq \chi^2_{aim}$, where $\chi^2_{aim}$ is a user specified reasonable limit on $\chi^2$ for a well fit model.  This has the theoretical advantage that a statistically reasonably $\chi^2_{aim}$ can be estimated a priori, while $\lambda$ cannot be.  If the data are sufficient to constrain the model, the solution will generally lie on the boundary $\chi^2 = \chi^2_{aim}$.  In an iterative fitting approach this is effectively maximizing $Q$ while dynamically updating $\lambda$, first allowing $\chi^2$ to reach $\chi^2_{aim}$, then modifying $\lambda$ so that $\chi^2$ remains at $\chi^2_{aim}$, while still allowing $S$ to increase, until the maximum of $S$ along this boundary is found.  

The optimal entropy definition to apply to the magnetic field is not obvious, since the classical Shannon entropy in the form $S = -\sum n \ln n$ is clearly not applicable.  Instead we require a definition of entropy that is applicable to positive and negative values, and we would like it to, in the absence of observational constraints, favor values of zero.  Thus we adopt the entropy of \citet{Hobson1998-entropy-positive-negative}, which allows both positive and negative values.  This is derived from the version of Shannon entropy used in \citet{Skilling1984-max-entropy-regularisation} (among other places), and is essentially based on considering the full distribution of parameters as being a combination of two positive distributions $\mathbf{h} = \mathbf{f} - \mathbf{g}$.  We use equation 8 of \citet{Hobson1998-entropy-positive-negative}:
\begin{equation}
  S = \sum_{i=1}^{N} \left\{ \psi_{i} - (m_{f})_{i} - (m_{g})_{i} - h_{i} \ln \left[ \frac{\psi_{i} + h_{i}}{2 (m_{f})_{i}} \right] \right\}
\end{equation}
where
\begin{equation}
\psi = \left[ h_{i}^{2} + 4 (m_{f})_{i} (m_{g})_{i}  \right]^{1/2}.
\end{equation}
Here $h_{i}$ are the values of the $N$ parameters over which entropy is calculated.  $(m_{f})_{i}$ and $(m_{g})_{i}$ represent the `default' values which the positive ($f$) and negative ($g$) parts of the distribution will tend to, or more rigorously, these represent Bayesian priors on the distributions.  Unlike for entropy on simple positive values, the maximum of the entropy function is not at $f_{i} = m_{i}$, but rather for $h_{i}$ between $(m_{f})_{i}$ and $-(m_{g})_{i}$.  For our application we wish to penalize positive and negative values equally, so we set $(m_{f})_{i} = (m_{g})_{i}$, and thus the maximum in entropy (for a given $i$) is at $h_{i} = 0$.  We also do not, a priori, know which values of $i$ should be preferred, so we use a constant $m_{i}$ for all $i$.  This entropy definition is then applied to the spherical harmonic coefficients $\alpha_{l m}$, $\beta_{l m}$, and $\gamma_{l m}$ (not the magnetic vectors), to derive the entropy for our fitting parameters.  We apply a weighting factor to the entropy values before summing that is equal to $l$, the degree of the spherical harmonic, in order to weight against small scale structures in the magnetic field.  Small scale structures are much more likely to be influenced by, or entirely due to, noise in the $V$ line profiles, thus to be conservative and avoid spurious structure we introduce this extra weighting.  This weighting may not be strictly necessary, in that reasonable looking magnetic maps can often be produced without it, or the weighting could be included by varying the values of $m_{i}$ with $i$ according to the $l$ value of the harmonic.  However this is the most direct way of ensuring our goal of caution in not overestimating magnetic field with ZDI.

For fitting brightness, the entropy definition is somewhat simpler.  The code currently provides two options for the form of entropy, one allowing for a model with both dark and bright spots on the star, the other restricting the model to only dark spots.  For allowing bright and dark spots we use a simple entropy \citep[e.g.][]{Skilling1984-max-entropy-regularisation, Hobson1998-entropy-positive-negative} in the form
\begin{equation}
  S = - \sum_{i=1}^{N} w_{i} \left[ m_{i} + f_{i} \left(  \ln \frac{f_{i}}{m_{i}} - 1 \right)  \right], 
\end{equation}
where $f_{i}$ is the relative brightness in pixel $i$ and $m_{i}$ is the `default' value for that brightness.  We use a constant $m_{i}$ for all pixels, usually 1.  We also weight the entropy of each pixel by the surface area of the pixel ($w_{i}$).  For allowing only dark spots, we include an entropy based on the filling factor formulation of \citet{CollierCameron1992-DI-modelling} \citep[see also][]{Unruh1995MNRAS-DI-filling-factor-tests}
\begin{equation}
  S = - \sum_{i=1}^{N} w_{i} \left[ f_{i} \ln \left( \frac{f_{i}}{m} \right) + (m_{\rm lim} - f_{i}) \ln \left( \frac{m_{\rm lim} - f_{i}}{m_{\rm lim} - m} \right) \right]. 
\end{equation}
Here again we use a constant default value $m$ for all pixels, $m_{\rm lim}$ is the upper limit allowed on the values $f_{i}$, and $w_{i}$ weights the pixels by their surface area.  Usually $m_{\rm lim}$ is set to one, and $m$ is set to a value slightly below one.  In that case, $f_{i}$ would represent spot filling factors if the spots were completely dark. 

The spherical harmonic description of the magnetic field combined with the weak field approximation for Stokes $V$ has a few useful properties.  Particularly, $\mathbf{B}$ is linear in the spherical harmonic coefficients, $B_{l}$ can be calculated from a simple dot product, and $V$ is linear in $B_{l}$.  Thus the Stokes $V$ profile is linear in $\alpha_{l m}$, $\beta_{l m}$, and $\gamma_{l m}$, and the partial derivative of $V$ with respect to one spherical harmonic coefficient is independent of the values of the spherical harmonic coefficients.  As a consequence, the response matrix for fitting a magnetic field with $V$ can be calculated initially, and then remains constant as the iterative fitting proceeds.  This provides a major saving in computation time.  Unfortunately the same convenient linearity does not exist for fitting brightness maps with Stokes $I$, due to the need to normalize by the varying continuum level.  However, parts of the expression for the response matrix can still be calculated in advance of iterative fitting.  

This ZDI code was implemented in Python, with heavy use of NumPy and SciPy.  While Python is not the most efficient language for numerical methods, with careful use of NumPy it can approach compiled C or Fortran in many situations, and Python is a much easier environment for implementing moderately complex programs.  Most of the operations involved in this program can be phrased as matrix operations, or operations across arrays, which can be performed efficiently with NumPy.  Thus the run-time of the program is not greatly different from the run-time of the ZDI program of \citet{Donati2006-tauSco}.  

This ZDI code was extensively tested against the code of \citet{Donati2006-tauSco}, as discussed in Sect.\ \ref{ZDI}.  The entire sample of stars for this paper was analyzed with both ZDI codes.  The resulting magnetic fields were always consistent to within $\sim$1 \%. Thus we consider the code to be well validated.

\section{Trends in magnetic geometry and resolution}
\label{Trends in magnetic geometry and resolution}

Here we consider some additional trends in large-scale geometry that are present in our magnetic field results.  We then investigate whether these trends could be due to systematic effects of varying resolution in our magnetic maps.  

\begin{figure*}
  \centering
  \includegraphics[width=3.4in]{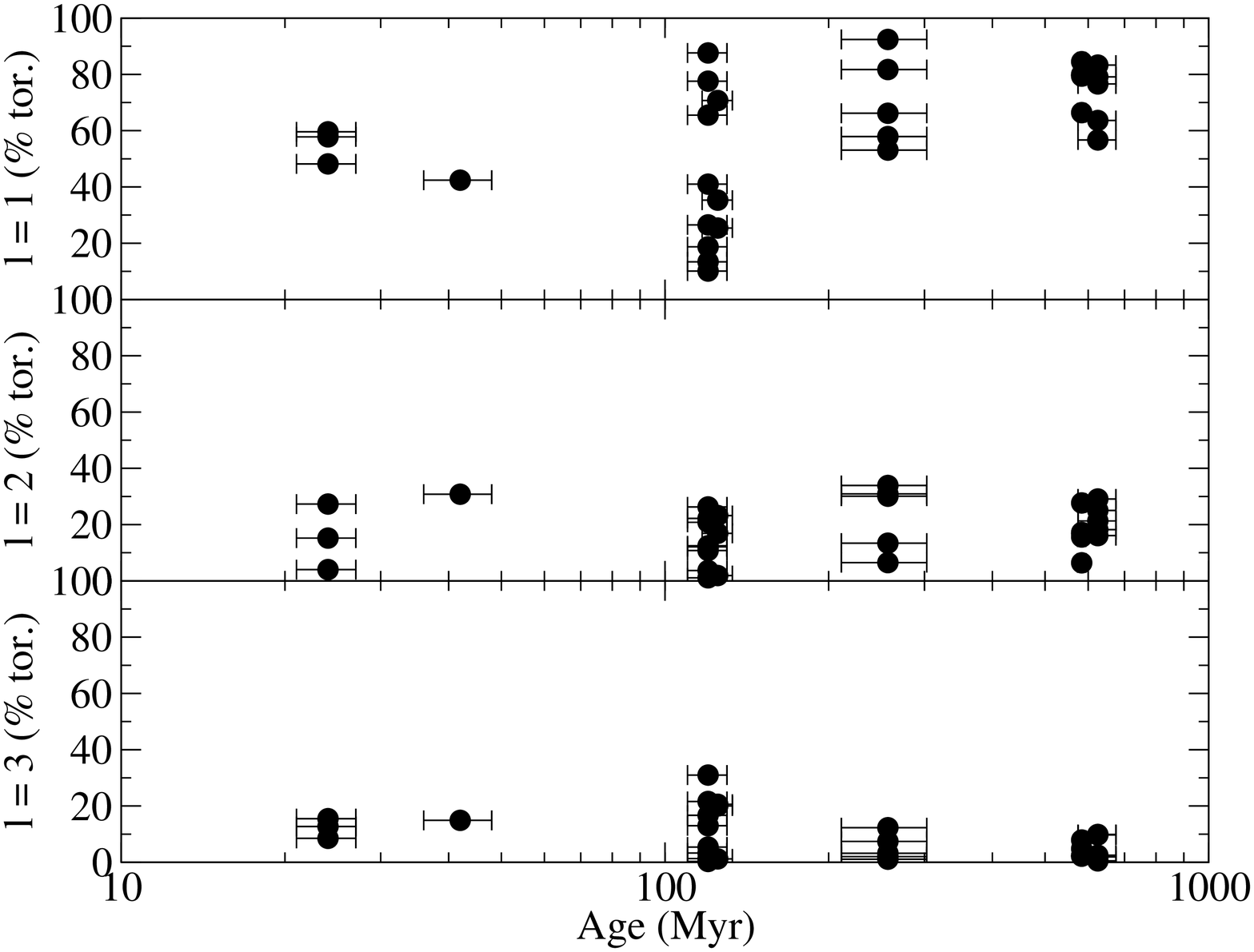}
  \includegraphics[width=3.4in]{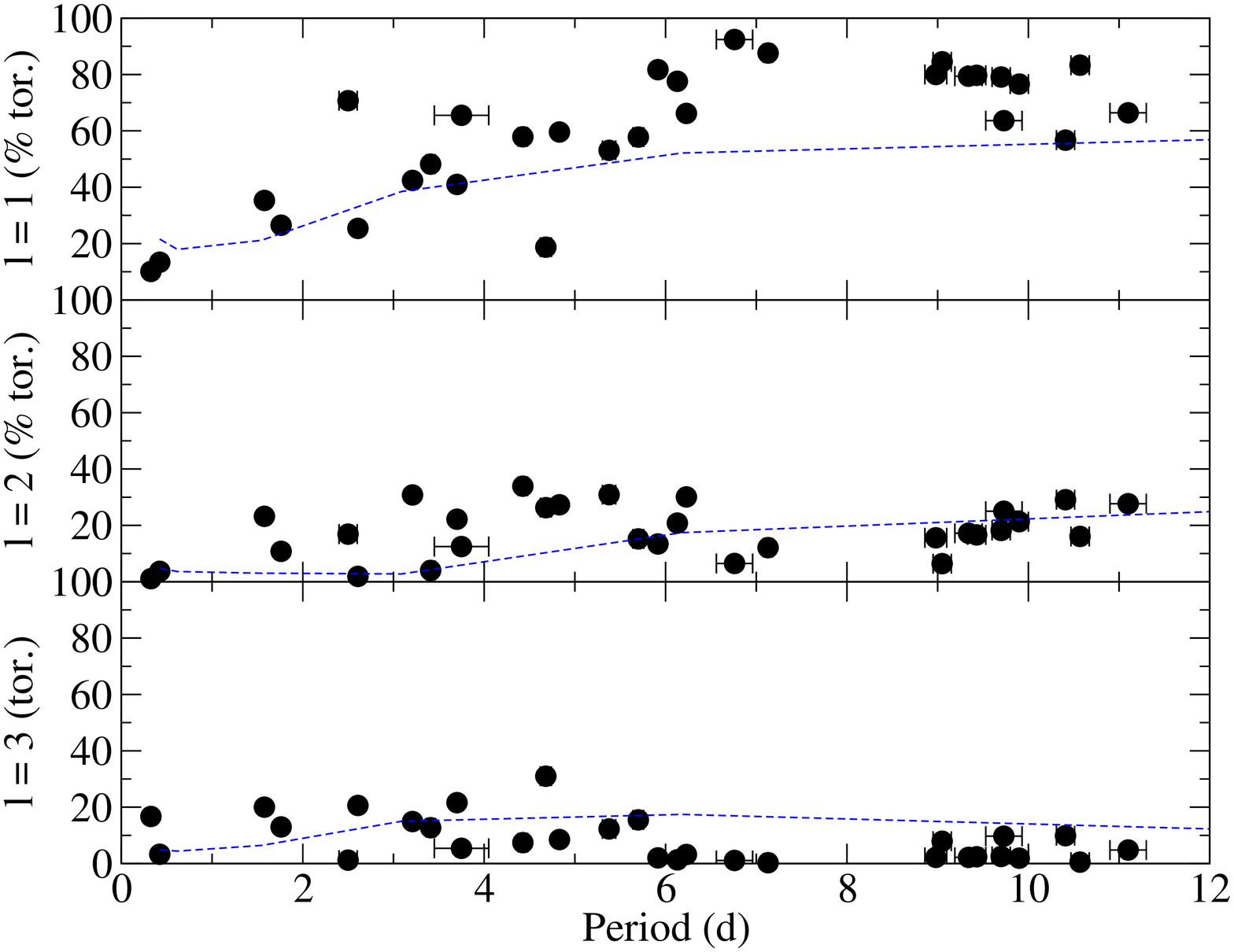}
  \caption{Fraction of toroidal magnetic energy in the $l=1$, 2, and 3 spherical harmonics.  Dashed lines are ZDI reconstructions of models based on the same magnetic geometry, reconstructed at different \vs. }
  \label{fig-trend-tor-l123}
\end{figure*}

Within the sample of stars in this paper and Paper I there is a trend towards the toroidal component of the magnetic field being dominantly in the $l=1$ spherical harmonic for older stars (see fig.~\ref{fig-trend-tor-l123}).  However, there is also a strong trend towards stars with long rotation periods having their toroidal field dominantly in the $l=1$ component.  This may well be an effect of lower resolution for the slower rotators, which would cause less of the magnetic field in higher order harmonics to be seen.  The fraction of the toroidal magnetic field in the $l=3$ component decreases with rotation period, which seems to support this.
It is well established, since the earliest studies of Doppler imaging in general and ZDI in particular, that the resolution of a map decreases as \vs\ decreases.  The ratio of \vs\ to the combination of the instrumental resolution and local line profile width effectively controls the number of resolution elements across a map.  Lower S/N can also lead to a lower resolution map, since small spacial scale features usually produce small features in line profiles, which get lost more easily as S/N decreases.  However, S/N is less of a concern for us, since S/N is roughly comparable across our study, and it does not correlate strongly with physical parameters of the stars. 

In order to investigate the impact of varying resolution in ZDI maps of stars with different \vs, we conducted a set of synthetic tests, reconstructing the same magnetic geometry using different model \vs\ values.  For this we used the magnetic geometry and strength of our map for LO Peg.  While this likely does not represent the magnetic fields of all stars, it is one of the highest resolution maps we have, and it has energy distributed over a wide range of spherical harmonic components.  Synthetic line profiles were generated from this map, at the phases of the real observations of LO Peg, and the inclination of LO Peg, but using a range of \vs\ (73, 50, 20, 10, 5, and 2 \kms).  Synthetic Gaussian noise was then added to the profiles at a level of $1.8\times10^{-4}$ of the continuum, matching the typical noise level in the observed LSD profiles of LO Peg.  We then used these synthetic profiles as input to ZDI, to reconstruct the magnetic map with degraded resolution.  The line profiles were fit to a $\chi_{\nu}^{2}$ of 1.0.  For comparison with other stars, the different \vs\ values were converted to periods and Rossby numbers, by assuming the same inclination, radius, and convective turnover time as LO Peg.  We used this comparison when considering trends with rotation rate or Rossby number in this section.

For the trend of toroidal energy in $l=1$ spherical harmonics, we find our synthetic test models have a significant trend towards larger fractions in this harmonic degree for longer periods.  Since the input magnetic geometry for the synthetic models was constant, this represents a systematic error.  Our models roughly follow the lower bound of the observed factions of toroidal energy in $l=1$, which suggests this trend in the observed stars may be largely a systematic consequence of varying resolution.  

\begin{figure*}
  \centering
  \includegraphics[width=3.4in]{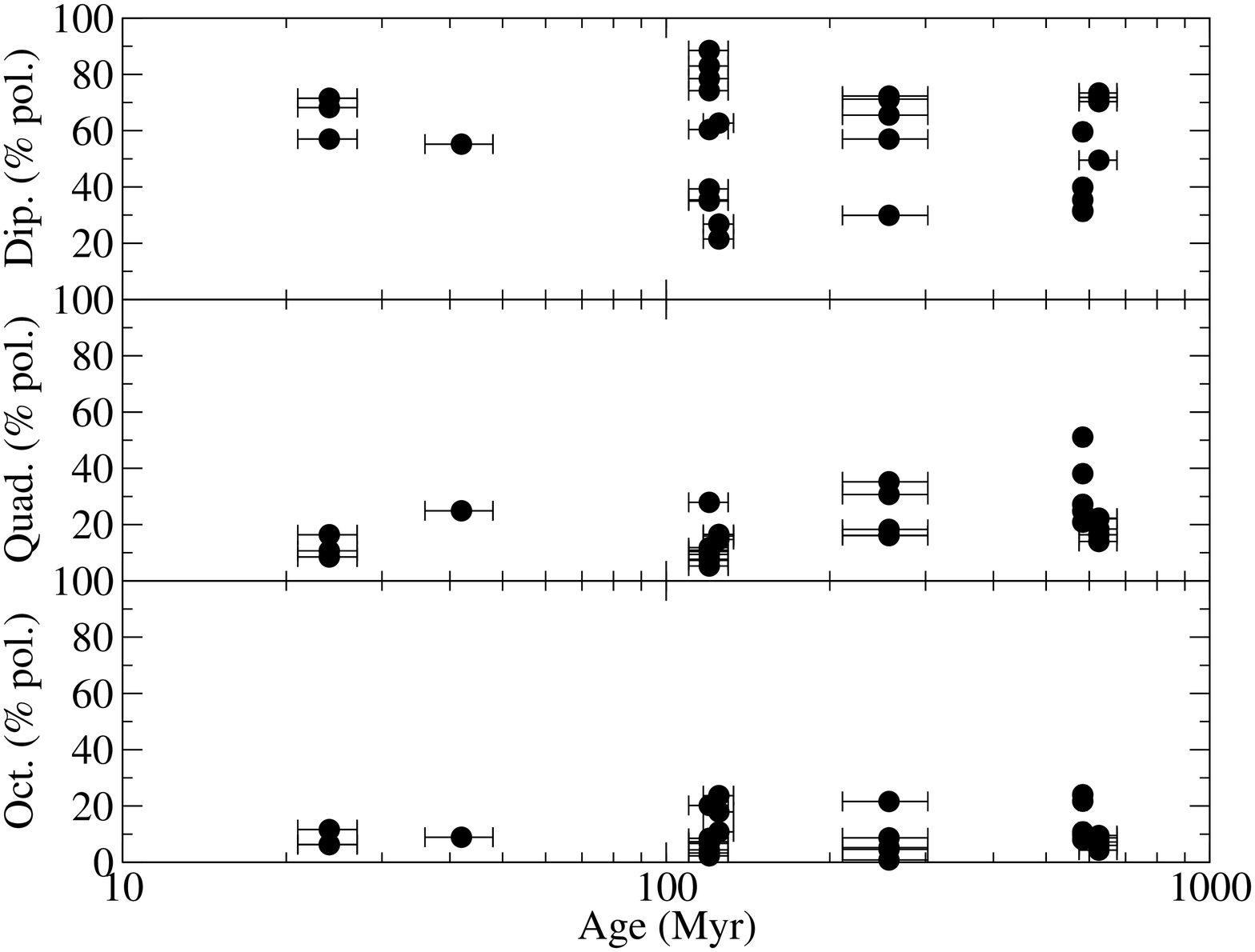}
  \includegraphics[width=3.4in]{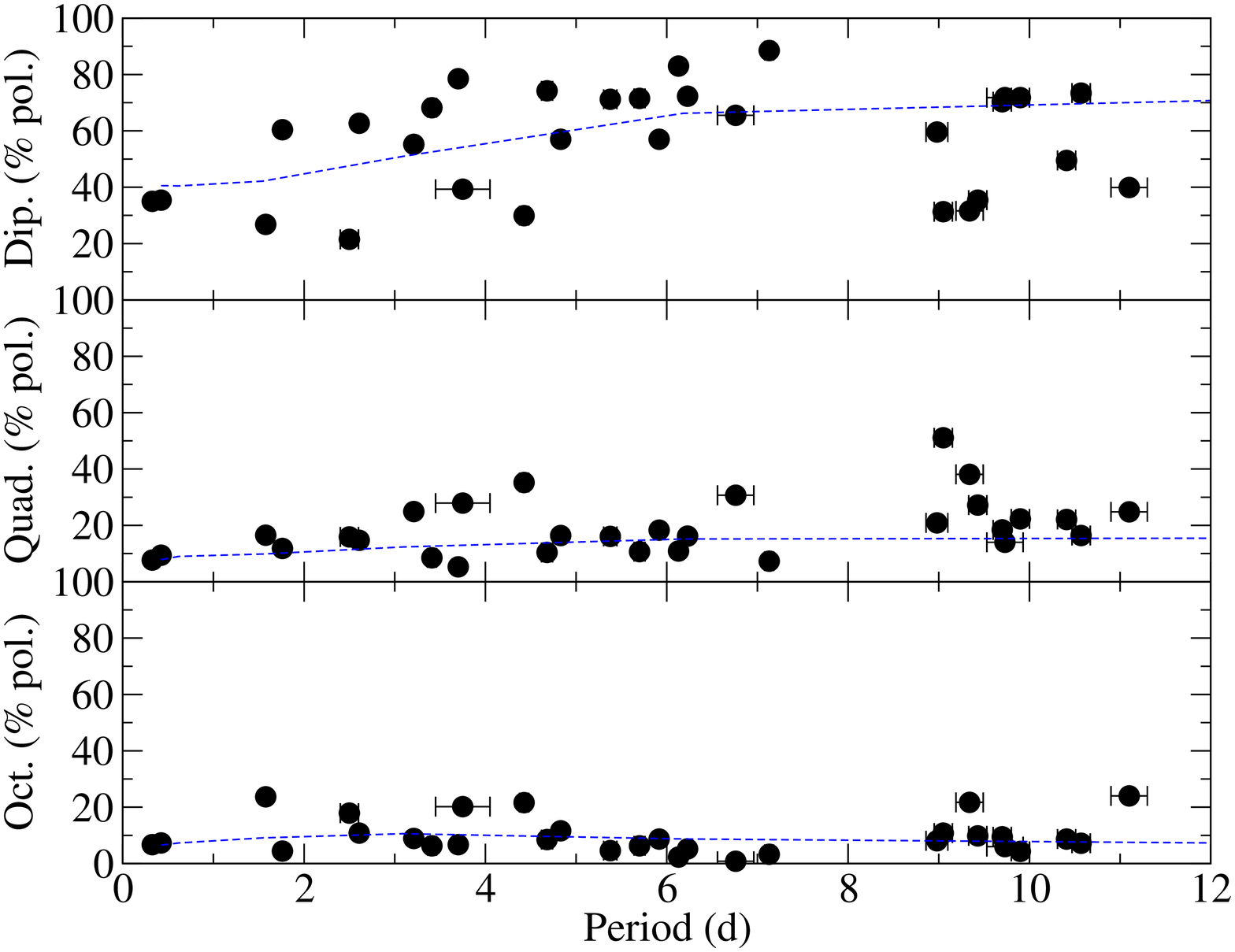}
  \caption{Fraction of poloidal magnetic energy in the dipole ($l=1$), quadrupole ($l=2$), and octupole ($l=3$) components.  Dashed lines are ZDI reconstructions of models based on the same magnetic geometry, reconstructed at different \vs.  }
  \label{fig-trend-pol-l123}
\end{figure*}

There is no clear trend for the fraction of poloidal magnetic energy in the dipolar component varying with rotation period, Rossby number, or age.  Similarly, there is no clear correlation with the fraction of energy in the octupolar component.  However, there is a weak trend towards an increasing quadrupolar fraction towards older ages, slower rotation periods, and larger Rossby numbers (shown in Fig.~\ref{fig-trend-pol-l123}).  This does not appear to be a consequence of changing resolution, since it does not impact the dipole or octupole components, nor does this trend appear in our synthetic tests.  However, a physical cause for this is not obvious, and the correlation is not particularly strong.  

\begin{figure*}
  \centering
  \includegraphics[width=5.0in]{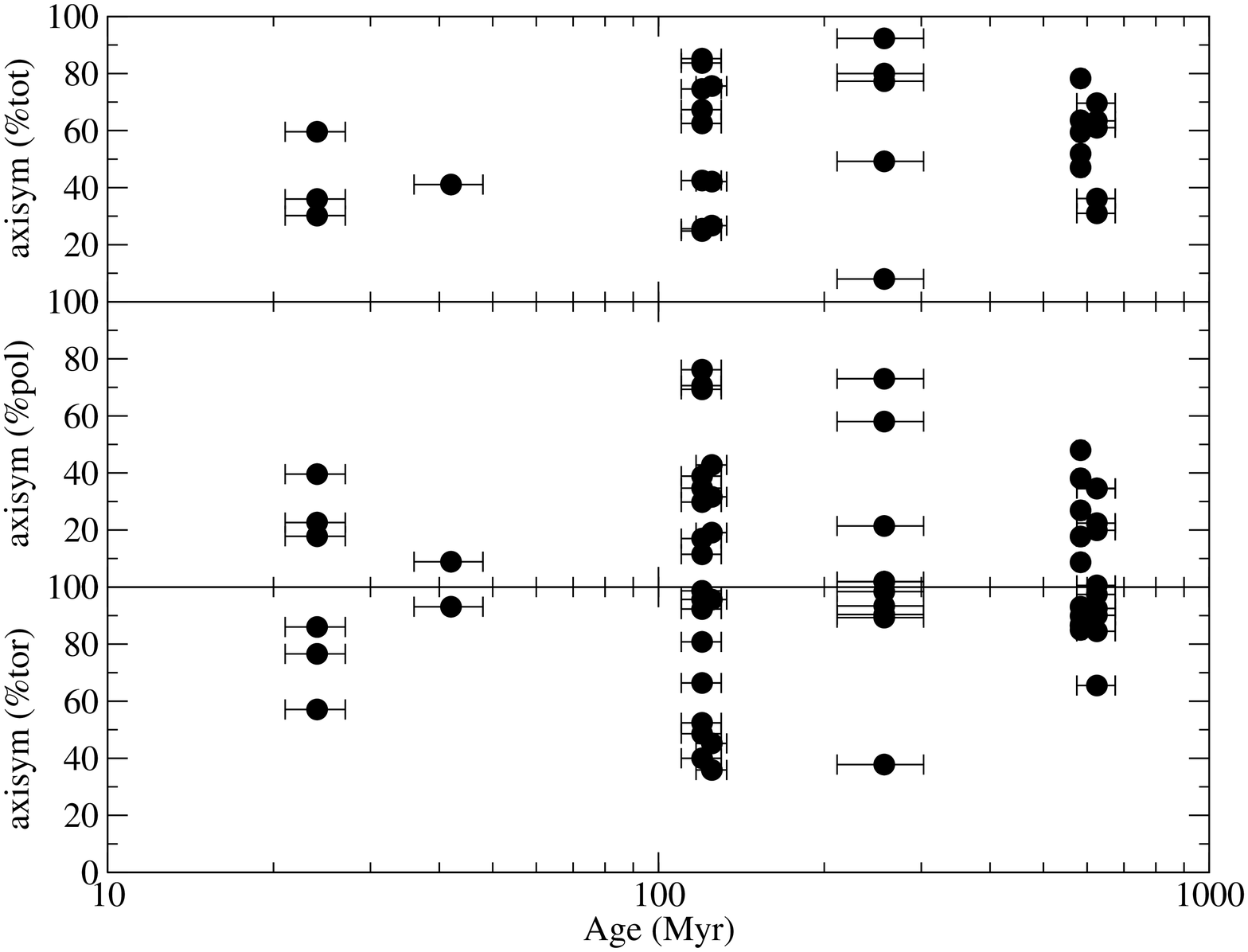}
  \includegraphics[width=5.0in]{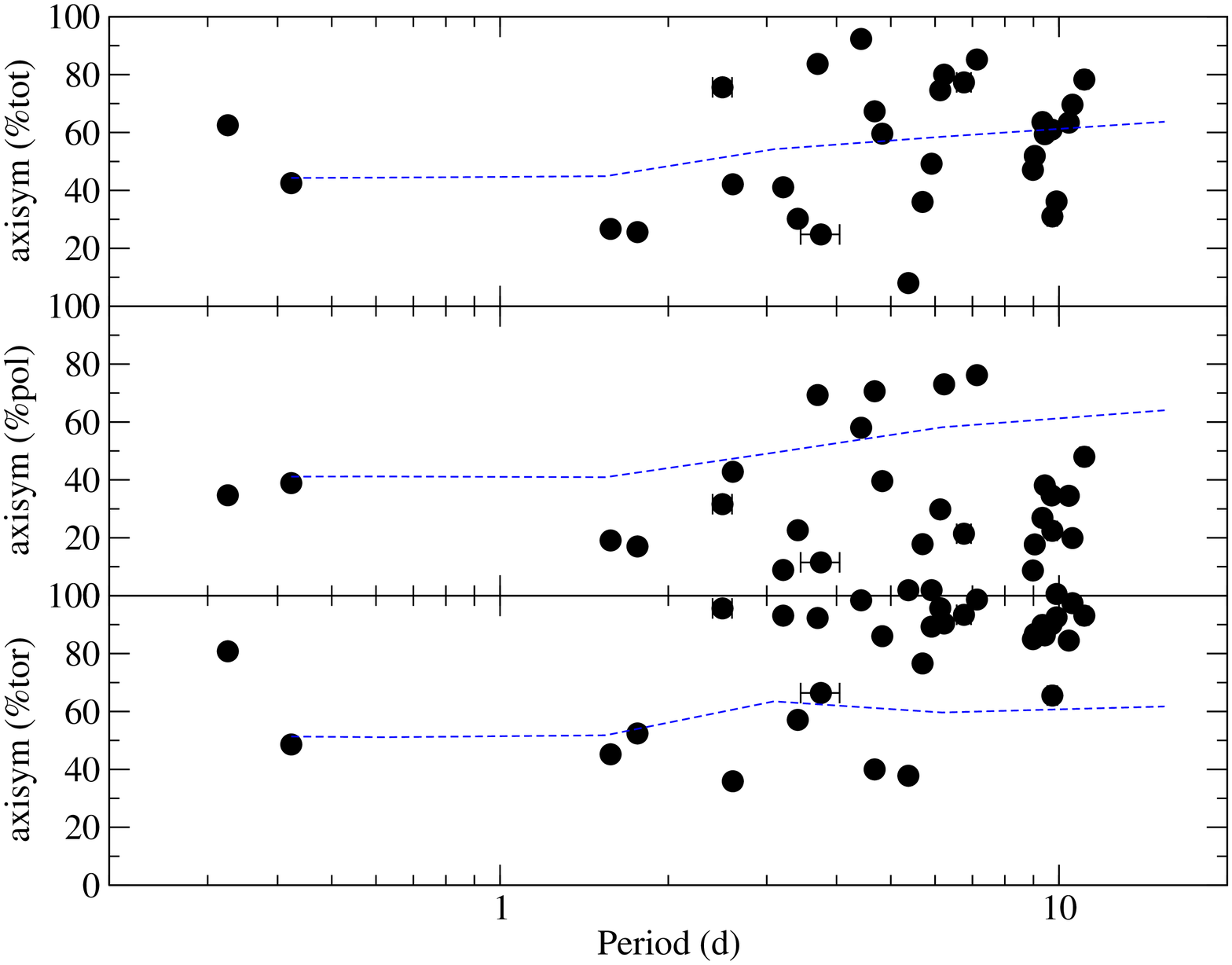}
  \caption{Axisymmetry of the total magnetic field (\%tot), poloidal magnetic field (\%pol) and toroidal magnetic field (\%tor), as a function of age (top) and rotation period (bottom).  The dashed line represents ZDI reconstructions of models, using the magnetic geometry, but at different \vs\ and hence rotation periods. }
  \label{fig-trend-axisym}
\end{figure*}

There is a general trend towards the oldest stars (>300 Myr) having dominantly axisymmetric toroidal fields, while the younger stars have a wider range of toroidal axisymmetries (see Fig~\ref{fig-trend-axisym}).  The poloidal field components have no clear trend in axisymmetry, but may be slightly more non-axisymmetric in the oldest stars in the sample.  The axisymmetry of the full field does not show a strong trend in age, apart from being less scattered for the oldest stars.  These trends in axisymmetry do not appear clearly in Rossby number, and thus may be age driven, although they are not strong.  However, the trend towards more axisymmetric toroidal fields is also seen for the longest rotation periods (see Fig~\ref{fig-trend-axisym}).  This raises the possibility that this is driven by varying resolution in the ZDI maps, thus we investigated this using the set of synthetic tests with the magnetic geometry of LO Peg.  In these tests no strong trend in axisymmetry was found when varying \vs\ and period, illustrated in Fig~\ref{fig-trend-axisym}.  This implies that we still have some sensitivity to non-axisymmetric toroidal fields at slow rotation rates, and suggests that this is not simply a systematic of the ZDI reconstruction. 

The two very fast rotators, LO Peg and BD-072388 both show largely axisymmetric dipolar components to their magnetic field.  The rest of the sample shows a wide dispersion in the axisymmetry of the dipolar components, and no correlations with age, rotation period, or Rossby number.  While this would be interesting if supported by further observations of very fast rotators in the saturated regime, with only two stars we cannot yet draw any conclusions.

\begin{figure*}
  \centering
  \includegraphics[width=5.0in]{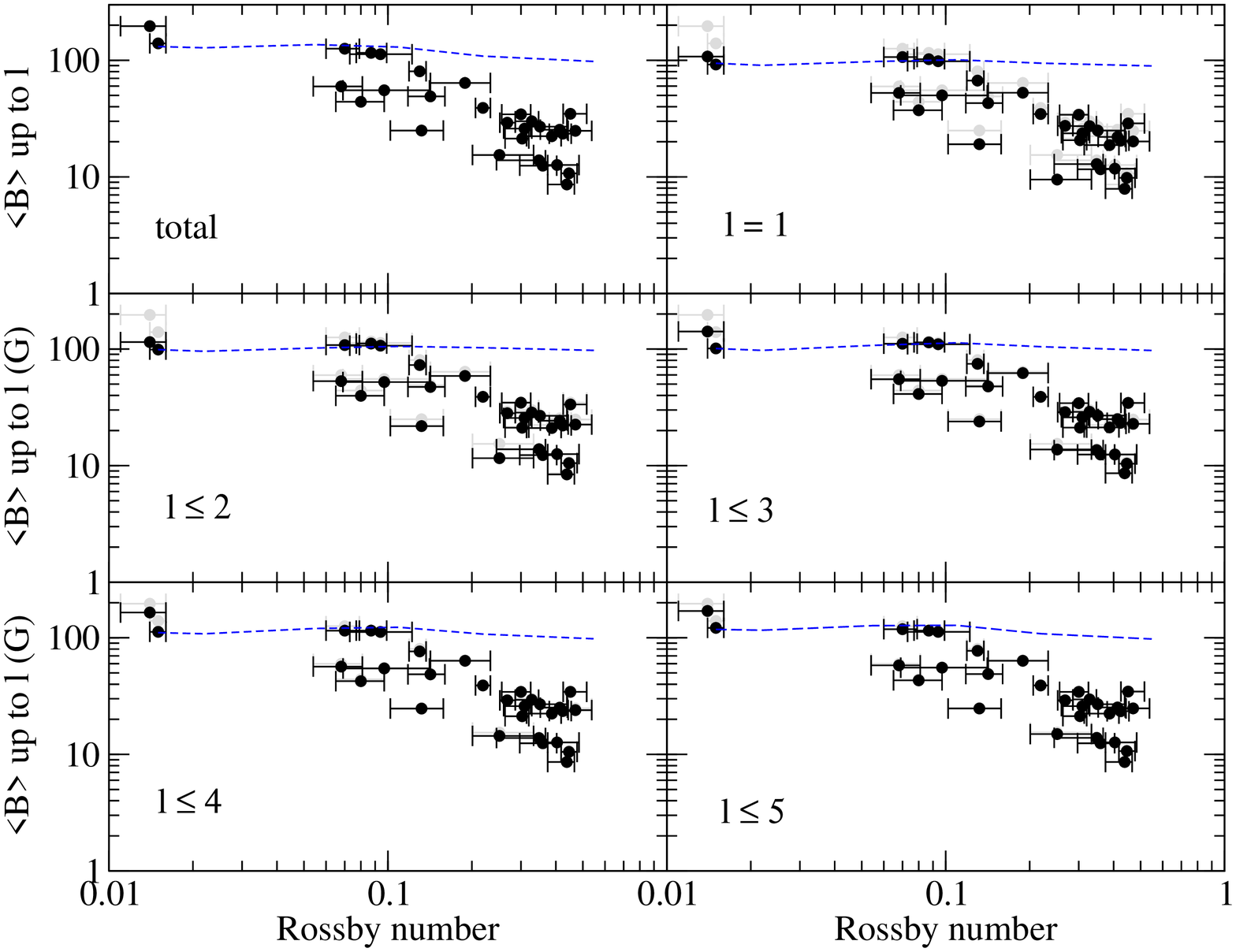}
  \caption{Unsigned magnetic field averaged over the surface of the star, evaluated from the lowest $l$ degree spherical harmonics, as a function of Rossby number.
    Grey points are \Bmean\ evaluated for the full magnetic map.  
    Dashed lines are models using the same magnetic geometry reconstructed at different \vs, and hence Rossby number. }
  \label{fig-trend-syn-tst}
\end{figure*}

In order to assess the impact of varying spacial resolution in the ZDI maps on the magnetic field strength we derive, we calculated the unsigned magnetic field strength, averaged over the surface, using limited spherical harmonic degrees.  We calculated \Bmean\ for just the $l=1$ harmonics, then the $l=1$ and 2 harmonics, and so on, varying the limiting $l_{max}$ up to 5.  This provides \Bmean\ values using only a fraction of the available information, limited by spacial resolution, plotted in Fig.\ \ref{fig-trend-syn-tst}.

The general trends in \Bmean\ with Rossby number are apparent across all limiting $l_{max}$.  Both the saturation at Rossby numbers much below 0.1, and the power law decrease in \Bmean\ with Rossby number are reproduced, even for $l_{max} = 1$.  The \Bmean\ value of BD-072388 and LO Peg are somewhat lower relative to the rest of the sample for $l_{max} = 1$ and 2.  This reflects the stars having an unusually low fraction of their magnetic energy in the two lowest degree spherical harmonics, however this is not enough to change the broad trends.

To further assess the systematic impact of varying resolution, we also include the synthetic tests using the magnetic geometry of LO Peg, reconstructed at different \vs\ and hence different Rossby number.  For the full magnetic field geometry, we see a very slight trend of decreasing \Bmean\ with increasing Rossby number.  This reflects the magnetic field in smaller spacial scales be lost as the resolution decreases.  However, this is a much weaker trend than that seen for for real sample of stars, implying that the decrease in observed \Bmean\ with increasing Rossby number is real and largely not a consequence of decreasing resolution.  When the magnetic field is evaluated using just the lowest couple degree spherical harmonics, there is essentially no trend in the synthetic LO Peg test values.  This indicates that the varying spacial resolution has a negligible impact on magnetic field in the lowest $l$ degree spherical harmonics, as expected.

\begin{figure*}
  \centering
  \includegraphics[width=5.0in]{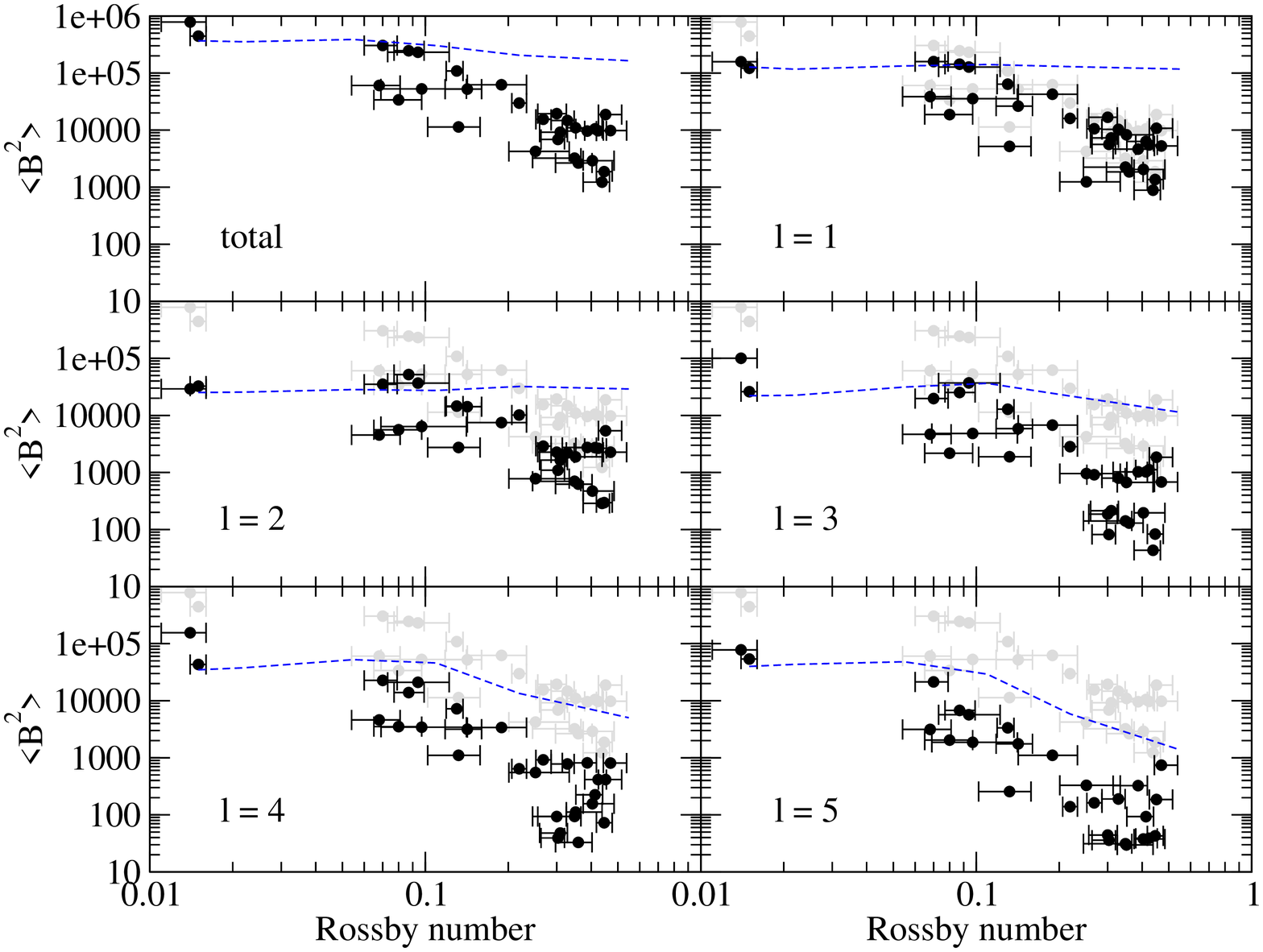}
  \caption{
    Magnetic field squared (proportional to magnetic energy), averaged over the star, evaluated from spherical harmonics only in degree $l$.
    Grey points are $\langle B^2 \rangle$ evaluated for the full magnetic map.  
    Dashed lines are ZDI reconstructions of models, calculated at different \vs\ and periods, based on the same magnetic geometry.  }
  \label{fig-trend-syn-tst2}
\end{figure*}

Another way of assessing the spacial distribution of magnetic field, and the impact of spacial resolution on the magnetic field, is to look at the averaged $B^2$ contained in each spherical harmonic degree.
The quantity $\langle B^2 \rangle$ is $\mathbf{B} \cdot  \mathbf{B}$ (proportional to the magnetic energy), averaged over the stellar surface (i.e. integrated over the surface, divided by surface area).  $\langle B^2 \rangle$ is convenient for this, since the total $\langle B_{tot}^2 \rangle$ is a sum of the values in individual degrees $\langle B_l^2 \rangle$.  This is plotted against Rossby number in Fig.\ \ref{fig-trend-syn-tst2}.  

The general trend of decreasing $\langle B^2 \rangle$ is seen across the sample for all $l$ degrees.  The slope of the decrease appears to steepen for higher $l$ degrees, above 3, and the dispersion of values at large Rossby number also increases for these higher degrees.  The change in slope and increased dispersion are likely both systematic effects of decreasing resolution.  At lower \vs, and larger Rossby number, we likely have difficulty resolving higher degree harmonics, and the extent to which we can partially reconstruct them becomes more sensitive to S/N and phase coverage.  

The synthetic tests based on LO Peg's magnetic geometry support the conclusion that the increase in slope for higher degree harmonics is a systematic effect.  For $l=1$ and 2, there is no change in the reconstructed $\langle B^2 \rangle$ with \vs, and for $l=3$ the trend is quite weak.  However from $l=4$ the reconstructed $\langle B^2 \rangle$ is lower for larger Rossby numbers (lower \vs).  Some of the general decrease in $\langle B^2 \rangle$ with $R_o$ still seems to be real in the $l=5$ harmonics, since $\langle B^2 \rangle$ for the real observations decreases with Rossby number faster than the synthetic LO Peg curve.  Even in the $l=6$ harmonics, the synthetic LO Peg curve remains consistently above the larger Rossby number, suggesting that there is still a small real trend, although the systematic trend becomes strong.

\section{Long term magnetic variability}
\label{Long term magnetic variability}

In order to characterize the long term (multi-year) variability of the large-scale magnetic field strengths seen in stars, we consider several multi-epoch ZDI studies of main sequence F, G and K stars.
A summary of this brief literature review is provided in Table \ref{table-multi-epoch-zdi}, with references.  
Summaries of much of this work are in \citet{Vidotto2014-magnetism-age-rot}, \citet{Vidotto2016-mag-geom-for-wind-activity}, and \citet{See2016-mag-geom-cycles-PolTor-Rossby}.  Many of the long term observations are available thanks to the BCool collaboration.  
The high resolution spectropolarimetric observations of these stars, and the analysis with ZDI, is comparable to the results we present in this paper.  
The results summarized here are for G, K, and a few late F stars, that are on or near the main sequence.  This sample covers a wider range of stellar parameters than our more focused sample, however they should have qualitatively similar internal structure, and should have solar-like dynamos operating.
The data sets all span greater than 6 months, up to 9 years in the longest case.  
In some cases we take the mean radial magnetic field from \citet{Vidotto2014-magnetism-age-rot} rather than the original references, as in some original references this quantity was not reported, and this provides a more uniform evaluation of this quantity. 

For each star, we consider the range of variability in the mean magnetic field strength from the ZDI maps, i.e. the difference between the maximum and minimum reported value (based on the absolute value of the magnetic field, i.e.\ `unsigned', averaged over the surface).  This amplitude of variability changes greatly from star to star, from less than 1 G to over 100 G.  We also consider the average value from the set of measurements.  Given the small sample sizes, the averages are not very robust, but a high degree of precision is not needed here.  There is a clear correlation between the amplitude of magnetic field variability and the average magnetic field strength.  However the ratio of the amplitude of variability to the mean field is much more consistent, from 0.2 (only in cases with poor time sampling) to 1.2.  Thus even in the cases with the largest fractional amplitude, the minimum and maximum $\langle B \rangle$ are only a bit less than half or a bit more than 1.5 times the average value.

This range of variability (from Table \ref{table-multi-epoch-zdi}) can be compared with the scatter in our Rossby number $\langle B \rangle$ relation, to estimate how much of the scatter could be due to long term variability.  The residuals for this relation are plotted in Fig \ref{fig-residualB-age}.  The absolute value of the residuals are in many cases larger than the ranges of variability in the mutli epoch stars, however the mean magnetic fields of our stars are on average larger.  The fractional residuals are generally comparable to the fractional variability of the multi-epoch stars.  
If we suppose that the scatter in our Rossby number $\langle B \rangle$ relation is only due to long term variability, then the residuals should correspond to the semi-amplitude of variability.  Averaging over the full sample of Toupies stars so far, the average fractional residual is 0.38 (0.37 without the saturated regime stars), and the standard deviation is 0.21.  The average of half the fractional amplitude in the multi-epoch stars is 0.32 (standard deviation 0.14).  Thus it seems possible that most of the dispersion in our Rossby number $\langle B \rangle$ relation could be explained by long term variability.  However, monitoring of more stars over longer time periods, together with single epoch observations of a larger sample of stars, is needed to robustly test this hypothesis.

\begin{table*}
\centering
\caption{Multi-epoch ZDI studies of F, G, and K stars near the main sequence.  The average of the multi-epoch set of mean magnetic field values, the range of variability in the mean magnetic field values (max-min), and the ratio of these quantities are presented.  
}
\begin{tabular}{lcccccl}
\hline
Star & Time-span & Num.   & Avg.          & Range        & Frac. & References \\
     &           & Epochs & $\langle B \rangle$ (G) & $\langle B \rangle$ (G) & var.  &       \\
\hline
HD 78366	& 2008-2011 & 3  &  5.2   &  5.0   &  0.833  &  \citep{Morgenthaler2011-cycles-ZDI} \\
$\xi$ Boo A 	& 2007-2011 & 7  &  13.9  &  10.4  &  0.694  &  \citep{Morgenthaler2012-xiBooA-ZDI-monitoring} \\
HD 190771 	& 2007-2010 & 4  &  7.8   &  8.9   &  0.994  &  \citep{Petit2009-ZDI-cycle-HD190771, Morgenthaler2011-cycles-ZDI} \\
HD 35296 	& 2007-2008 & 2  &  16.6  &  6.3   &  0.381  &  \citep{Waite2015-2-young-solar-B} \\
EK Dra  	& 2006-2012 & 5  &  75.2  &  38    &  0.521  &  \citep{Waite2017-young-solar-EKDra} \\
NZ Lup  	& 2007-2010 & 3  &  36.6  &  18    &  0.489  &  \citep{Marsden2011-hd141943-sempol-hd101412} \\
$\tau$ Boo	& 2007-2015 & 10 &  2.5   &  2.7   &  1.059  &  \citep{Fares2009-ZDI-tauBoo-reversal, Mengel2016-tauBoo-mag-update} \\
HD 179949	& 2007-2009 & 2  &  2.0   &  0.86  &  0.439  &  \citep{Fares2012-ZDI-HD179949-planet-host} \\
HD 189733	& 2006-2015 & 9  &  31.6  &  24    &  0.800  &  \citep{Fares2010-ZDI-HD189733-planet-host, Fares2017-ZDI-HD189733-planet-host-monitoring} \\
61 Cyg A	& 2007-2015 & 6  &  8.2   &  9     &  1.200  &  \citep{BoroSaikia2016-61CygA-solar-like} \\
HN Peg  	& 2007-2013 & 6  &  17.5  &  13    &  0.743  &  \citep{BoroSaikia2015-HNPeg} \\
$\epsilon$ Eri 	& 2007-2013 & 6  &  14.3  &  10    &  0.667  &  \citep{Jeffers2014-epsEri-mag-var} \\
II Peg  	& 2012-2013 & 2  &  364   &  148   &  0.407  &  \citep{Rosen2015ApJ-ZDI-IIPeg-4Stokes} \\
$\chi^1$ Ori	& 2007-2011 & 4  &  16.0  &  7     &  0.424  &  \citep{Rosen2016-mag-young-solar-twins} \\
$\kappa^1$ Cet	& 2012-2013 & 2  &  23.5  &  5     &  0.213  &  \citep{Rosen2016-mag-young-solar-twins} \\
\hline
\end{tabular} 
\label{table-multi-epoch-zdi} 
\end{table*}

\bsp	
\label{lastpage}
\end{document}